\documentclass[12pt]{article}
\usepackage{amsmath,amssymb}
\usepackage{graphicx}
\newcommand{\eqdef}{\stackrel{\text{def}}{=}}
\newcommand{\eqdefrm}{\stackrel{\text{\rm def}}{=}}
\newcommand{\n}{\nonumber\\}
\newcommand{\bm}{\boldsymbol}

\newcommand{\ignore}[1]{}
\numberwithin{equation}{section}
\newcommand{\Romannumeral}[1]{\uppercase\expandafter{\romannumeral#1}}
\newcommand{\I}{\text{\Romannumeral{1}}}
\newcommand{\II}{\text{\Romannumeral{2}}}
\newcommand{\III}{\text{\Romannumeral{3}}}
\newcommand{\IV}{\text{\Romannumeral{4}}}
\newcommand{\V}{\text{\Romannumeral{5}}}
\newtheorem{thm}{\bf Theorem}
\newtheorem{conj}{\bf Conjecture}
\newtheorem{prop}{\bf Proposition}
\newtheorem{lemma}{\bf Lemma}

\allowdisplaybreaks[4]

\begin{document}

\baselineskip=20pt

\newfont{\elevenmib}{cmmib10 scaled\magstep1}
\newcommand{\preprint}{
    \begin{flushright}\normalsize \sf
     DPSU-18-1\\
   \end{flushright}}
\newcommand{\Title}[1]{{\baselineskip=26pt
   \begin{center} \Large \bf #1 \\ \ \\ \end{center}}}
\newcommand{\Author}{\begin{center}
   \large \bf Satoru Odake \end{center}}
\newcommand{\Address}{\begin{center}
     Faculty of Science, Shinshu University,\\
     Matsumoto 390-8621, Japan
   \end{center}}
\newcommand{\Accepted}[1]{\begin{center}
   {\large \sf #1}\\ \vspace{1mm}{\small \sf Accepted for Publication}
   \end{center}}

\preprint
\thispagestyle{empty}

\Title{Recurrence Relations of\\
the Multi-Indexed Orthogonal Polynomials $\V$ :\\
Racah and $q$-Racah types}

\Author

\Address
\vspace{1cm}

\begin{abstract}
In previous papers, we discussed the recurrence relations of the
multi-indexed orthogonal polynomials of the Laguerre, Jacobi, Wilson and
Askey-Wilson types.
In this paper we explore those of the Racah and $q$-Racah types.
For the $M$-indexed ($q$-)Racah polynomials, we derive
$3+2M$ term recurrence relations with variable dependent coefficients
and $1+2L$ term ($L\geq M+1$) recurrence relations with
constant coefficients.
Based on the latter, the generalized closure relations and
the creation and annihilation operators of the quantum mechanical systems
described by the multi-indexed ($q$-)Racah polynomials are obtained.

In appendix we present a proof and some data of the recurrence relations
with constant
coefficients for the multi-indexed Wilson and Askey-Wilson polynomials.
\end{abstract}

\section{Introduction}
\label{intro}

Ordinary orthogonal polynomials in one variable are characterized
by the three term
recurrence relations and those satisfying second order differential or
difference equations are severely restricted by Bochner's theorem and its
generalizations \cite{bochner,szego}.
The exceptional and multi-indexed orthogonal polynomials
$\{\mathcal{P}_n(\eta)|n\in\mathbb{Z}_{\geq 0}\}$
\cite{gkm08}--\cite{detmiop} are new types of orthogonal polynomials.
They satisfy second order differential or difference equations and
form a complete set of orthogonal basis in an appropriate Hilbert space
in spite of missing degrees.
This degree missing is a characteristic feature of them.
Instead of the three term recurrence relations, they satisfy some recurrence
relations with more terms \cite{stz10}--\cite{rrmiop4}, and the constraints
by Bochner's theorem are avoided.
We distinguish the following two cases;
the set of missing degrees $\mathcal{I}=\mathbb{Z}_{\geq 0}\backslash
\{\deg\mathcal{P}_n|n\in\mathbb{Z}_{\geq 0}\}$ is
case-(1): $\mathcal{I}=\{0,1,\ldots,\ell-1\}$, or
case-(2): $\mathcal{I}\neq\{0,1,\ldots,\ell-1\}$, where $\ell$ is a positive
integer. The situation of case-(1) is called stable in \cite{gkm11}.
Our approach to orthogonal polynomials is based on the quantum mechanical
formulations: ordinary quantum mechanics (oQM), discrete quantum mechanics
with pure imaginary shifts (idQM) \cite{os13}--\cite{os24} and discrete
quantum mechanics with real shifts (rdQM) \cite{os12}--\cite{os34}.
The Askey scheme of the (basic) hypergeometric orthogonal polynomials
\cite{kls} is well matched to these quantum mechanical formulations:
the Jacobi polynomial etc.\ in oQM, the Askey-Wilson polynomial etc.\ in
idQM and the $q$-Racah polynomial etc.\ in rdQM.
A new type of orthogonal polynomials are obtained by applying the Darboux
transformations with appropriate seed solutions to the exactly solvable
quantum mechanical systems described by the classical orthogonal polynomials
in the Askey scheme.
When the virtual state wavefunctions are used as seed solutions, the case-(1)
multi-indexed orthogonal polynomials are obtained \cite{os25,os27,os26}.
When the eigenstate and/or pseudo virtual state wavefunctions are used as
seed solutions, the case-(2) multi-indexed orthogonal polynomials are
obtained \cite{os29}--\cite{casoidrdqm}.

In previous papers \cite{rrmiop,rrmiop2,rrmiop3,rrmiop4},
the recurrence relations for the case-(1) multi-indexed polynomials
(Laguerre (L) and Jacobi (J) types in oQM, Wilson (W) and Askey-Wilson (AW)
types in idQM) were studied.
There are two kinds of recurrence relations:
with variable dependent coefficients \cite{rrmiop} and
with constant coefficients \cite{rrmiop2,rrmiop3}.
The recurrence relations with variable dependent coefficients have been proved
for L, J, W and AW types, but those with constant coefficients have been
conjectured
for L, J, W and AW types and proved only for L and J types.

In this paper we explore the recurrence relations for the case-(1)
multi-indexed polynomials of Racah (R) and $q$-Racah ($q$R) types in rdQM.
By similar methods used in idQM case, we derive two kinds of recurrence
relations: with variable dependent coefficients and with constant coefficients.
We present examples of the latter.
Through the process of deriving the recurrence relations with constant
coefficients and their examples, we have noticed that similar techniques can
be applied to W and AW types.
In appendix \ref{app:proofAW} and \ref{app:exAW}, we present a proof and
some explicit form of the
recurrence relations with constant coefficients for the multi-indexed
(Askey-)Wilson polynomials.
The recurrence relations with constant coefficients are closely related to
the generalized closure relations \cite{rrmiop4}.
The generalized closure relations provide the exact Heisenberg operator solution
of a certain operator, from which the creation and annihilation operators of
the system are obtained.

This paper is organized as follows.
In section \ref{sec:miop} the essence of the multi-indexed ($q$-)Racah
polynomials are recapitulated.
In section \ref{sec:rr_var} we derive the recurrence relations with variable
dependent coefficients.
In section \ref{sec:rr_const} we derive the recurrence relations with
constant coefficients and present some explicit examples.
In section \ref{sec:gcr} the generalized closure relations and the creation
and annihilation operators are presented.
Section \ref{sec:summary} is for a summary and comments.
In Appendix \ref{app:data} some basic data of the multi-indexed
($q$-)Racah polynomials are summarized.
In Appendix \ref{app:proofAW} we prove the recurrence relations with constant
coefficients for the multi-indexed (Askey-)Wilson polynomials.
In Appendix \ref{app:exAW} some data of the recurrence relations with constant
coefficients for the multi-indexed (Askey-)Wilson polynomials are presented.

\section{Multi-indexed ($q$-)Racah Orthogonal Polynomials}
\label{sec:miop}

In this section we recapitulate the multi-indexed Racah (R) and
$q$-Racah ($q$R) orthogonal polynomials \cite{os26}.
Various quantities depend on a set of parameters
$\bm{\lambda}=(\lambda_1,\lambda_2,\ldots)$ and their dependence is
expressed like, $f=f(\bm{\lambda})$, $f(x)=f(x;\bm{\lambda})$.
(We sometimes omit writing $\bm{\lambda}$-dependence, when it does not
cause confusion.)
The parameter $q$ is $0<q<1$ and $q^{\bm{\lambda}}$ stands for
$q^{(\lambda_1,\lambda_2,\ldots)}=(q^{\lambda_1},q^{\lambda_2},\ldots)$.

\subsection{($q$-)Racah polynomials}
\label{sec:org_qR}

The set of parameters $\bm{\lambda}=(\lambda_1,\lambda_2,\lambda_3,\lambda_4)$,
its shift $\bm{\delta}$ and $\kappa$ are
\begin{equation}
  \begin{array}{rl}
  \text{R}:&\bm{\lambda\,}=(a,b,c,d),\quad \bm{\delta}=(1,1,1,1),
  \quad\kappa=1,\\[5pt]
  \text{$q$R}:&q^{\bm{\lambda}}=(a,b,c,d),\quad \bm{\delta}=(1,1,1,1),
  \quad\kappa=q^{-1}.
  \end{array}
\end{equation}
For $N\in\mathbb{Z}_{>0}$, we take $n_{\text{max}}=x_{\text{max}}=N$ and
\begin{equation}
  \text{R}:\ a=-N,\qquad
  \text{$q$R}:\ a=q^{-N},
  \label{a=-N}
\end{equation}
and assume the following parameter ranges:
\begin{equation}
  \text{R}:\ 0<d<a+b,\ \ 0<c<1+d,\qquad
  \text{$q$R}:\ 0<ab<d<1,\ \ qd<c<1.
  \label{qRpara}
\end{equation}

The ($q$-)Racah polynomials $P_n(\eta)$ ($n=0,1,\ldots,n_{\text{max}}$) are
\begin{align}
  &\check{P}_n(x;\bm{\lambda})
  \eqdef P_n\bigl(\eta(x;\bm{\lambda});\bm{\lambda}\bigr)=\left\{
  \begin{array}{ll}
  {\displaystyle
  {}_4F_3\Bigl(
  \genfrac{}{}{0pt}{}{-n,\,n+\tilde{d},\,-x,\,x+d}
  {a,\,b,\,c}\Bigm|1\Bigr)}&:\text{R}\\[8pt]
  {\displaystyle
  {}_4\phi_3\Bigl(
  \genfrac{}{}{0pt}{}{q^{-n},\,\tilde{d}q^n,\,q^{-x},\,dq^x}
  {a,\,b,\,c}\Bigm|q\,;q\Bigr)}&:\text{$q$R}
  \end{array}\right.
  \label{cP}\\
  &\phantom{\check{P}_n(x;\bm{\lambda})
  =P_n\bigl(\eta(x;\bm{\lambda});\bm{\lambda}\bigr)}=\left\{
  \begin{array}{ll}
  {\displaystyle
  R_n\bigl(\eta(x;\bm{\lambda});a-1,\tilde{d}-a,c-1,d-c\bigr)}
  &:\text{R}\\[4pt]
  {\displaystyle
  R_n\bigl(\eta(x;\bm{\lambda})+1+d\,;
  aq^{-1},\tilde{d}a^{-1},cq^{-1},dc^{-1}|q\bigr)}&:\text{$q$R}
  \end{array}\right.,\n
  &\quad
  \eta(x;\bm{\lambda})\eqdef\left\{
  \begin{array}{ll}
  x(x+d)&:\text{R}\\[2pt]
  (q^{-x}-1)(1-dq^x)&:\text{$q$R}
  \end{array}\right.,\quad
  \tilde{d}\eqdef\left\{
  \begin{array}{ll}
  a+b+c-d-1&:\text{R}\\[2pt]
  abcd^{-1}q^{-1}&:\text{$q$R}
  \end{array}\right.,
  \label{eta}
\end{align}
where $R_n\bigl(x(x+\gamma+\delta+1);\alpha,\beta,\gamma,\delta\bigr)$ and
$R_n(q^{-x}+\gamma\delta q^{x+1};\alpha,\beta,\gamma,\delta|q)$ are the Racah
and $q$-Racah polynomials in the conventional parametrization \cite{kls},
respectively.
Our parametrization respects the correspondence between the ($q$-)Racah and
(Askey-)Wilson polynomials, and symmetries in $(a,b,c,d)$ are transparent.
Note that the sinusoidal coordinates $\eta(x;\bm{\lambda})$ depend on
parameters $\bm{\lambda}$ only through $d$.
The normalization of $\eta(x)$ and $P_n(\eta)$ is
\begin{equation}
  \eta(0;\bm{\lambda})=0,\quad
  \check{P}_n(0;\bm{\lambda})=P_n(0;\bm{\lambda})=1.
  \label{Pn(0)=1}
\end{equation}
The three term recurrence relations are
\begin{equation}
  \eta P_n(\eta;\bm{\lambda})
  =A_n(\bm{\lambda})P_{n+1}(\eta;\bm{\lambda})
  +B_n(\bm{\lambda})P_n(\eta;\bm{\lambda})
  +C_n(\bm{\lambda})P_{n-1}(\eta;\bm{\lambda}),
  \label{3trr}
\end{equation}
where $A_n$, $B_n$ and $C_n$ are given in \eqref{AnBnCn}.
As a consequence of \eqref{Pn(0)=1}, $B_n$ is equal to $-A_n-C_n$.

In the quantum mechanical formulation \cite{os12}, the polynomials
$P_n(\eta)$ appear in the eigenvectors $\phi_n(x)$,
\begin{equation}
  \phi_n(x;\bm{\lambda})=\phi_0(x;\bm{\lambda})\check{P}_n(x;\bm{\lambda})
  \ \ \Bigl(\begin{array}{ll}
  n=0,1,\ldots,n_{\text{max}}\\
  x=0,1,\ldots,x_{\text{max}}
  \end{array}\Bigr),
\end{equation}
and the orthogonality relations are
\begin{equation}
  \sum_{x=0}^{x_{\text{max}}}\phi_n(x;\bm{\lambda})\phi_m(x;\bm{\lambda})
  =\frac{\delta_{nm}}{d_n(\bm{\lambda})^2}
  \ \ (n,m=0,1,\ldots,n_{\text{max}}),
  \label{ortho}
\end{equation}
where the ground state eigenvector $\phi_0(x)$ and the normalization
constant $d_n(\bm{\lambda})$ are given in \eqref{phi0} and \eqref{dn},
respectively.
The Hamiltonian of this rdQM system is a tridiagonal matrix
$\mathcal{H}=(\mathcal{H}_{x,y})_{0\leq x,y\leq x_{\text{max}}}$,
\begin{equation}
  \mathcal{H}=-\sqrt{B(x)}\,e^{\partial}\sqrt{D(x)}
  -\sqrt{D(x)}\,e^{-\partial}\sqrt{B(x)}+B(x)+D(x),
  \label{H}
\end{equation}
where potential functions $B(x;\bm{\lambda})$ and $D(x;\bm{\lambda})$
are given in \eqref{B,D} and
matrices $e^{\pm\partial}$ are $(e^{\pm\partial})_{x,y}=\delta_{x\pm 1,y}$
and the unit matrix $\bm{1}=(\delta_{x,y})$ is suppressed.
The notation $f(x)Ag(x)$, where $f(x)$ and $g(x)$ are functions of $x$ and
$A$ is a matrix $A=(A_{x,y})$, stands for a matrix whose $(x,y)$-element
is $f(x)A_{x,y}g(y)$.
The Schr\"{o}dinger equation is
\begin{equation}
  \mathcal{H}(\bm{\lambda})\phi_n(x;\bm{\lambda})
  =\mathcal{E}_n(\bm{\lambda})\phi_n(x;\bm{\lambda})
  \ \ (n=0,1,\ldots,n_{\text{max}}),
  \label{Hphin=}
\end{equation}
where the energy eigenvalue $\mathcal{E}_n$ is given in \eqref{En}
($0=\mathcal{E}_0<\mathcal{E}_1<\cdots<\mathcal{E}_{n_\text{max}}$).
By similarity transformation, \eqref{Hphin=} is rewritten as
\begin{align}
  &\widetilde{\mathcal{H}}(\bm{\lambda})
  \eqdef\phi_0(x;\bm{\lambda})^{-1}\circ\mathcal{H}(\bm{\lambda})\circ
  \phi_0(x;\bm{\lambda})
  =B(x;\bm{\lambda})\bigl(1-e^{\partial}\bigr)+
  D(x;\bm{\lambda})\bigl(1-e^{-\partial}\bigr),\\
  &\widetilde{\mathcal{H}}(\bm{\lambda})\check{P}_n(x;\bm{\lambda})
  =\mathcal{E}_n(\bm{\lambda})\check{P}_n(x;\bm{\lambda})
  \ \ (n=0,1,\ldots,n_{\text{max}}),
  \label{tHcPn=}
\end{align}
namely ($q$-)Racah polynomials $\check{P}_n(x)$ satisfy second order
difference equations.
The three term recurrence relations of $P_n(\eta)$ \eqref{3trr} imply
those of the eigenvectors $\phi_n(x)$,
\begin{equation}
  \eta(x;\bm{\lambda})\phi_n(x;\bm{\lambda})
  =A_n(\bm{\lambda})\phi_{n+1}(x;\bm{\lambda})
  +B_n(\bm{\lambda})\phi_n(x;\bm{\lambda})
  +C_n(\bm{\lambda})\phi_{n-1}(x;\bm{\lambda}).
  \label{3trrphi}
\end{equation}

Let $\mathcal{R}$ be the ring of polynomials in $x$ (the Racah case) or the ring
of Laurent polynomials in $q^x$ (the $q$-Racah case). Let us introduce
automorphisms $\mathcal{I}_{\bm{\lambda}}$ in $\mathcal{R}$ by
\begin{equation}
  \mathcal{I}_{\bm{\lambda}}(x)=-x-d \quad :\text{R},\qquad
  \mathcal{I}_{\bm{\lambda}}(q^x)=q^{-x}d^{-1} \quad :q\text{R},
  \label{autom}
\end{equation}
which are involutions $\mathcal{I}_{\bm{\lambda}}^2=\text{id}$.
We have the following lemma \cite{os26}.
\begin{lemma}
If a (Laurent) polynomial $\check{f}$ in $x$ ($q^x$)
is invariant under $\mathcal{I}_{\bm{\lambda}}$, it is a polynomial in the
sinusoidal coordinate $\eta(x;\bm{\lambda})$:
\begin{equation}
  \mathcal{I}_{\bm{\lambda}}\bigl(\check{f}(x)\bigr)=\check{f}(x)
  \ \Leftrightarrow\ \check{f}(x)=f\bigl(\eta(x;\bm{\lambda})\bigr),
  \ \ \text{$f(\eta)$ : a polynomial in $\eta$}.
  \label{invI}
\end{equation}
\label{lem:invI}
\end{lemma}
\vspace*{-8mm}
Note that the involutions $\mathcal{I}_{\bm{\lambda}}$ depend on
parameters $\bm{\lambda}$ only through $d$.

\subsection{Multi-indexed ($q$-)Racah polynomials}
\label{sec:miop_qR}

Let us introduce the twist operation $\mathfrak{t}$ and the twisted shift
$\tilde{\bm{\delta}}$,
\begin{equation}
  \mathfrak{t}(\bm{\lambda})\eqdef
  (\lambda_4-\lambda_1+1,\lambda_4-\lambda_2+1,\lambda_3,\lambda_4),\quad
  \tilde{\bm{\delta}}\eqdef(0,0,1,1).
\end{equation}
Note that $\eta\bigl(x;\mathfrak{t}(\bm{\lambda})\bigr)=\eta(x;\bm{\lambda})$
and $\eta(x;\bm{\lambda}+\beta\tilde{\bm{\delta}})
=\eta(x;\bm{\lambda}+\beta\bm{\delta})$ ($\beta\in\mathbb{R}$).
The virtual state polynomial $\xi_{\text{v}}(\eta)$ is defined as
\begin{equation}
  \check{\xi}_{\text{v}}(x;\bm{\lambda})\eqdef
  \xi_{\text{v}}\bigl(\eta(x;\bm{\lambda});\bm{\lambda}\bigr)\eqdef
  \check{P}_{\text{v}}\bigl(x;\mathfrak{t}(\bm{\lambda})\bigr)
  =P_{\text{v}}\bigl(\eta(x;\bm{\lambda});
  \mathfrak{t}(\bm{\lambda})\bigr).
\end{equation}
Let $\mathcal{D}=\{d_1,d_2,\ldots,d_M\}$
($d_1<d_2<\cdots<d_M$, $d_j\in\mathbb{Z}_{\geq 1}$) be the
multi-index set, which specifies the virtual state vectors used in the
$M$-step Darboux transformations.
(Although this notation $d_j$ conflicts with the notation of the normalization
constant $d_n(\bm{\lambda})$ in \eqref{ortho},
we think this does not cause any confusion
because the latter appears as $\frac{1}{d_n(\bm{\lambda})^2}\,\delta_{nm}$.)
We restrict the parameter range for $M$ virtual states deletion,
\begin{equation}
  \text{R}:\ d+\max(\mathcal{D})+1<a+b,\qquad
  \text{$q$R}:\ ab<dq^{\max(\mathcal{D})+1}.
  \label{Mrange}
\end{equation}
Although these parameter ranges are important for the well-definedness of the
quantum systems, they are irrelevant to the recurrence relations considered in
this paper, which are polynomial equations and valid independent of the
parameter ranges (except for the zeros of the denominators).
So we do not bother about the range of parameters (except for orthogonality
relations, positivity of some quantities and some part of \S\,\ref{sec:gcr}).

The denominator polynomials $\Xi_{\mathcal{D}}(\eta)$ and the multi-indexed
($q$-)Racah polynomials $P_{\mathcal{D},n}(\eta)$
($n=0,1,\ldots,n_{\text{max}}$) are defined as
\begin{align}
  \check{\Xi}_{\mathcal{D}}(x;\bm{\lambda})
  &\eqdef \Xi_{\mathcal{D}}\bigl(\eta(x;\bm{\lambda}+(M-1)\bm{\delta});
  \bm{\lambda}\bigr)\n
  &\eqdef\mathcal{C}_{\mathcal{D}}(\bm{\lambda})^{-1}
  \varphi_M(x;\bm{\lambda})^{-1}
  \det\bigl(\check{\xi}_{d_k}(x_j;\bm{\lambda})\bigr)_{1\leq j,k\leq M},
  \label{XiD}\\
  \check{P}_{\mathcal{D},n}(x;\bm{\lambda})
  &\eqdef P_{\mathcal{D},n}\bigl(\eta(x;\bm{\lambda}+M\bm{\delta});
  \bm{\lambda}\bigr)\n
  &\eqdef\mathcal{C}_{\mathcal{D},n}(\bm{\lambda})^{-1}
  \varphi_{M+1}(x;\bm{\lambda})^{-1}\n
  &\quad\times\left|
  \begin{array}{cccc}
  \check{\xi}_{d_1}(x_1;\bm{\lambda})&\cdots&\check{\xi}_{d_M}(x_1;\bm{\lambda})
  &r_1(x_1)\check{P}_n(x_1;\bm{\lambda})\\
  \check{\xi}_{d_1}(x_2;\bm{\lambda})&\cdots&\check{\xi}_{d_M}(x_2;\bm{\lambda})
  &r_2(x_2)\check{P}_n(x_2;\bm{\lambda})\\
  \vdots&\cdots&\vdots&\vdots\\
  \check{\xi}_{d_1}(x_{M+1};\bm{\lambda})&\cdots
  &\check{\xi}_{d_M}(x_{M+1};\bm{\lambda})
  &r_{M+1}(x_{M+1})\check{P}_n(x_{M+1};\bm{\lambda})\\
  \end{array}\right|,
  \label{PDn}
\end{align}
where $x_j\eqdef x+j-1$ and $r_j(x_j)=r_j(x_j;\bm{\lambda},M)$
($1\leq j\leq M+1$) are given in \eqref{rj}.
The constants $\mathcal{C}_{\mathcal{D}}(\bm{\lambda})$
\eqref{CD} and $\mathcal{C}_{\mathcal{D},n}(\bm{\lambda})$ \eqref{CDn}
correspond to the normalization
\begin{equation}
  \check{\Xi}_{\mathcal{D}}(0;\bm{\lambda})
  =\Xi_{\mathcal{D}}(0;\bm{\lambda})=1,\quad
  \check{P}_{\mathcal{D},n}(0;\bm{\lambda})
  =P_{\mathcal{D},n}(0;\bm{\lambda})=1.
  \label{PDn(0)=1}
\end{equation}
The denominator polynomial $\Xi_{\mathcal{D}}(\eta;\bm{\lambda})$ and the
multi-indexed orthogonal polynomial $P_{\mathcal{D},n}(\eta;\bm{\lambda})$
are polynomials in $\eta$ and their degrees are $\ell_{\mathcal{D}}$ and
$\ell_{\mathcal{D}}+n$, respectively (we assume $c^{\Xi}_{\mathcal{D}}$
\eqref{cXiD} and $c^{P}_{\mathcal{D},n}$ \eqref{cPDn} do not vanish).
Here $\ell_{\mathcal{D}}$ is
\begin{equation}
  \ell_{\mathcal{D}}\eqdef\sum_{j=1}^{M}d_j-\tfrac12M(M-1).
\end{equation}
Note that
\begin{equation}
  \check{P}_{\mathcal{D},0}(x;\bm{\lambda})
  =\check{\Xi}_{\mathcal{D}}(x;\bm{\lambda}+\bm{\delta}),
  \label{PD0=Xi} 
\end{equation}
as a consequence of the shape invariance of the system \cite{os25,os27,os26}.
Other determinant expressions of $\check{P}_{\mathcal{D},n}(x;\bm{\lambda})$
can be found in \cite{detmiop}.

The isospectral deformation is realized by multi-step Darboux
transformations with virtual state vectors as seed solutions.
The multi-indexed polynomials $P_{\mathcal{D},n}(\eta)$ appear in the
eigenvectors $\phi_{\mathcal{D}\,n}(x)$ of the deformed Hamiltonian,
which is also a tridiagonal matrix
$\mathcal{H}_{\mathcal{D}}
=(\mathcal{H}_{\mathcal{D};x,y})_{0\leq x,y\leq x_{\text{max}}}$,
\begin{align}
  &\phi_{\mathcal{D}\,n}(x;\bm{\lambda})
  \eqdef\psi_{\mathcal{D}}(x;\bm{\lambda})
  \check{P}_{\mathcal{D},n}(x;\bm{\lambda})
  \ \ \Bigl(\begin{array}{ll}
  n=0,1,\ldots,n_{\text{max}}\\
  x=0,1,\ldots,x_{\text{max}}
  \end{array}\Bigr),
  \label{phiDn}\\
  &\psi_{\mathcal{D}}(x;\bm{\lambda})\eqdef
  \sqrt{\check{\Xi}_{\mathcal{D}}(1;\bm{\lambda})}\,
  \frac{\phi_0(x;\bm{\lambda}+M\tilde{\bm{\delta}})}
  {\sqrt{\check{\Xi}_{\mathcal{D}}(x;\bm{\lambda})\,
  \check{\Xi}_{\mathcal{D}}(x+1;\bm{\lambda})}},\\
  &\mathcal{H}_{\mathcal{D}}
  =-\sqrt{B_{\mathcal{D}}(x)}\,e^{\partial}\sqrt{D_{\mathcal{D}}(x)}
  -\sqrt{D_{\mathcal{D}}(x)}\,e^{-\partial}\sqrt{B_{\mathcal{D}}(x)}
  +B_{\mathcal{D}}(x)+D_{\mathcal{D}}(x),
  \label{HD}\\
  &\mathcal{H}_{\mathcal{D}}(\bm{\lambda})\phi_{\mathcal{D}\,n}(x;\bm{\lambda})
  =\mathcal{E}_n(\bm{\lambda})\phi_{\mathcal{D}\,n}(x;\bm{\lambda})
  \ \ (n=0,1,\ldots,n_{\text{max}}),
  \label{HDphiDn=}
\end{align}
where potential functions $B_{\mathcal{D}}(x;\bm{\lambda})$ and
$D_{\mathcal{D}}(x;\bm{\lambda})$ are given in \eqref{BD,DD}.
The orthogonality relations are
\begin{equation}
  \sum_{x=0}^{x_{\text{max}}}
  \frac{\psi_{\mathcal{D}}(x;\bm{\lambda})^2}
  {\check{\Xi}_{\mathcal{D}}(1;\bm{\lambda})}
  \check{P}_{\mathcal{D},n}(x;\bm{\lambda})
  \check{P}_{\mathcal{D},m}(x;\bm{\lambda})
  =\frac{\delta_{nm}}{d_{\mathcal{D},n}(\bm{\lambda})^2}
  \ \ (n,m=0,1,\ldots,n_{\text{max}}),
  \label{orthoPDn}
\end{equation}
where the normalization constant $d_{\mathcal{D},n}(\bm{\lambda})$ is
given in \eqref{dDn}.
The normalization of $\psi_{\mathcal{D}}(x)$ and $\phi_{\mathcal{D}\,n}(x)$
is
$\psi_{\mathcal{D}}(0;\bm{\lambda})=\phi_{\mathcal{D}\,n}(0;\bm{\lambda})=1$.
By similarity transformation, \eqref{HDphiDn=} is rewritten as
\begin{align}
  &\widetilde{\mathcal{H}}_{\mathcal{D}}(\bm{\lambda})
  \eqdef\psi_{\mathcal{D}}(x;\bm{\lambda})^{-1}\circ
  \mathcal{H}_{\mathcal{D}}(\bm{\lambda})\circ
  \psi_{\mathcal{D}}(x;\bm{\lambda})\n
  &\phantom{\widetilde{\mathcal{H}}_{\mathcal{D}}(\bm{\lambda})}
  =B(x;\bm{\lambda}+M\tilde{\bm{\delta}})\,
  \frac{\check{\Xi}_{\mathcal{D}}(x;\bm{\lambda})}
  {\check{\Xi}_{\mathcal{D}}(x+1;\bm{\lambda})}
  \biggl(\frac{\check{\Xi}_{\mathcal{D}}(x+1;\bm{\lambda}+\bm{\delta})}
  {\check{\Xi}_{\mathcal{D}}(x;\bm{\lambda}+\bm{\delta})}-e^{\partial}
  \biggr)\n
  &\phantom{\widetilde{\mathcal{H}}_{\mathcal{D}}(\bm{\lambda})}
  \quad+D(x;\bm{\lambda}+M\tilde{\bm{\delta}})\,
  \frac{\check{\Xi}_{\mathcal{D}}(x+1;\bm{\lambda})}
  {\check{\Xi}_{\mathcal{D}}(x;\bm{\lambda})}
  \biggl(\frac{\check{\Xi}_{\mathcal{D}}(x-1;\bm{\lambda}+\bm{\delta})}
  {\check{\Xi}_{\mathcal{D}}(x;\bm{\lambda}+\bm{\delta})}-e^{-\partial}
  \biggr),
  \label{tHD}\\
  &\widetilde{\mathcal{H}}_{\mathcal{D}}(\bm{\lambda})
  \check{P}_{\mathcal{D},n}(x;\bm{\lambda})
  =\mathcal{E}_n(\bm{\lambda})\check{P}_{\mathcal{D},n}(x;\bm{\lambda})
  \ \ (n=0,1,\ldots,n_{\text{max}}),
  \label{tHDcPDn=}
\end{align}
namely multi-indexed ($q$-)Racah polynomials $\check{P}_{\mathcal{D},n}(x)$
satisfy second order difference equations.

In the following we set
\begin{equation}
  P_n(\eta;\bm{\lambda})=P_{\mathcal{D},n}(\eta;\bm{\lambda})=0\ \ (n<0),
  \label{Pn=0}
\end{equation}
and $A_{-1}(\bm{\lambda})=0$.
We remark that the coefficients of $P_n(\eta)$ and $P_{\mathcal{D},n}(\eta)$
are rational functions of the parameters $(a,b,c,d)$, in which the number $N$
appears only through the parameter $a$ \eqref{a=-N}.
If we treat the parameter $a$ as an indeterminate, $\check{P}_n(x)$ and
$\check{P}_{\mathcal{D},n}(x)$ are defined for $n\in\mathbb{Z}_{\geq 0}$
and $x\in\mathbb{C}$.
For the choice \eqref{a=-N}, however, $\check{P}_n(x)$ and
$\check{P}_{\mathcal{D},n}(x)$ are ill-defined for $n>n_{\text{max}}$ and
$x\in\mathbb{C}\backslash\{0,1,\ldots,x_{\text{max}}\}$,
because $\check{P}_n(x)$ \eqref{cP} have the form
${}_4F_3(\cdots)=\sum\limits_{k=0}^n\frac{(-x)_k}{(a)_k}(\cdots)$ and
${}_4\phi_3(\cdots)=\sum\limits_{k=0}^n\frac{(q^{-x};q)_k}{(a;q)_k}(\cdots)$.
For the choice \eqref{a=-N} (we take the limit from an indeterminate $a$ to $a$
in \eqref{a=-N}), $\check{P}_{\mathcal{D},n}(x)$ are well-defined for
$n\in\{0,1,\ldots,n_{\text{max}}\}$ and $x\in\mathbb{C}$, or
$n\in\mathbb{Z}_{>n_{\text{max}}}$ and $x\in\{0,1,\ldots,x_{\text{max}}\}$,
for which the factors $(x+a)_{j-1}$ or $(aq^x;q)_{j-1}$ in
$r_j(x_j)$ \eqref{rj} contribute.
In order for $\check{P}_n(x)$ and $\check{P}_{\mathcal{D},n}(x)$ to be
orthogonal polynomials, parameters should satisfy \eqref{a=-N}--\eqref{qRpara}
and \eqref{Mrange}, and $n$ should be $n\in\{0,1,\ldots,n_{\max}\}$.

\section{Recurrence Relations with Variable Dependent Coefficients}
\label{sec:rr_var}

In this section we present $3+2M$ term recurrence relations with variable
dependent coefficients.
The quantum mechanical formulation is used to derive them.
For simplicity of the arguments, we assume that $N$ is sufficiently large
($N\gg n$), or the parameter $a$ is treated as an indeterminate.

For the discrete quantum mechanics with real shifts, the multi-step Darboux
transformations in terms of the virtual state vectors were given in \cite{os26}.
For the ($q$-)Racah systems, the general expression for the
eigenvector of the deformed system is (eq.(3.36) in \cite{os26})
\begin{equation}
  \phi^{\text{gen}}_{\mathcal{D}\,n}(x;\bm{\lambda})
  =\frac{(-1)^M\kappa^{\frac14M(M-1)}}
  {\sqrt{\check{\Xi}_{\mathcal{D}}(1;\bm{\lambda})}}
  \frac{\mathcal{C}_{\mathcal{D},n}(\bm{\lambda})}
  {\mathcal{C}_{\mathcal{D}}(\bm{\lambda})}
  \sqrt{\prod_{j=1}^M\alpha(\bm{\lambda})
  B'\bigl(0;\bm{\lambda}+(j-1)\tilde{\bm{\delta}}\bigr)}
  \,\times\phi_{\mathcal{D}\,n}(x;\bm{\lambda}),
  \label{phiDngen}
\end{equation}
where $\alpha(\bm{\lambda})$ and $B'(x;\bm{\lambda})$ are
given in \eqref{Etv} and \eqref{B'D'}, respectively.
Let us denote $\phi^{[s]}_n(x;\bm{\lambda})
\eqdef\phi^{\text{gen}}_{d_1\ldots d_s\,n}(x;\bm{\lambda})$.
Starting from the original eigenvectors $\phi^{[0]}_n(x)=\phi_n(x)$, the
multi-step Darboux transformations give the eigenvectors of the deformed
systems,
\begin{equation}
  \phi^{[s]}_n(x)=\hat{\mathcal{A}}_{d_1\ldots d_s}\phi^{[s-1]}_n(x)
  \ \ (s\geq 1),
  \label{phisn=hA..}
\end{equation}
where the matrix $\hat{\mathcal{A}}_{d_1\ldots d_s}$ is
$\hat{\mathcal{A}}_{d_1\ldots d_s}=\sqrt{\hat{B}_{d_1\ldots d_s}(x)}
-e^{\partial}\sqrt{\hat{D}_{d_1\ldots d_s}(x)}$.
Potential functions $\hat{B}_{d_1\ldots d_s}(x)$ and
$\hat{D}_{d_1\ldots d_s}(x)$ are given in \eqref{BdsDdsform}.

First we note that the matrix
$\hat{\mathcal{A}}=\sqrt{\hat{B}(x)}-e^{\partial}\sqrt{\hat{D}(x)}$
with $\hat{D}(x_{\text{max}}+1)=0$ acts on a vector $\psi(x)$ defined by
the product of two vectors $\psi(x)=f(x)\phi(x)$ as
\begin{align}
  &\quad\hat{\mathcal{A}}\bigl(f(x)\phi(x)\bigr)
  =\sqrt{\hat{B}(x)}\,f(x)\phi(x)
  -\sqrt{\hat{D}(x+1)}\,f(x+1)\phi(x+1)\n
  &=f(x)\Bigl(\sqrt{\hat{B}(x)}\,\phi(x)
  -\sqrt{\hat{D}(x+1)}\,\phi(x+1)\Bigr)
  +\bigl(f(x)-f(x+1)\bigr)\sqrt{\hat{D}(x+1)}\,\phi(x+1)\n
  &=f(x)\hat{\mathcal{A}}\phi(x)
  +\bigl(f(x)-f(x+1)\bigr)\sqrt{\hat{D}(x+1)}\,\phi(x+1).
  \label{Afphi=..}
\end{align}

Let us define $\check{R}^{[s]}_{n,k}(x)$ ($n,k\in\mathbb{Z}$,
$s\in\mathbb{Z}_{\geq -1}$) as follows:
\begin{align}
  &\check{R}^{[s]}_{n,k}(x)=0\ \ (|k|>s+1\ \text{or}\ n+k<0),\quad
  \check{R}^{[-1]}_{n,0}(x)=1\ \ (n\geq 0),\n
  &\check{R}^{[s]}_{n,k}(x)
  =A_n\check{R}^{[s-1]}_{n+1,k-1}(x)
  +\bigl(B_n-\eta(x+s)\bigr)\check{R}^{[s-1]}_{n,k}(x)
  +C_n\check{R}^{[s-1]}_{n-1,k+1}(x)\ \ (s\geq 0).
  \label{Rcdef}
\end{align}
Here $A_n$, $B_n$ and $C_n$ are the coefficients of the three term recurrence
relations \eqref{3trr} with $A_{-1}=0$ and we regard $A_{-1}\times(\cdots)=0$,
which implies that $A_n$ ($n<-1$), $B_n$ ($n<0$) and $C_n$ ($n<0$) do not
appear.
For example, non-trivial $\check{R}^{[s]}_{n,k}(x)$ ($n+k\geq 0$)
for $s=0,1$ are
\begin{align*}
  s=0:\quad&\check{R}^{[0]}_{n,1}(x)=A_n,
  \ \ \check{R}^{[0]}_{n,0}(x)=B_n-\eta(x),
  \ \ \check{R}^{[0]}_{n,-1}(x)=C_n,\\
  s=1:\quad&\check{R}^{[1]}_{n,2}(x)=A_nA_{n+1},
  \ \ \check{R}^{[1]}_{n,1}(x)=A_n\bigl(B_n+B_{n+1}-\eta(x)-\eta(x+1)\bigr),\\
  &\check{R}^{[1]}_{n,0}(x)=A_nC_{n+1}+A_{n-1}C_n
  +\bigl(B_n-\eta(x)\bigr)\bigl(B_n-\eta(x+1)\bigr),\\
  &\check{R}^{[1]}_{n,-2}(x)=C_nC_{n-1},
  \ \ \check{R}^{[1]}_{n,-1}(x)=C_n\bigl(B_n+B_{n-1}-\eta(x)-\eta(x+1)\bigr).
\end{align*}
Note that $\check{R}^{[s]}_{n,\pm(s+1)}(x)$ are $x$-independent.
By induction in $s$, we can show that
\begin{equation}
  \check{R}^{[s]}_{n,k}(x+1)-\check{R}^{[s]}_{n,k}(x)
  =\bigl(\eta(x)-\eta(x+s+1)\bigr)\check{R}^{[s-1]}_{n,k}(x+1)
  \ \ (s\geq 0).
  \label{Rprop2}
\end{equation}

We will show the following proposition, the $3+2s$ term recurrence relations
of $\phi^{[s]}_n(x)$.
\begin{prop}
\label{prop:rr_phis}
\begin{equation}
  \sum_{k=-s-1}^{s+1}\check{R}^{[s]}_{n,k}(x)\phi^{[s]}_{n+k}(x)=0
  \ \ (s\geq 0\,;\,n\in\mathbb{Z}).
  \label{RRphi}
\end{equation}
\end{prop}
Proof: We prove this proposition by induction in $s$.\\
\underline{first step} :
For $s=0$, \eqref{RRphi} is
\begin{equation*}
  A_n\phi_{n+1}(x)+\bigl(B_n-\eta(x)\bigr)\phi_n(x)+C_n\phi_{n-1}(x)=0,
\end{equation*}
which is the three term recurrence relation itself \eqref{3trrphi}.
Therefore $s=0$ case holds.

\noindent
\underline{second step} :
Assume that \eqref{RRphi} holds till $s$ ($s\ge 0$),
we will show that it also holds for $s+1$.\\
By applying $\hat{\mathcal{A}}_{d_1\ldots d_{s+1}}$ to \eqref{RRphi}
and using \eqref{phisn=hA..} and \eqref{Afphi=..},
we have
\begin{equation*}
  0=\sum_{k=-s-1}^{s+1}\check{R}^{[s]}_{n,k}(x)\phi^{[s+1]}_{n+k}(x)
  +\sum_{k=-s-1}^{s+1}
  \bigl(\check{R}^{[s]}_{n,k}(x)-\check{R}^{[s]}_{n,k}(x+1)\bigr)
  \sqrt{\hat{D}_{d_1\ldots d_{s+1}}(x+1)}\,\phi^{[s]}_{n+k}(x+1).
\end{equation*}
By using \eqref{Rprop2} this is rewritten as
\begin{equation}
  \sum_{k=-s-1}^{s+1}\check{R}^{[s]}_{n,k}(x)\phi^{[s+1]}_{n+k}(x)
  =\bigl(\eta(x)-\eta(x+s+1)\bigr)G^{[s+1]}_n(x),
  \label{sono2}
\end{equation}
where
\begin{equation}
  G^{[s+1]}_n(x)\eqdef\sqrt{\hat{D}_{d_1\ldots d_{s+1}}(x+1)}
  \sum_{k=-s}^s\check{R}^{[s-1]}_{n,k}(x+1)\phi^{[s]}_{n+k}(x+1).
  \label{Gn}
\end{equation}
We remark that when there is a factor $\check{R}^{[s]}_{n,k}$ in a sum
$\sum\limits_{k=-s-1}^{s+1}$, the range of the sum can be extended to
$\sum\limits_{k\in\mathbb{Z}}$ due to the definition
$\check{R}^{[s]}_{n,k}(x)=0$ $(|k|>s+1)$,
which will be abbreviated as $\sum\limits_k$.
Then we have
\begin{align*}
  &\quad A_nG^{[s+1]}_{n+1}(x)+\bigl(B_n-\eta(x+s)\bigr)G^{[s+1]}_n(x)
  +C_nG^{[s+1]}_{n-1}(x)\\
  &\stackrel{(\text{\romannumeral1})}{=}
  \sqrt{\hat{D}_{d_1\ldots d_{s+1}}(x+1)}\,\Bigl(
  A_n\sum_k\check{R}^{[s-1]}_{n+1,k}(x+1)
  \phi^{[s]}_{n+1+k}(x+1)\\
  &\qquad
  +\bigl(B_n-\eta(x+s)\bigr)\sum_k\check{R}^{[s-1]}_{n,k}(x+1)
  \phi^{[s]}_{n+k}(x+1)
  +C_n\sum_k\check{R}^{[s-1]}_{n-1,k}(x+1)
  \phi^{[s]}_{n-1+k}(x+1)\Bigr)\\
  &\stackrel{(\text{\romannumeral2})}{=}
  \sqrt{\hat{D}_{d_1\ldots d_{s+1}}(x+1)}\,\sum_k\Bigl(
  A_n\check{R}^{[s-1]}_{n+1,k-1}(x+1)\\
  &\qquad
  +\bigl(B_n-\eta(x+s)\bigr)\check{R}^{[s-1]}_{n,k}(x+1)
  +C_n\check{R}^{[s-1]}_{n-1,k+1}(x+1)\Bigr)
  \phi^{[s]}_{n+k}(x+1)\\
  &\stackrel{(\text{\romannumeral3})}{=}
  \sqrt{\hat{D}_{d_1\ldots d_{s+1}}(x+1)}\,\sum_k\Bigl(
  \check{R}^{[s]}_{n,k}(x+1)
  +\bigl(\eta(x+s+1)-\eta(x+s)\bigr)\check{R}^{[s-1]}_{n,k}(x+1)\Bigr)
  \phi^{[s]}_{n+k}(x+1)\\
  &\stackrel{(\text{\romannumeral4})}{=}
  \bigl(\eta(x+s+1)-\eta(x+s)\bigr)
  \sqrt{\hat{D}_{d_1\ldots d_{s+1}}(x+1)}\,\sum_k
  \check{R}^{[s-1]}_{n,k}(x+1)\phi^{[s]}_{n+k}(x+1)\\
  &\stackrel{(\text{\romannumeral5})}{=}
  \bigl(\eta(x+s+1)-\eta(x+s)\bigr)G^{[s+1]}_n(x),
\end{align*}
((\romannumeral1): \eqref{Gn},
(\romannumeral2): shift of $k$,
(\romannumeral3): \eqref{Rcdef},
(\romannumeral4): assumption \eqref{RRphi},
(\romannumeral5): \eqref{Gn} are used),
namely,
\begin{equation}
  A_nG^{[s+1]}_{n+1}(x)+\bigl(B_n-\eta(x+s+1)\bigr)G^{[s+1]}_n(x)
  +C_nG^{[s+1]}_{n-1}(x)=0.
  \label{sono3}
\end{equation}
{}From \eqref{sono2} and \eqref{sono3} we obtain
\begin{align*}
  0&=A_n\sum_k\check{R}^{[s]}_{n+1,k}(x)\phi^{[s+1]}_{n+1+k}(x)
  +\bigl(B_n-\eta(x+s+1)\bigr)
  \sum_k\check{R}^{[s]}_{n,k}(x)\phi^{[s+1]}_{n+k}(x)\n
  &\quad+C_n\sum_k\check{R}^{[s]}_{n-1,k}(x)\phi^{[s+1]}_{n-1+k}(x)\n
  &=\sum_k\Bigl(A_n\check{R}^{[s]}_{n+1,k-1}(x)
  +\bigl(B_n-\eta(x+s+1)\bigr)\check{R}^{[s]}_{n,k}(x)
  +C_n\check{R}^{[s]}_{n-1,k+1}(x)\Bigr)\phi^{[s+1]}_{n+k}(x)\n
  &=\sum_k\check{R}^{[s+1]}_{n,k}(x)\phi^{[s+1]}_{n+k}(x)
  =\sum_{k=-s-2}^{s+2}\check{R}^{[s+1]}_{n,k}(x)\phi^{[s+1]}_{n+k}(x),
\end{align*}
which shows \eqref{RRphi} with $s\to s+1$.
This concludes the induction proof of \eqref{RRphi}.
\hfill\fbox{}\\
Note that $\check{R}^{[s]}_{n,k}(x)$ does not depend on the specific values
of $d_j$'s.

Since $\phi^{[s]}_n(x)$ has the form \eqref{phiDngen}, the recurrence
relations of $\phi^{[s]}_n(x)$ \eqref{RRphi} imply those of the
multi-indexed orthogonal polynomials
$\check{P}_{\mathcal{D},n}(x;\bm{\lambda})$ for $s=M$,
\begin{equation}
  \sum_{k=-M-1}^{M+1}\mathcal{C}_{\mathcal{D},n+k}(\bm{\lambda})
  \check{R}^{[M]}_{n,k}(x;\bm{\lambda})
  \check{P}_{\mathcal{D},n+k}(x;\bm{\lambda})=0,
  \label{RRcPDn1}
\end{equation}
where $\mathcal{C}_{\mathcal{D},n+k}(\bm{\lambda})$ depends on the specific
values of $d_j$'s.
More explicitly, by dividing them by
$\mathcal{C}_{\mathcal{D},n}(\bm{\lambda})$, the recurrence relations become
\begin{align}
  &\sum_{k=-M-1}^{M+1}\prod_{j=1}^M
  \frac{\mathcal{E}_{n+k}(\bm{\lambda})
  -\tilde{\mathcal{E}}_{d_j}(\bm{\lambda})}
  {\mathcal{E}_n(\bm{\lambda})-\tilde{\mathcal{E}}_{d_j}(\bm{\lambda})}\cdot
  \check{R}^{[M]}_{n,k}(x;\bm{\lambda})
  \check{P}_{\mathcal{D},n+k}(x;\bm{\lambda})=0,
  \label{RRcPDn2}\\
  &\mathcal{E}_n(\bm{\lambda})-\tilde{\mathcal{E}}_{\text{v}}(\bm{\lambda})
  =\left\{
  \begin{array}{ll}
  (n+\text{v}+c)(n-\text{v}+\tilde{d}-c)&:\text{R}\\[2pt]
  q^{-n}(1-cq^{n+\text{v}})(1-c^{-1}\tilde{d}q^{n-\text{v}})&:\text{$q$R}
  \end{array}\right..
\end{align}

We can check that $\check{R}^{[s]}_{n,k}(x;\bm{\lambda})$ are symmetric
polynomials in
$\eta(x;\bm{\lambda}),\eta(x+1;\bm{\lambda}),\ldots,\eta(x+s;\bm{\lambda})$.
The elementary symmetric polynomials $e_k$ in
$\eta(x;\bm{\lambda}),\eta(x+1;\bm{\lambda}),\ldots,\eta(x+s;\bm{\lambda})$
are polynomials in $\eta(x;\bm{\lambda}+s\bm{\delta})$, because
\begin{align*}
  &\quad\sum_{k=0}^{s+1}(-1)^ke_kt^{s+1-k}
  =\prod_{j=0}^s\bigl(t-\eta(x+j;\bm{\lambda})\bigr),\qquad
  A=\left\{
  \begin{array}{ll}
  t-\eta(x+\frac{s}{2};\bm{\lambda})&:\text{$s$ even}\\
  1&:\text{$s$ odd}
  \end{array}\right.,\\
  &=A\times\prod_{j=0}^{[\frac{s-1}{2}]}\bigl(t-\eta(x+j;\bm{\lambda})\bigr)
  \bigl(t-\eta(x+s-j;\bm{\lambda})\bigr)\\
  &=A\times\prod_{j=0}^{[\frac{s-1}{2}]}\Bigl(t^2
  -\bigl(\eta(x+j;\bm{\lambda})+\eta(x+s-j;\bm{\lambda})\bigr)t
  +\eta(x+j;\bm{\lambda})\eta(x+s-j;\bm{\lambda})\Bigr),
\end{align*}
and
\begin{align*}
  &\eta(x+\tfrac{s}{2};\bm{\lambda})
  =B\eta(x;\bm{\lambda}+s\bm{\delta})+\eta(\tfrac{s}{2};\bm{\lambda}),\\
  &\eta(x+j;\bm{\lambda})+\eta(x+s-j;\bm{\lambda})
  =B'\eta(x;\bm{\lambda}+s\bm{\delta})
  +\eta(j;\bm{\lambda})+\eta(s-j;\bm{\lambda}),\\
  &\eta(x+j;\bm{\lambda})\eta(x+s-j;\bm{\lambda})
  =B''\eta(x;\bm{\lambda}+s\bm{\delta})^2
  +B'''\eta(x;\bm{\lambda}+s\bm{\delta})
  +\eta(j;\bm{\lambda})\eta(s-j;\bm{\lambda}),\\
  &B=\left\{
  \begin{array}{ll}
  1&:\text{R}\\
  q^{-\frac{s}{2}}&:\text{$q$R}
  \end{array}\right.,\quad
  B'=\left\{
  \begin{array}{ll}
  2&:\text{R}\\
  q^{-j}+q^{j-s}&:\text{$q$R}
  \end{array}\right.,\quad
  B''=\left\{
  \begin{array}{ll}
  1&:\text{R}\\
  q^{-s}&:\text{$q$R}
  \end{array}\right.,\\[2pt]
  &B'''=\left\{
  \begin{array}{ll}
  2j(s-j)+sd&:\text{R}\\[2pt]
  q^{-s}\bigl(2(1+dq^s)-(q^j+q^{s-j})(1+d)\bigr)&:\text{$q$R}
  \end{array}\right..
\end{align*}
This implies that any symmetric polynomial in
$\eta(x;\bm{\lambda}),\eta(x+1;\bm{\lambda}),\ldots,\eta(x+s;\bm{\lambda})$
is expressed as a polynomial in $\eta(x;\bm{\lambda}+s\bm{\delta})$.
Therefore we obtain
\begin{align}
  &\check{R}^{[s]}_{n,k}(x;\bm{\lambda})
  \eqdef R^{[s]}_{n,k}\bigl(\eta(x;\bm{\lambda}+s\bm{\delta});
  \bm{\lambda}\bigr)
  \ \ (|k|\leq s+1)\n
  &R^{[s]}_{n,k}(\eta;\bm{\lambda})
  \ :\text{a polynomial of degree $s+1-|k|$ in $\eta$}.
  \label{Rdef2}
\end{align}
By rewriting the recurrence relations \eqref{RRcPDn1} and \eqref{RRcPDn2},
we obtain the following theorem.
\begin{thm}
\label{thm:rr_var}
The multi-indexed ($q$-)Racah polynomials satisfy the $3+2M$ term recurrence
relations with variable dependent coefficients:
\begin{align}
  &\sum_{k=-M-1}^{M+1}\mathcal{C}_{\mathcal{D},n+k}(\bm{\lambda})
  R^{[M]}_{n,k}(\eta;\bm{\lambda})
  P_{\mathcal{D},n+k}(\eta;\bm{\lambda})=0,
  \label{RRPDn1}\\
  \text{or}\quad
  &\sum_{k=-M-1}^{M+1}\prod_{j=1}^M
  \frac{\mathcal{E}_{n+k}(\bm{\lambda})
  -\tilde{\mathcal{E}}_{d_j}(\bm{\lambda})}
  {\mathcal{E}_n(\bm{\lambda})-\tilde{\mathcal{E}}_{d_j}(\bm{\lambda})}\cdot
  R^{[M]}_{n,k}(\eta;\bm{\lambda})
  P_{\mathcal{D},n+k}(\eta;\bm{\lambda})=0.
  \label{RRPDn2}
\end{align}
\end{thm}
{\bf Remark}\,
We have assumed that $N$ is sufficiently large ($N\gg n$), or the parameter
$a$ is treated as an indeterminate.

According to the same argument for the multi-indexed Laguerre, Jacobi, Wilson and
Askey-Wilson polynomials \cite{rrmiop}, the multi-indexed ($q$-)Racah
polynomials $P_{\mathcal{D},n}(\eta;\bm{\lambda})$ ($n\geq M+1$) are
determined by the $3+2M$ term recurrence relations \eqref{RRPDn1} with
$M+1$ ``initial data''
\begin{equation}
  P_{\mathcal{D},0}(\eta;\bm{\lambda}),\,P_{\mathcal{D},1}(\eta;\bm{\lambda}),
  \,\ldots\,,\,P_{\mathcal{D},M}(\eta;\bm{\lambda}).
  \label{initdata}
\end{equation}
After calculating the initial data \eqref{initdata} by \eqref{PDn}, we can
obtain $P_{\mathcal{D},n}(\eta;\bm{\lambda})$ through the $3+2M$ term recurrence
relations \eqref{RRPDn1}.
The calculation cost of this method is much less than the original determinant
expression \eqref{PDn} for large $M$.

\section{Recurrence Relations with Constant Coefficients}
\label{sec:rr_const}

In this section we present $1+2L$ term recurrence relations with constant
coefficients.
Depending on whether the parameter $a$ is an indeterminate or \eqref{a=-N},
we have two different kinds of recurrence relations.

We want to find $X(\eta)=X(\eta;\bm{\lambda})$, which is a polynomial of
degree $L$ in $\eta$ and gives the following expansion:
\begin{equation*}
  X(\eta)P_{\mathcal{D},n}(\eta)
  =\sum_{k=-n}^Lr_{n,k}^{X,\mathcal{D}}P_{\mathcal{D},n+k}(\eta)\quad
  \text{or}\quad\check{X}(x)\check{P}_{\mathcal{D},n}(x)
  =\sum_{k=-n}^Lr_{n,k}^{X,\mathcal{D}}\check{P}_{\mathcal{D},n+k}(x),
\end{equation*}
where $r_{n,k}^{X,\mathcal{D}}\,$'s are constants and
$\check{X}(x)=\check{X}(x;\bm{\lambda})$ is defined by
\begin{equation}
  \check{X}(x;\bm{\lambda})\eqdef
  X\bigl(\eta(x;\bm{\lambda}+M\bm{\delta});\bm{\lambda}\bigr).
  \label{cX}
\end{equation}
Since the multi-indexed polynomials $P_{\mathcal{D},n}(\eta)$ are
orthogonal polynomials, the above recurrence relations with constant
coefficients are expressed as (see Lemma 1 in \cite{rrmiop2})
\begin{equation}
  X(\eta)P_{\mathcal{D},n}(\eta)
  =\sum_{k=-L}^Lr_{n,k}^{X,\mathcal{D}}P_{\mathcal{D},n+k}(\eta),
  \label{XP}
\end{equation}
under the convention \eqref{Pn=0}.
Here we have assumed $N$ is sufficiently large.
Unlike the multi-indexed polynomials in \cite{rrmiop2}, there is a
maximal value of $n$ for $P_{\mathcal{D},n}(\eta)$, $n_{\text{max}}=N$.
So \eqref{XP} is expected to be modified as
\begin{equation}
  \check{X}(x)\check{P}_{\mathcal{D},n}(x)
  =\sum_{k=-\min(L,n)}^{\min(L,N-n)}
  r_{n,k}^{X,\mathcal{D}}\check{P}_{\mathcal{D},n+k}(x)
  \ \ \Bigl(\begin{array}{ll}
  n=0,1,\ldots,n_{\text{max}}\\
  x=0,1,\ldots,x_{\text{max}}
  \end{array}\Bigr),
  \label{XP2}
\end{equation}
where $\check{P}_{\mathcal{D},n}(x)$ with $n<0$ or $n>n_{\max}$ does not appear.
The overall normalization and the constant term of $X(\eta)$ are not important,
because the change of the former induces that of the overall normalization
of $r_{n,k}^{X,\mathcal{D}}$ and the shift of the latter induces that of
$r_{n,0}^{X,\mathcal{D}}$.
Without loss of generality, we take the constant term of $X(\eta)$ as $X(0)=0$.

\subsection{Parameter $a$ : indeterminate}
\label{sec:a:indet}

In this subsection we assume that the parameter $a$ is an indeterminate,
$x$ and $\eta$ are continuous variables ($x,\eta\in\mathbb{C}$) and other
parameters $\bm{\lambda}$ ($b,c,d$) are generic.

\subsubsection{step 0}
\label{sec:step_0}

The sinusoidal coordinates $\eta(x;\bm{\lambda})$ \eqref{eta} have the
following property \cite{os14,rrmiop2}
\begin{equation}
  \frac{\eta(x;\bm{\lambda})^{n+1}-\eta(x-1;\bm{\lambda})^{n+1}}
  {\eta(x;\bm{\lambda})-\eta(x-1;\bm{\lambda})}
  =\sum_{k=0}^ng_n^{\prime\,(k)}(\bm{\lambda})
  \eta(x;\bm{\lambda}-\bm{\delta})^{n-k}
  \ \ (n\in\mathbb{Z}_{\geq 0}),
  \label{eta_gpnk}
\end{equation}
where $g_n^{\prime\,(k)}$ is given by
\begin{align}
  \text{R}:\ \ &g_n^{\prime\,(k)}(\bm{\lambda})\eqdef
  \sum_{r=0}^k\sum_{l=0}^{k-r}
  \genfrac{(}{)}{0pt}{}{n+1}{r}\genfrac{(}{)}{0pt}{}{n-r-l}{n-k}
  (-1)^{r+l}\Bigl(\frac{d}{2}\Bigr)^{2r}
  \Bigl(\frac{d-1}{2}\Bigr)^{2(k-r-l)}
  g_{n-r}^{\prime\,(l)\,\text{W}},\\
  \text{$q$R}:\ \ &g_n^{\prime\,(k)}(\bm{\lambda})\eqdef
  \sum_{r=0}^k\sum_{l=0}^{k-r}
  \genfrac{(}{)}{0pt}{}{n+1}{r}\genfrac{(}{)}{0pt}{}{n-r-l}{n-k}
  (-1)^r\bigl(2d^{\frac12}\bigr)^lq^{\frac12(n-r-l)}
  \bigl(1+d\bigr)^r\bigl(1+dq^{-1}\bigr)^{k-r-l}\n
  &\phantom{g_n^{\prime\,(k)}(\bm{\lambda})\eqdef}\times
  g_{n-r}^{\prime\,(l)\,\text{AW}}.
\end{align}
Here $g_n^{\prime\,(k)\,\text{W}}$ and $g_n^{\prime\,(k)\,\text{AW}}$ are
given by \cite{os32,rrmiop2}
\begin{align*}
  &g_n^{\prime\,(k)\,\text{W}}\eqdef
  \frac{(-1)^k}{2^{2k+1}}\genfrac{(}{)}{0pt}{}{2n+2}{2k+1},\n
  &g_n^{\prime\,(k)\,\text{AW}}\eqdef
  \theta(\text{$k$\,:\,even})\frac{(n+1)!}{2^k}
  \sum_{r=0}^{\frac{k}{2}}\genfrac{(}{)}{0pt}{}{n-k+r}{r}
  \frac{(-1)^rq^{-\frac12(n-k+2r)}}{(\frac{k}{2}-r)!\,(n-\frac{k}{2}+1+r)!}
  \frac{1-q^{n-k+1+2r}}{1-q},
\end{align*}
and $\theta(P)$ is a step function (an indicator function) for a proposition $P$,
$\theta(P)=1$ ($P$ : true), $0$ ($P$ : false).

For a polynomial $p(\eta)$ in $\eta$, let us define a polynomial in $\eta$,
$I_{\bm{\lambda}}[p](\eta)$, as follows:
\begin{equation}
  p(\eta)=\sum_{k=0}^na_k\eta^k\mapsto
  I_{\bm{\lambda}}[p](\eta)\eqdef\sum_{k=0}^{n+1}b_k\eta^k,
  \label{mapI}
\end{equation}
where $b_k$'s are defined by
\begin{equation}
  b_{k+1}=\frac{1}{g_k^{\prime\,(0)}(\bm{\lambda})}
  \Bigl(a_k-\sum_{j=k+1}^ng_j^{\prime\,(j-k)}(\bm{\lambda})b_{j+1}\Bigr)
  \ \ (k=n,n-1,\ldots,1,0),\quad
  b_0=0.
\end{equation}
The constant term of $I_{\bm{\lambda}}[p](\eta)$ is chosen to be zero.
Note that $a_k=\sum\limits_{j=k}^ng_j^{\prime\,(j-k)}(\bm{\lambda})b_{j+1}$.
It is easy to show that this polynomial $I_{\bm{\lambda}}[p](\eta)=P(\eta)$
satisfies
\begin{equation}
  \frac{P\bigl(\eta(x;\bm{\lambda})\bigr)-P\bigl(\eta(x-1;\bm{\lambda})\bigr)}
  {\eta(x;\bm{\lambda})-\eta(x-1;\bm{\lambda})}
  =p\bigl(\eta(x;\bm{\lambda}-\bm{\delta})\bigr).
  \label{P-P=p}
\end{equation}
So we can call $P(\eta)$ the `primitive polynomial' of $p(\eta)$.
The above equations are valid for $x,\eta\in\mathbb{C}$, but
\eqref{P-P=p} with $P(0)=0$ gives the following expression:
\begin{equation}
  P\big(\eta(x;\bm{\lambda}))
  =\sum_{j=1}^x\bigl(\eta(j;\bm{\lambda})-\eta(j-1;\bm{\lambda})\bigr)
  p\bigl(\eta(j;\bm{\lambda}-\bm{\delta})\bigr)
  \ \ (x\in\mathbb{Z}_{\geq 0}).
  \label{P=(eta-eta)p}
\end{equation}
It is nontrivial to show directly that the RHS is a polynomial in
$\eta(x;\bm{\lambda})$, but it is so by construction.
Note that the maps $I_{\bm{\lambda}}$ depend on
parameters $\bm{\lambda}$ only through $d$
because coefficients $g^{\prime\,(k)}_n(\bm{\lambda})$ depend on $d$ only.

\subsubsection{step 1}
\label{sec:step_1}

Let us define the set of finite linear combinations of
$P_{\mathcal{D},n}(\eta)$, $\mathcal{U}_{\mathcal{D}}\subset\mathbb{C}[\eta]$,
by
\begin{equation}
  \mathcal{U}_{\mathcal{D}}
  \eqdef\text{Span}\{P_{\mathcal{D},n}(\eta)\bigm|n\in\mathbb{Z}_{\geq 0}\}.
  \label{UD}
\end{equation}
Since the degree of $P_{\mathcal{D},n}(\eta)$ is $\ell_{\mathcal{D}}+n$,
it is trivial that $p(\eta)\in\mathcal{U}_{\mathcal{D}}\Rightarrow
\deg\,p\geq\ell_{\mathcal{D}}$, except for $p(\eta)=0$.
Corresponding to \eqref{tHDcPDn=}, the multi-indexed ($q$-)Racah polynomials
$\check{P}_{\mathcal{D},n}(x)$ with $x\in\mathbb{C}$ satisfy the second order
difference equations,
\begin{equation}
  \widetilde{\mathcal{H}}^{\text{cont}}_{\mathcal{D}}(\bm{\lambda})
  \check{P}_{\mathcal{D},n}(x;\bm{\lambda})
  =\mathcal{E}_n(\bm{\lambda})\check{P}_{\mathcal{D},n}(x;\bm{\lambda})
  \ \ (n\in\mathbb{Z}_{\geq 0}),
  \label{cont:tHDcPDn=}
\end{equation}
where $\widetilde{\mathcal{H}}^{\text{cont}}_{\mathcal{D}}(\bm{\lambda})$ is
obtained from $\widetilde{\mathcal{H}}_{\mathcal{D}}(\bm{\lambda})$
\eqref{tHD} by replacing the matrices $e^{\pm\partial}$ with the shift
operators $e^{\pm\frac{d}{dx}}$,
\begin{align}
  \widetilde{\mathcal{H}}^{\text{cont}}_{\mathcal{D}}(\bm{\lambda})
  &=B(x;\bm{\lambda}+M\tilde{\bm{\delta}})\,
  \frac{\check{\Xi}_{\mathcal{D}}(x;\bm{\lambda})}
  {\check{\Xi}_{\mathcal{D}}(x+1;\bm{\lambda})}
  \biggl(\frac{\check{\Xi}_{\mathcal{D}}(x+1;\bm{\lambda}+\bm{\delta})}
  {\check{\Xi}_{\mathcal{D}}(x;\bm{\lambda}+\bm{\delta})}-e^{\frac{d}{dx}}
  \biggr)\n
  &\quad+D(x;\bm{\lambda}+M\tilde{\bm{\delta}})\,
  \frac{\check{\Xi}_{\mathcal{D}}(x+1;\bm{\lambda})}
  {\check{\Xi}_{\mathcal{D}}(x;\bm{\lambda})}
  \biggl(\frac{\check{\Xi}_{\mathcal{D}}(x-1;\bm{\lambda}+\bm{\delta})}
  {\check{\Xi}_{\mathcal{D}}(x;\bm{\lambda}+\bm{\delta})}-e^{-\frac{d}{dx}}
  \biggr).
  \label{conttHD}
\end{align}
For $p(\eta)\in\mathbb{C}[\eta]$,
$\widetilde{\mathcal{H}}^{\text{cont}}_{\mathcal{D}}(\bm{\lambda})$ acts on
$\check{p}(x)\eqdef p\bigl(\eta(x;\bm{\lambda}+M\bm{\delta})\bigr)$ as
\begin{align}
  \widetilde{\mathcal{H}}^{\text{cont}}_{\mathcal{D}}(\bm{\lambda})
  \check{p}(x)
  &=B(x;\bm{\lambda}+M\tilde{\bm{\delta}})\,
  \frac{\check{\Xi}_{\mathcal{D}}(x;\bm{\lambda})}
  {\check{\Xi}_{\mathcal{D}}(x+1;\bm{\lambda})}
  \biggl(\frac{\check{\Xi}_{\mathcal{D}}(x+1;\bm{\lambda}+\bm{\delta})}
  {\check{\Xi}_{\mathcal{D}}(x;\bm{\lambda}+\bm{\delta})}\check{p}(x)
  -\check{p}(x+1)
  \biggr)\n
  &\quad+D(x;\bm{\lambda}+M\tilde{\bm{\delta}})\,
  \frac{\check{\Xi}_{\mathcal{D}}(x+1;\bm{\lambda})}
  {\check{\Xi}_{\mathcal{D}}(x;\bm{\lambda})}
  \biggl(\frac{\check{\Xi}_{\mathcal{D}}(x-1;\bm{\lambda}+\bm{\delta})}
  {\check{\Xi}_{\mathcal{D}}(x;\bm{\lambda}+\bm{\delta})}\check{p}(x)
  -\check{p}(x-1)
  \biggr).
  \label{conttHDp}
\end{align}
Let zeros of $\Xi_{\mathcal{D}}(\eta;\bm{\lambda})$ and
$\Xi_{\mathcal{D}}(\eta;\bm{\lambda}+\bm{\delta})$ be $\beta^{(\eta)}_j$ and
$\beta^{\prime\,(\eta)}_j$ ($j=1,2,\ldots,\ell_{\mathcal{D}}$), respectively,
which are simple for generic parameters
(This property can be verified by numerical calculation but we do not have
its analytical proof. We assume this property in the following.).
We define $\beta_j$ and $\beta'_j$ as
$\beta^{(\eta)}_j=\eta(\beta_j;\bm{\lambda}+M\bm{\delta})$ and
$\beta^{\prime\,(\eta)}_j=\eta(\beta'_j;\bm{\lambda}+M\bm{\delta})$.
(For $x\in\mathbb{C}$, $\eta=\eta(x;\bm{\lambda})$ are not one-to-one functions
($\eta(x;\bm{\lambda})=\eta(-x-d;\bm{\lambda})$ for R,
$\eta(x;\bm{\lambda})=\eta(-x-\lambda_4;\bm{\lambda})
=\eta(x+\frac{2\pi i}{\log q};\bm{\lambda})$ for $q$R), but it
does not cause any problems in the following argument,
because we need \eqref{condcp} and it is determined by the values of 
$\eta(\beta_j:\bm{\lambda}+M\bm{\delta})$ and
$\eta(\beta_j-1:\bm{\lambda}+M\bm{\delta})$
instead of multi-valued $\beta_j$.)

Let us consider the condition such that
$\widetilde{\mathcal{H}}^{\text{cont}}_{\mathcal{D}}(\bm{\lambda})
\check{p}(x)$ \eqref{conttHDp} is a polynomial in
$\eta(x;\bm{\lambda}+M\bm{\delta})$.
The poles at $x=\beta'_j,\beta_j,\beta_j-1$ in \eqref{conttHDp} should be
canceled.
First we consider $x=\beta'_j$.
Since $\check{\Xi}_{\mathcal{D}}(x;\bm{\lambda}+\bm{\delta})
=\check{P}_{\mathcal{D},0}(x;\bm{\lambda})$ and
$\check{P}_{\mathcal{D},n}(x;\bm{\lambda})$ ($n>0$) do not have common roots
for generic parameters
(This property can be verified by numerical calculation but we do not have
its analytical proof. We assume this property in the following.),
\eqref{conttHDp} with
$\check{p}(x)=\check{P}_{\mathcal{D},n}(x)$ implies that the poles at
$x=\beta'_j$ are canceled, namely,
\begin{align}
  &\quad B(\beta'_j;\bm{\lambda}+M\tilde{\bm{\delta}})\,
  \frac{\check{\Xi}_{\mathcal{D}}(\beta'_j;\bm{\lambda})}
  {\check{\Xi}_{\mathcal{D}}(\beta'_j+1;\bm{\lambda})}\,
  \check{\Xi}_{\mathcal{D}}(\beta'_j+1;\bm{\lambda}+\bm{\delta})\n
  &+D(\beta'_j;\bm{\lambda}+M\tilde{\bm{\delta}})\,
  \frac{\check{\Xi}_{\mathcal{D}}(\beta'_j+1;\bm{\lambda})}
  {\check{\Xi}_{\mathcal{D}}(\beta'_j;\bm{\lambda})}\,
  \check{\Xi}_{\mathcal{D}}(\beta'_j-1;\bm{\lambda}+\bm{\delta})
  =0\ \ (j=1,2,\ldots,\ell_{\mathcal{D}}).
  \label{beta'j=0}
\end{align}
This relation implies that we do not need bother the poles at $x=\beta'_j$
in \eqref{conttHDp} for general $p(\eta)$.
Next we consider $x=\beta_j,\beta_j-1$. For generic parameters,
$\check{\Xi}_{\mathcal{D}}(x;\bm{\lambda})$ and
$\check{\Xi}_{\mathcal{D}}(x+1;\bm{\lambda})$ do not have common roots,
and the numerators of $B(x;\bm{\lambda}+M\tilde{\bm{\delta}})$ and
$D(x;\bm{\lambda}+M\tilde{\bm{\delta}})$ do not cancel the poles coming
from $\check{\Xi}_{\mathcal{D}}(x;\bm{\lambda})$ and
$\check{\Xi}_{\mathcal{D}}(x+1;\bm{\lambda})$,
and zeros of the denominators of $B(x;\bm{\lambda}+M\tilde{\bm{\delta}})$ and
$D(x;\bm{\lambda}+M\tilde{\bm{\delta}})$ do not coincide with $\beta_j$ and
$\beta_j-1$.
The residue of the first term of \eqref{conttHDp} at $x=\beta_j-1$ is
\begin{equation*}
  B(\beta_j-1;\bm{\lambda}+M\tilde{\bm{\delta}})\,
  \frac{\check{\Xi}_{\mathcal{D}}(\beta_j-1;\bm{\lambda})}
  {\frac{d}{dx}\check{\Xi}_{\mathcal{D}}(x+1;\bm{\lambda})|_{x=\beta_j-1}}
  \biggl(\frac{\check{\Xi}_{\mathcal{D}}(\beta_j;\bm{\lambda}+\bm{\delta})}
  {\check{\Xi}_{\mathcal{D}}(\beta_j-1;\bm{\lambda}+\bm{\delta})}\,
  \check{p}(\beta_j-1)
  -\check{p}(\beta_j)
  \biggr),
\end{equation*}
and that of the second term of \eqref{conttHDp} at $x=\beta_j$ is
\begin{equation*}
  D(\beta_j;\bm{\lambda}+M\tilde{\bm{\delta}})\,
  \frac{\check{\Xi}_{\mathcal{D}}(\beta_j+1;\bm{\lambda})}
  {\frac{d}{dx}\check{\Xi}_{\mathcal{D}}(x;\bm{\lambda})|_{x=\beta_j}}
  \biggl(\frac{\check{\Xi}_{\mathcal{D}}(\beta_j-1;\bm{\lambda}+\bm{\delta})}
  {\check{\Xi}_{\mathcal{D}}(\beta_j;\bm{\lambda}+\bm{\delta})}\,
  \check{p}(\beta_j)
  -\check{p}(\beta_j-1)
  \biggr).
\end{equation*}
These residues should be vanished. So we obtain the conditions:
\begin{equation}
  \frac{\check{\Xi}_{\mathcal{D}}(\beta_j;\bm{\lambda}+\bm{\delta})}
  {\check{\Xi}_{\mathcal{D}}(\beta_j-1;\bm{\lambda}+\bm{\delta})}\,
  \check{p}(\beta_j-1)
  =\check{p}(\beta_j)
  \ \ (j=1,2,\ldots,\ell_{\mathcal{D}}).
  \label{condcp}
\end{equation}

Let us assume $\deg p(\eta)<\ell_{\mathcal{D}}$.
Without loss of generality, we take $p(\eta)$ as a monic polynomial.
Then the number of adjustable coefficients of $p(\eta)$ is $\deg p(\eta)$.
On the other hand, the number of conditions \eqref{condcp} is
$\ell_{\mathcal{D}}$. Therefore the conditions \eqref{condcp} can not be
satisfied for generic parameters, except for $p(\eta)=0$.

Since any polynomial $p(\eta)$ is expanded as
\begin{equation*}
  p(\eta)=\!\!\!\sum_{n=0}^{\deg p-\ell_{\mathcal{D}}}\!\!\!
  a_nP_{\mathcal{D},n}(\eta)+r(\eta),
  \ \ \deg r(\eta)<\ell_{\mathcal{D}}
  \ \ \bigl(\text{$p(\eta)=r(\eta)$ for $\deg\,p<\ell_{\mathcal{D}}$}\bigr),
\end{equation*}
we have
\begin{align*}
  &\phantom{\Leftrightarrow\ \,}
  \text{$\widetilde{\mathcal{H}}^{\text{cont}}_{\mathcal{D}}(\bm{\lambda})
  \check{p}(x)$ : a polynomial in $\eta(x;\bm{\lambda}+M\bm{\delta})$}\\
  &\Leftrightarrow
  \text{$\widetilde{\mathcal{H}}^{\text{cont}}_{\mathcal{D}}(\bm{\lambda})
  \check{r}(x)$ : a polynomial in $\eta(x;\bm{\lambda}+M\bm{\delta})$}
  \Leftrightarrow
  r(\eta)=0.
\end{align*}
Therefore we obtain the following proposition:
\begin{prop}
\label{prop:pinUD}
For $p(\eta)\in\mathbb{C}[\eta]$, the following holds:
\begin{equation}
  p(\eta)\in\mathcal{U}_{\mathcal{D}}\Leftrightarrow
  \widetilde{\mathcal{H}}^{\text{\rm cont}}_{\mathcal{D}}(\bm{\lambda})
  \check{p}(x):\text{a polynomial in $\eta(x;\bm{\lambda}+M\bm{\delta})$}.
  \label{pinUD}
\end{equation}
\end{prop}

\subsubsection{step 2}
\label{sec:step_2}

Let us consider a polynomial $X(\eta)$ satisfying \eqref{XP}.
{}From Proposition\,\ref{prop:pinUD}, $X(\eta)$ in \eqref{XP} should satisfy
\begin{equation*}
  \text{$\widetilde{\mathcal{H}}^{\text{cont}}_{\mathcal{D}}(\bm{\lambda})
  \bigl(\check{X}(x)\check{P}_{\mathcal{D},n}(x;\bm{\lambda})\bigr)$
  : a polynomial in $\eta(x;\bm{\lambda}+M\bm{\delta})$}.
\end{equation*}
Action of $\widetilde{\mathcal{H}}^{\text{cont}}_{\mathcal{D}}(\bm{\lambda})$
on $\check{X}(x)\check{P}_{\mathcal{D},n}(x;\bm{\lambda})$ is
\begin{align*}
  &\quad\widetilde{\mathcal{H}}^{\text{cont}}_{\mathcal{D}}(\bm{\lambda})
  \bigl(\check{X}(x)\check{P}_{\mathcal{D},n}(x;\bm{\lambda})\bigr)\\
  &=\check{X}(x)
  \widetilde{\mathcal{H}}^{\text{cont}}_{\mathcal{D}}(\bm{\lambda})
  \check{P}_{\mathcal{D},n}(x;\bm{\lambda})
  -B(x;\bm{\lambda}+M\tilde{\bm{\delta}})\,
  \frac{\check{\Xi}_{\mathcal{D}}(x;\bm{\lambda})}
  {\check{\Xi}_{\mathcal{D}}(x+1;\bm{\lambda})}
  \bigl(\check{X}(x+1)-\check{X}(x)\bigr)
  \check{P}_{\mathcal{D},n}(x+1;\bm{\lambda})\\
  &\phantom{=\check{X}(x)
  \widetilde{\mathcal{H}}^{\text{cont}}_{\mathcal{D}}(\bm{\lambda})
  \check{P}_{\mathcal{D},n}(x;\bm{\lambda})}
  -D(x;\bm{\lambda}+M\tilde{\bm{\delta}})\,
  \frac{\check{\Xi}_{\mathcal{D}}(x+1;\bm{\lambda})}
  {\check{\Xi}_{\mathcal{D}}(x;\bm{\lambda})}
  \bigl(\check{X}(x-1)-\check{X}(x)\bigr)
  \check{P}_{\mathcal{D},n}(x-1;\bm{\lambda}),
\end{align*}
namely,
\begin{equation}
  \widetilde{\mathcal{H}}^{\text{cont}}_{\mathcal{D}}(\bm{\lambda})
  \bigl(\check{X}(x)\check{P}_{\mathcal{D},n}(x;\bm{\lambda})\bigr)
  =\mathcal{E}_n(\bm{\lambda})\check{X}(x)
  \check{P}_{\mathcal{D},n}(x;\bm{\lambda})+F(x).
  \label{tHDcXcPDn}
\end{equation}
Here $F(x)$ is
\begin{align}
  F(x)&=-B(x;\bm{\lambda}+M\tilde{\bm{\delta}})\,
  \frac{\check{\Xi}_{\mathcal{D}}(x;\bm{\lambda})}
  {\check{\Xi}_{\mathcal{D}}(x+1;\bm{\lambda})}
  \bigl(\check{X}(x+1)-\check{X}(x)\bigr)
  \check{P}_{\mathcal{D},n}(x+1;\bm{\lambda})\n
  &\quad-D(x;\bm{\lambda}+M\tilde{\bm{\delta}})\,
  \frac{\check{\Xi}_{\mathcal{D}}(x+1;\bm{\lambda})}
  {\check{\Xi}_{\mathcal{D}}(x;\bm{\lambda})}
  \bigl(\check{X}(x-1)-\check{X}(x)\bigr)
  \check{P}_{\mathcal{D},n}(x-1;\bm{\lambda}).
  \label{F}
\end{align}
Equations \eqref{eta_gpnk} and \eqref{cX} imply
\begin{align*}
  \check{X}(x)-\check{X}(x-1)
  &=\bigl(\eta(x;\bm{\lambda}+M\bm{\delta})
  -\eta(x-1;\bm{\lambda}+M\bm{\delta})\bigr)\\
  &\quad\times\Bigl(\text{a polynomial in
  $\eta\bigl(x;\bm{\lambda}+(M-1)\bm{\delta}\bigr)$}\Bigr).
\end{align*}
In order to cancel the zeros of $\check{\Xi}_{\mathcal{D}}(x;\bm{\lambda})
=\Xi_{\mathcal{D}}\bigl(\eta(x;\bm{\lambda}+(M-1)\bm{\delta});
\bm{\lambda}\bigr)$ in \eqref{F}, the polynomial appeared in the above
expression should have the following form,
\begin{equation}
  \check{X}(x)-\check{X}(x-1)
  =\bigl(\eta(x;\bm{\lambda}+M\bm{\delta})
  -\eta(x-1;\bm{\lambda}+M\bm{\delta})\bigr)
  \check{\Xi}_{\mathcal{D}}(x;\bm{\lambda})
  Y\bigl(\eta(x;\bm{\lambda}+(M-1)\bm{\delta})\bigl),
  \label{X-X}
\end{equation}
where $Y(\eta)$ is an arbitrary polynomial in $\eta$.
Note that this $X(\eta)$ can be expressed in terms of the map
$I_{\bm{\lambda}}$ \eqref{mapI} by \eqref{P-P=p},
\begin{equation}
  X(\eta)=I_{\bm{\lambda}+M\bm{\delta}}
  \bigl[\Xi_{\mathcal{D}}Y\bigr](\eta).
\end{equation}
Then $F(x)$ \eqref{F} becomes
\begin{align}
  F(x)&=-B(x;\bm{\lambda}+M\tilde{\bm{\delta}})
  \bigl(\eta(x+1;\bm{\lambda}+M\bm{\delta})
  -\eta(x;\bm{\lambda}+M\bm{\delta})\bigr)\n
  &\qquad\times
  \check{\Xi}_{\mathcal{D}}(x;\bm{\lambda})
  Y\bigl(\eta(x+1;\bm{\lambda}+(M-1)\bm{\delta})\bigl)
  \check{P}_{\mathcal{D},n}(x+1;\bm{\lambda})\n
  &\quad-D(x;\bm{\lambda}+M\tilde{\bm{\delta}})
  \bigl(\eta(x-1;\bm{\lambda}+M\bm{\delta})
  -\eta(x;\bm{\lambda}+M\bm{\delta})\bigr)
  \label{F2}\\
  &\qquad\times
  \check{\Xi}_{\mathcal{D}}(x+1;\bm{\lambda})
  Y\bigl(\eta(x;\bm{\lambda}+(M-1)\bm{\delta})\bigl)
  \check{P}_{\mathcal{D},n}(x-1;\bm{\lambda}).
  \nonumber
\end{align}
{}From the explicit forms of $B(x;\bm{\lambda})$ and $D(x;\bm{\lambda})$
\eqref{B,D}, we have
\begin{align}
  &\quad B(x;\bm{\lambda}+M\tilde{\bm{\delta}})
  \bigl(\eta(x+1;\bm{\lambda}+M\bm{\delta})
  -\eta(x;\bm{\lambda}+M\bm{\delta})\bigr)\n
  &=\left\{
  \begin{array}{ll}
  {\displaystyle
  -\frac{(x+a)(x+b)(x+M+c)(x+M+d)}{2x+M+d}}&:\text{R}\\[10pt]
  {\displaystyle -(1-q)\frac{(1-aq^x)(1-bq^x)(1-cq^{x+M})(1-dq^{x+M})}
  {q^{x+1}(1-dq^{2x+M})}}&:\text{$q$R}
  \end{array}\right.,
  \label{B(eta-eta)}\\[4pt]
  &\quad D(x;\bm{\lambda}+M\tilde{\bm{\delta}})
  \bigl(\eta(x-1;\bm{\lambda}+M\bm{\delta})
  -\eta(x;\bm{\lambda}+M\bm{\delta})\bigr)\n
  &=\left\{
  \begin{array}{ll}
  {\displaystyle
  \frac{(x+M+d-a)(x+M+d-b)(x+d-c)x}{2x+M+d}}&:\text{R}\\[10pt]
  {\displaystyle (1-q)\frac{abc}{d}
  \frac{(1-a^{-1}dq^{x+M})(1-b^{-1}dq^{x+M})(1-c^{-1}dq^x)(1-q^x)}
  {q^{x+1}(1-dq^{2x+M})}}&:\text{$q$R}
  \end{array}\right..
  \label{D(eta-eta)}
\end{align}
These denominators vanish at $x=-\frac12(M+\lambda_4)\eqdef x_0$
and their residues are related as
\begin{align*}
  &\quad\text{Res}_{x=x_0}\Bigl(B(x;\bm{\lambda}+M\tilde{\bm{\delta}})
  \bigl(\eta(x+1;\bm{\lambda}+M\bm{\delta})
  -\eta(x;\bm{\lambda}+M\bm{\delta})\bigr)\Bigr)\\
  &=-\text{Res}_{x=x_0}\Bigl(D(x;\bm{\lambda}+M\tilde{\bm{\delta}})
  \bigl(\eta(x-1;\bm{\lambda}+M\bm{\delta})
  -\eta(x;\bm{\lambda}+M\bm{\delta})\bigr)\Bigr).
\end{align*}
(For $q$R, \eqref{B(eta-eta)}--\eqref{D(eta-eta)} are rational functions of
$z=q^x$, and it is better to consider the residue with respect to $z$ at
$z=\pm(dq^M)^{-\frac12}$.)
At $x=x_0$, we have
\begin{equation*}
  \eta\bigl(x_0;\bm{\lambda}+(M-1)\bm{\delta}\bigr)
  =\eta\bigl(x_0+1;\bm{\lambda}+(M-1)\bm{\delta}\bigr),
  \ \ \eta(x_0+1;\bm{\lambda}+M\bm{\delta})
  =\eta(x_0-1;\bm{\lambda}+M\bm{\delta}).
\end{equation*}
Combining these and \eqref{F2}, we obtain
\begin{equation*}
  \text{Res}_{x=x_0}F(x)=0.
\end{equation*}
Therefore $F(x)$ \eqref{F2} is a (Laurent) polynomial in $x$ ($q^x$).
For the involution $\mathcal{I}_{\bm{\lambda}}$ \eqref{autom}, we have
\begin{align*}
  &\quad
  \mathcal{I}_{\bm{\lambda}+M\bm{\delta}}\Bigl(
  B(x;\bm{\lambda}+M\tilde{\bm{\delta}})
  \bigl(\eta(x+1;\bm{\lambda}+M\bm{\delta})
  -\eta(x;\bm{\lambda}+M\bm{\delta})\bigr)\Bigr)\\
  &=D(x;\bm{\lambda}+M\tilde{\bm{\delta}})
  \bigl(\eta(x-1;\bm{\lambda}+M\bm{\delta})
  -\eta(x;\bm{\lambda}+M\bm{\delta})\bigr),\\
  &\quad\mathcal{I}_{\bm{\lambda}+M\bm{\delta}}\Bigl(
  \eta\bigl(x;\bm{\lambda}+(M-1)\bm{\delta}\bigr)\Bigr)
  =\eta\bigl(x+1;\bm{\lambda}+(M-1)\bm{\delta}\bigr),\\
  &\quad\mathcal{I}_{\bm{\lambda}+M\bm{\delta}}\bigl(
  \eta(x+1;\bm{\lambda}+M\bm{\delta})\bigr)
  =\eta(x-1;\bm{\lambda}+M\bm{\delta}).
\end{align*}
Hence $F(x)$ \eqref{F2} satisfies
$\mathcal{I}_{\bm{\lambda}+M\bm{\delta}}\bigl(F(x)\bigr)=F(x)$.
By Lemma\,\ref{lem:invI}, $F(x)$ is a polynomial in
$\eta(x;\bm{\lambda}+M\bm{\delta})$.
Therefore, from \eqref{tHDcXcPDn}, we have shown that
$\widetilde{\mathcal{H}}^{\text{cont}}_{\mathcal{D}}(\bm{\lambda})
\bigl(\check{X}(x)\check{P}_{\mathcal{D},n}(x;\bm{\lambda})\bigr)$
is a polynomial in $\eta(x;\bm{\lambda}+M\bm{\delta})$.

\subsubsection{step 3}
\label{sec:step_3}

Let us summarize the result.
For the denominator polynomial
$\Xi_{\mathcal{D}}(\eta)=\Xi_{\mathcal{D}}(\eta;\bm{\lambda})$ and
a polynomial in $\eta$, $Y(\eta)(\neq 0)$, we set $X(\eta)=X(\eta;\bm{\lambda})
=X^{\mathcal{D},Y}(\eta;\bm{\lambda})$ as
\begin{equation}
  X(\eta)=I_{\bm{\lambda}+M\bm{\delta}}
  \bigl[\Xi_{\mathcal{D}}Y\bigr](\eta),\quad
  \deg X(\eta)=L=\ell_{\mathcal{D}}+\deg Y(\eta)+1,
  \label{X=I[XiY]}
\end{equation}
where $\Xi_{\mathcal{D}}Y$ means a polynomial
$(\Xi_{\mathcal{D}}Y)(\eta)=\Xi_{\mathcal{D}}(\eta)Y(\eta)$.
Note that $L\geq M+1$ because of $\ell_{\mathcal{D}}\geq M$.
The minimal degree one, which corresponds to $Y(\eta)=1$, is
\begin{equation}
  X_{\text{min}}(\eta)=I_{\bm{\lambda}+M\bm{\delta}}
  \bigl[\Xi_{\mathcal{D}}\bigr](\eta),\quad
  \deg X_{\text{min}}(\eta)=\ell_{\mathcal{D}}+1.
  \label{Xmin}
\end{equation}
Then we have the following theorem.
\begin{thm}
\label{thm:rr_indet_a}
Let the parameter $a$ be an indeterminate.
For any polynomial $Y(\eta)(\neq 0)$, we take
$X(\eta)=X^{\mathcal{D},Y}(\eta)$ as \eqref{X=I[XiY]}.
Then the multi-indexed ($q$-)Racah polynomials $P_{\mathcal{D},n}(\eta)$
satisfy $1+2L$ term recurrence relations with constant coefficients:
\begin{align}
  &X(\eta)P_{\mathcal{D},n}(\eta)
  =\sum_{k=-L}^Lr_{n,k}^{X,\mathcal{D}}P_{\mathcal{D},n+k}(\eta)
  \ \ (n\in\mathbb{Z}_{\geq 0}\,;\,\eta\in\mathbb{C}),
  \label{XPthm}\\
  \text{or}\quad&\check{X}(x)\check{P}_{\mathcal{D},n}(x)
  =\sum_{k=-L}^Lr_{n,k}^{X,\mathcal{D}}\check{P}_{\mathcal{D},n+k}(x)
  \ \ (n\in\mathbb{Z}_{\geq 0}\,;\,x\in\mathbb{C}).
  \label{cXPthm}
\end{align}
\end{thm}
{\bf Remark 1}\,
We have assumed the convention \eqref{Pn=0}. If we replace
$\sum\limits_{k=-L}^L$ with $\sum\limits_{k=-\min(L,n)}^L$, it is unnecessary.

\noindent
{\bf Remark 2}\,
As shown near \eqref{X-X}, any polynomial $X(\eta)$ giving the recurrence
relations with constant coefficients must have the form \eqref{X=I[XiY]}.

\noindent
{\bf Remark 3}\,
Direct verification of this theorem is rather straightforward for lower
$M$ and smaller $d_j$, $n$ and $\deg Y$, by a computer algebra system,
e.g.\! Mathematica.
The coefficients $r_{n,k}^{X,\mathcal{D}}$ are explicitly obtained for
small $d_j$ and $n$. However, to obtain the closed expression of
$r_{n,k}^{X,\mathcal{D}}$ for general $n$ is not an easy task even for
small $d_j$, and it is a different kind of problem.
We present some examples in \S\,\ref{sec:ex}.

\noindent
{\bf Remark 4}\,
Explicit examples (see \S\,\ref{sec:ex}) suggest that, for $1\leq k\leq L$,
the coefficients $r_{n,k}^{X,\mathcal{D}}$ have the factor
$\left\{\begin{array}{ll}(a+n)_k&:\text{\rm R}\\
(aq^n;q)_k&:\text{\rm $q$R}\end{array}\right.$ and
$r_{n,-k}^{X,\mathcal{D}}$ have the factor
$\left\{\begin{array}{ll}(n-k+1)_k&:\text{\rm R}\\
(q^{n-k+1};q)_k&:\text{\rm $q$R}\end{array}\right.$.
\vspace*{2mm}

\noindent
{\bf Remark 5}\,
Since $Y(\eta)$ is arbitrary, we obtain infinitely many recurrence relations.
However not all of them are independent. The relations among them are unclear.
For `$M=0$ case' (namely, ordinary orthogonal polynomials), it is trivial that
recurrence relations obtained from arbitrary $Y(\eta)$ ($\deg Y\geq 1$)
are derived by the three term recurrence relations.

\subsection{Parameter $a$ : \eqref{a=-N}}
\label{sec:a=-N}

In this subsection we assume that the parameter $a$ is given by \eqref{a=-N}.
We write $a$, $\check{\Xi}_{\mathcal{D}}(x)$, $\check{P}_{\mathcal{D},n}(x)$,
$\check{X}(x)$, $Y(\eta)$, $r_{n,k}^{X,\mathcal{D}}$ etc.\ in
\S\,\ref{sec:a:indet} as those with bar: $\bar{a}$,
$\check{\bar{\Xi}}_{\mathcal{D}}(x)$, $\check{\bar{P}}_{\mathcal{D},n}(x)$,
$\check{\bar{X}}(\eta)$, $\bar{Y}(\eta)$,
$\bar{r}_{n,k}^{\bar{X},\mathcal{D}}$ etc.
This notation is used in this subsection only.
In the limit $\bar{a}\to a$, the quantities with bar reduce to the quantities
without bar, if they exist.
As remarked in the end of \S\,\ref{sec:miop}, the $\bar{a}\to a$ limit of
$\check{\bar{P}}_{\mathcal{D},n}(x)$ exists for
$n\in\{0,1,\ldots,n_{\text{max}}\}$ and $x\in\mathbb{C}$, or
$n\in\mathbb{Z}_{>n_{\text{max}}}$ and $x\in\{0,1,\ldots,x_{\text{max}}\}$,
but does not exist for $n\in\mathbb{Z}_{>n_{\text{max}}}$ and
$x\in\mathbb{C}\backslash\{0,1,\ldots,x_{\text{max}}\}$,
for which $\check{\bar{P}}_{\mathcal{D},n}(x)$ behaves as
$\sim\frac{1}{\bar{a}-a}$.
Recall the relation \eqref{PD0=Xi}.
We take $\bar{Y}(\eta)$ such that its $\bar{a}\to a$ limit exists,
$\lim\limits_{\bar{a}\to a}\bar{Y}(\eta)=Y(\eta)$.
Then $\check{\bar{X}}(x)$ also has a finite limit,
$\lim\limits_{\bar{a}\to a}\check{\bar{X}}(x)=\check{X}(x)$.
In \eqref{cXPthm} with $x\in\{0,1,\ldots,x_{\text{max}}\}$,
$\check{\bar{X}}(x)$, $\check{\bar{P}}_{\mathcal{D},n}(x)$ and
$\check{\bar{P}}_{\mathcal{D},n+k}(x)$ have finite $\bar{a}\to a$ limits.
So the coefficients $\bar{r}_{n,k}^{\bar{X},\mathcal{D}}$ also have finite
limits, $\lim\limits_{\bar{a}\to a}\bar{r}_{n,k}^{\bar{X},\mathcal{D}}
=r_{n,k}^{X,\mathcal{D}}$.

For $n\in\{0,1,\ldots,n_{\text{max}}\}$, the $\bar{a}\to a$ limit of
\eqref{cXPthm} gives
\begin{equation}
  \check{X}(x)\check{P}_{\mathcal{D},n}(x)
  =\!\!\sum_{k=-\min(L,n)}^{\min(L,N-n)}\!\!\!r_{n,k}^{X,\mathcal{D}}
  \check{P}_{\mathcal{D},n+k}(x)
  +\lim_{\bar{a}\to a}\sum_{k=\min(L,N-n)+1}^L\!\!\!\!\!
  \bar{r}_{n,k}^{\bar{X},\mathcal{D}}\check{\bar{P}}_{\mathcal{D},n+k}(x)
  \ \ (x\in\mathbb{C}),
  \label{cXP_lim}
\end{equation}
where the second sum is zero unless $N-L+1\leq n\leq N$.
So, unless $N-L+1\leq n\leq N$, we have obtained the recurrence relations
for $a$ \eqref{a=-N}.
Let us consider the case $N-L+1\leq n\leq N$.
This relation implies
\begin{equation*}
  \lim_{\bar{a}\to a}\sum_{k=\min(L,N-n)+1}^L\!\!\!\!\!
  \bar{r}_{n,k}^{\bar{X},\mathcal{D}}\check{\bar{P}}_{\mathcal{D},n+k}(x)
  :\text{finite}
  \ \ (x\in\mathbb{C}).
\end{equation*}
Since this is a polynomial, this means
\begin{equation*}
  \lim_{\bar{a}\to a}
  \bar{r}_{n,k}^{\bar{X},\mathcal{D}}\check{\bar{P}}_{\mathcal{D},n+k}(x)
  :\text{finite}
  \ \ (\min(L,N-n)+1\leq k\leq L\,;\,x\in\mathbb{C}).
\end{equation*}
Note that $\min(L,N-n)+1\leq k\leq L\Leftrightarrow N-n+1\leq k\leq L$.
By setting $x\in\mathbb{C}\backslash\{0,1,\ldots,x_{\text{max}}\}$,
for which $\check{\bar{P}}_{\mathcal{D},n+k}(x)$ diverge in the $\bar{a}\to a$
limit, we obtain
\begin{equation}
  r_{n,k}^{X,\mathcal{D}}
  =\lim_{\bar{a}\to a}\bar{r}_{n,k}^{\bar{X},\mathcal{D}}=0
  \ \ (N-n+1\leq k\leq L).
\end{equation}
Note that Remark\,4 below Theorem\,\ref{thm:rr_indet_a} is
consistent with this.
The relation \eqref{cXP_lim} with $x\in\{0,1,\ldots,x_{\text{max}}\}$,
for which $\check{\bar{P}}_{\mathcal{D},n+k}(x)$ have finite $\bar{a}\to a$
limits, gives the following theorem.
\begin{thm}
\label{thm:rr_a=-N}
Let the parameter $a$ be \eqref{a=-N}.
For any polynomial $Y(\eta)(\neq 0)$, we take
$X(\eta)=X^{\mathcal{D},Y}(\eta)$ as \eqref{X=I[XiY]}.
Then the multi-indexed ($q$-)Racah polynomials $\check{P}_{\mathcal{D},n}(x)$
satisfy $1+2L$ term recurrence relations with constant coefficients:
\begin{equation}
  \check{X}(x)\check{P}_{\mathcal{D},n}(x)
  =\sum_{k=-\min(L,n)}^{\min(L,N-n)}
  r_{n,k}^{X,\mathcal{D}}\check{P}_{\mathcal{D},n+k}(x)
  \ \ \Bigl(\begin{array}{ll}
  n=0,1,\ldots,n_{\text{max}}\\
  x=0,1,\ldots,x_{\text{max}}
  \end{array}\Bigr).
  \label{cXcP}
\end{equation}
\end{thm}
{\bf Remark 1}\,
For $L>\frac12N$, the number of terms is not $1+2L$ but $N+1$.

\noindent
{\bf Remark 2}\,
Unless $N-L+1\leq n\leq N$, \eqref{cXcP} is an equation as a polynomial,
namely it holds for $x\in\mathbb{C}$.
On the other hand, for $N-L+1\leq n\leq N$, \eqref{cXcP} holds only for
$x=0,1,\ldots,x_{\text{max}}$.

\noindent
{\bf Remark 3}\,
If we set $r_{n,k}^{X,\mathcal{D}}=0$ unless $0\leq n+k\leq n_{\text{max}}$
(see Remark\,3 below Theorem\,\ref{thm:rr_indet_a}),
the sum $\sum\limits_{k=-\min(L,n)}^{\min(L,N-n)}$ in \eqref{cXcP} can
be rewritten as $\sum\limits_{k=-L}^L$,
\begin{equation}
  \check{X}(x;\bm{\lambda})\check{P}_{\mathcal{D},n}(x;\bm{\lambda})
  =\sum_{k=-L}^Lr_{n,k}^{X,\mathcal{D}}(\bm{\lambda})
  \check{P}_{\mathcal{D},n+k}(x;\bm{\lambda})
  \ \ \Bigl(\begin{array}{ll}
  n=0,1,\ldots,n_{\text{max}}\\
  x=0,1,\ldots,x_{\text{max}}
  \end{array}\Bigr).
  \label{cXcP2}
\end{equation}

\noindent
{\bf Remark 4}\,
By \eqref{P=(eta-eta)p}, $\check{X}(x;\bm{\lambda})$ is expressed as
\begin{align}
  \check{X}(x;\bm{\lambda})
  &=\sum_{j=1}^x\bigl(\eta(j;\bm{\lambda}+M\bm{\delta})
  -\eta(j-1;\bm{\lambda}+M\bm{\delta})\bigr)\n
  &\qquad\quad\times\check{\Xi}_{\mathcal{D}}(j;\bm{\lambda})
  Y\bigl(\eta(j;\bm{\lambda}+(M-1)\bm{\delta})\bigr)
  \ \ (x\in\mathbb{Z}_{\geq 0}).
\end{align}

For later use, we provide a conjecture about $r_{n,k}^{X,\mathcal{D}}$.
\begin{conj}
The coefficients $r_{n,k}^{X,\mathcal{D}}$ are rational functions of $n$
(for R) or $q^n$ (for $q$R).
They satisfy
\begin{equation}
  \text{\rm R}:\ r_{n,k}^{X,\mathcal{D}}(\bm{\lambda})
  \Bigl|_{n\to-n-\tilde{d}}
  =r_{n,-k}^{X,\mathcal{D}}(\bm{\lambda}),\quad
  \text{\rm $q$R}:\ r_{n,k}^{X,\mathcal{D}}(\bm{\lambda})
  \Bigl|_{q^n\to q^{-n}\tilde{d}^{-1}}
  =r_{n,-k}^{X,\mathcal{D}}(\bm{\lambda}),
  \label{rnk|n->}
\end{equation}
for $1\leq k\leq L$.
Therefore $r_{n,k}^{X,\mathcal{D}}(\bm{\lambda})+
r_{n,-k}^{X,\mathcal{D}}(\bm{\lambda})$ $(1\leq k\leq L)$ is a rational
function of $\mathcal{E}_n(\bm{\lambda})$ and let this rational function be
$I_k(z)=I_k(z;\bm{\lambda})$, namely
$I_k\bigl(\mathcal{E}_n(\bm{\lambda});\bm{\lambda}\bigr)=
r_{n,k}^{X,\mathcal{D}}(\bm{\lambda})+r_{n,-k}^{X,\mathcal{D}}(\bm{\lambda})$.
The following function $I(z)=I(z;\bm{\lambda})$,
\begin{equation}
  I(z)\eqdefrm
  \prod_{j=1}^L\alpha_j(z)\alpha_{2L+1-j}(z)\times\sum_{k=1}^LI_k(z),
  \label{Iz}
\end{equation}
is a polynomial of degree $2L$ in $z$.
Here $\alpha_j(z)\alpha_{2L+1-j}(z)$ will be given in \eqref{al*al}.
\label{conj_rnk2}
\end{conj}

The recurrence relations \eqref{XPthm} or \eqref{cXcP} with $\eta=0$ or $x=0$
and the normalization \eqref{PDn(0)=1} give
\begin{equation}
  r_{n,0}^{X,\mathcal{D}}
  =-\sum_{k=1}^L\bigl(r_{n,k}^{X,\mathcal{D}}+r_{n,-k}^{X,\mathcal{D}}\bigr).
  \label{rn0}
\end{equation}
Hence \eqref{rnk|n->} also holds for $k=0$ and the second factor of $I(z)$
\eqref{Iz} corresponds to $-r_{n,0}^{X,\mathcal{D}}$.
By the recurrence relations \eqref{cXcP} and orthogonality relations
\eqref{orthoPDn} (with an appropriate range of parameters \eqref{Mrange}),
we have
\begin{align*}
  &\quad\sum_{x=0}^{x_{\text{max}}}\frac{\psi_{\mathcal{D}}(x;\bm{\lambda})^2}
  {\check{\Xi}_{\mathcal{D}}(1;\bm{\lambda})}
  \Bigl(\check{X}(x;\bm{\lambda})
  \check{P}_{\mathcal{D},n}(x;\bm{\lambda})\Bigr)
  \check{P}_{\mathcal{D},n+k}(x;\bm{\lambda})
  =\frac{r_{n,k}^{X,\mathcal{D}}(\bm{\lambda})}
  {d_{\mathcal{D},n+k}(\bm{\lambda})^2}\\
  &=\sum_{x=0}^{x_{\text{max}}}\frac{\psi_{\mathcal{D}}(x;\bm{\lambda})^2}
  {\check{\Xi}_{\mathcal{D}}(1;\bm{\lambda})}
  \check{P}_{\mathcal{D},n}(x;\bm{\lambda})
  \Bigl(\check{X}(x;\bm{\lambda})
  \check{P}_{\mathcal{D},n+k}(x;\bm{\lambda})\Bigr)
  =\frac{r_{n+k,-k}^{X,\mathcal{D}}(\bm{\lambda})}
  {d_{\mathcal{D},n}(\bm{\lambda})^2},
\end{align*}
for $1\leq k\leq L$ and $n+k\leq n_{\text{max}}$.
So we obtain the relations among the coefficients $r_{n,k}^{X,\mathcal{D}}$,
\begin{equation}
  r_{n+k,-k}^{X,\mathcal{D}}(\bm{\lambda})
  =\frac{d_{\mathcal{D},n}(\bm{\lambda})^2}
  {d_{\mathcal{D},n+k}(\bm{\lambda})^2}\,
  r_{n,k}^{X,\mathcal{D}}(\bm{\lambda})
  \ \ (1\leq k\leq L\,;\,n+k\leq n_{\text{max}}),
\end{equation}
which are valid for any parameter ranges (except for the zeros of the
denominators).
Therefore it is sufficient to find $r_{n,k}^{X,\mathcal{D}}$ ($1\leq k\leq L$).
For sufficiently large $N$ (or treating $a$ as an indeterminate),
the top coefficient $r_{n,L}^{X,\mathcal{D}}$ is easily
obtained by comparing the highest degree terms,
\begin{equation}
  r_{n,L}^{X,\mathcal{D}}
  =\frac{c^Xc^P_{\mathcal{D},n}}{c^P_{\mathcal{D},n+L}},
\end{equation}
where $c^X$ is the coefficient of the highest term of
$X(\eta)=c^X\eta^L+(\text{lower order terms})$ and
$c^P_{\mathcal{D},n}$ is given by \eqref{cPDn}.

\subsection{Examples}
\label{sec:ex}

For illustration, we present some examples of the coefficients
$r_{n,k}^{X,\mathcal{D}}$ of the recurrence relations \eqref{XP} for
$X(\eta)=X_{\text{min}}(\eta)$ and small $d_j$.
The parameter $a$ is treated as an indeterminate.
Since the overall normalization of $X(\eta)$ is not important,
we multiply $X(\eta)$ \eqref{X=I[XiY]} by an appropriate factor.

\subsubsection{multi-indexed Racah polynomials}
\label{sec:ex_R}

We set $\sigma_1=a+b$, $\sigma_2=ab$, $\sigma'_1=c+d$ and $\sigma'_2=cd$.\\
\noindent
\underline{Ex.1} $\mathcal{D}=\{1\}$, $Y(\eta)=1$
($\Rightarrow X(\eta)=X_{\text{min}}(\eta)$): 5-term recurrence relations
\begin{align}
  X(\eta)&=2c(d-a+1)(d-b+1)
  I_{\bm{\lambda}+\bm{\delta}}[\Xi_{\mathcal{D}}](\eta)\n
  &=\eta\bigl((2-\sigma_1+\sigma'_1)\eta
  -\sigma_1(2c+d+2\sigma'_2)+2\sigma_2c+2\sigma'_1
  +\sigma'_2(5+2d)+d^2\bigr),\n
  r_{n,2}^{X,\mathcal{D}}&=
  \frac{(2-\sigma_1+\sigma'_1)(c+n)(c+n+3)(a+n,b+n,\tilde{d}+n)_2}
  {(\tilde{d}+2n)_4},\n
  r_{n,-2}^{X,\mathcal{D}}&=
  \frac{(2-\sigma_1+\sigma'_1)(\tilde{d}-c+n-3)(\tilde{d}-c+n)
  (\tilde{d}-a+n-1,\tilde{d}-b+n-1,n-1)_2}
  {(\tilde{d}+2n-3)_4},\n
  r_{n,1}^{X,\mathcal{D}}&=
  \frac{2(a+n)(b+n)(c+n)(c+n+2)(\tilde{d}-c+n)(\tilde{d}+n)}
  {(\tilde{d}+2n+3)(\tilde{d}+2n-1)_3}\n
  &\quad\times
  \Bigl(-2(2-\sigma_1+\sigma'_1)n(n+\tilde{d}+1)
  +2(1-\tilde{d})(1+c-\sigma_2)+d(1-\tilde{d}^2)\Bigr),\\
  r_{n,-1}^{X,\mathcal{D}}&=
  \frac{2n(\tilde{d}-a+n)(\tilde{d}-b+n)(c+n)(\tilde{d}-c+n-2)(\tilde{d}-c+n)}
  {(\tilde{d}+2n-3)(\tilde{d}+2n-1)_3}\n
  &\quad\times
  \Bigl(-2(2-\sigma_1+\sigma'_1)n(n+\tilde{d}-1)
  +2(1+c-\sigma_2)+2(\sigma_2+c-\tilde{d})\tilde{d}+d(1-\tilde{d}^2)\Bigr),\n
  r_{n,0}^{X,\mathcal{D}}&=
  -\sum_{k=1}^2\bigl(r_{n,-k}^{X,\mathcal{D}}+r_{n,k}^{X,\mathcal{D}}\bigr).
  \nonumber
\end{align}
\ignore{
\begin{align}
  X(\eta)&=X_{\text{min}}(\eta)
  =\frac{\eta}{2c(d-a+1)(d-b+1)}\n
  &\quad\times\Bigl((2-\sigma_1+\sigma'_1)\eta
  -\sigma_1(2c+d+2\sigma'_2)+2\sigma_2c+2\sigma'_1
  +\sigma'_2(5+2d)+d^2\Bigr),\n
  r_{n,2}^{X,\mathcal{D}}&=
  \frac{(2-\sigma_1+\sigma'_1)(c+n)(c+n+3)(a+n,b+n,\tilde{d}+n)_2}
  {2c(d-a+1)(d-b+1)(\tilde{d}+2n)_4},\n
  r_{n,-2}^{X,\mathcal{D}}&=
  \frac{(2-\sigma_1+\sigma'_1)(\tilde{d}-c+n-3)(\tilde{d}-c+n)
  (\tilde{d}-a+n-1,\tilde{d}-b+n-1,n-1)_2}
  {2c(d-a+1)(d-b+1)(\tilde{d}+2n-3)_4},\n
  r_{n,1}^{X,\mathcal{D}}&=
  \frac{(a+n)(b+n)(c+n)(c+n+2)(\tilde{d}-c+n)(\tilde{d}+n)}
  {c(d-a+1)(d-b+1)(\tilde{d}+2n+3)(\tilde{d}+2n-1)_3}\n
  &\quad\times
  \Bigl(-2(2-\sigma_1+\sigma'_1)n(n+\tilde{d}+1)
  +2(1-\tilde{d})(1+c-\sigma_2)+d(1-\tilde{d}^2)\Bigr),\\
  r_{n,-1}^{X,\mathcal{D}}&=
  \frac{n(\tilde{d}-a+n)(\tilde{d}-b+n)(c+n)(\tilde{d}-c+n-2)(\tilde{d}-c+n)}
  {c(d-a+1)(d-b+1)(\tilde{d}+2n-3)(\tilde{d}+2n-1)_3}\n
  &\quad\times
  \Bigl(-2(2-\sigma_1+\sigma'_1)n(n+\tilde{d}-1)
  +2(1+c-\sigma_2)+2(\sigma_2+c-\tilde{d})\tilde{d}+d(1-\tilde{d}^2)\Bigr),\n
  r_{n,0}^{X,\mathcal{D}}&=
  -\sum_{k=1}^2\bigl(r_{n,-k}^{X,\mathcal{D}}+r_{n,k}^{X,\mathcal{D}}\bigr).
  \nonumber
\end{align}
}
Direct calculation shows that $I(z)$ \eqref{Iz} is a polynomial of degree
4 in $z$. 
Its explicit form is somewhat lengthy and we omit it.

We have also obtained 7-term recurrence relations for
$\mathcal{D}=\{2\},\{1,2\}$ with $X(\eta)=X_{\text{min}}(\eta)$ and
$\mathcal{D}=\{1\}$ with non-minimal $X(\eta)$ ($Y(\eta)=\eta$).
Since the explicit forms of $r_{n,k}^{X,\mathcal{D}}$ are somewhat lengthy,
we do not write down them here.

\subsubsection{multi-indexed $q$-Racah polynomials}
\label{sec:ex_qR}

We set $\sigma_1=a+b$, $\sigma_2=ab$, $\sigma'_1=c+d$ and $\sigma'_2=cd$.\\
\noindent
\underline{Ex.1} $\mathcal{D}=\{1\}$, $Y(\eta)=1$
($\Rightarrow X(\eta)=X_{\text{min}}(\eta)$): 5-term recurrence relations
\begin{align}
  X(\eta)&=(1+q)(1-c)(1-a^{-1}dq)(1-b^{-1}dq)
  I_{\bm{\lambda}+\bm{\delta}}[\Xi_{\mathcal{D}}](\eta)\n
  &=\eta\Bigl((1-\sigma_2^{-1}\sigma'_2q^2)\eta
  +\sigma_2^{-1}q^2(1+q-2cq)d^2\n
  &\phantom{\quad\times\Bigl(}
  -\sigma_2^{-1}\bigl(\sigma_1q(1+q)(1-c)+(1-q)(\sigma_2+cq^2)\bigr)d
  +2-c(1+q)\Bigr),\n
  r_{n,2}^{X,\mathcal{D}}&=
  \frac{(1-\sigma_2^{-1}\sigma'_2q^2)(1-cq^n)(1-cq^{n+3})
  (aq^n,bq^n,\tilde{d}q^n;q)_2}
  {(\tilde{d}q^{2n};q)_4},\n
  r_{n,-2}^{X,\mathcal{D}}&=
  \frac{d^2q^2(1-\sigma_2^{-1}\sigma'_2q^2)
  (1-c^{-1}\tilde{d}q^{n-3})(1-c^{-1}\tilde{d}q^n)
  (a^{-1}\tilde{d}q^{n-1},b^{-1}\tilde{d}q^{n-1},q^{n-1};q)_2}
  {(\tilde{d}q^{2n-3};q)_4},\n
  r_{n,1}^{X,\mathcal{D}}&=
  \frac{(1+q)(1-aq^n)(1-bq^n)(1-cq^n)(1-cq^{n+2})(1-c^{-1}\tilde{d}q^n)
  (1-\tilde{d}q^n)}
  {\sigma_2d
  (1-\tilde{d}q^{2n+3})(\tilde{d}q^{2n-1};q)_3}\n
  &\quad\times
  \Bigl(-\bigl(\sigma_2\sigma'_1+\sigma_1(1-c)dq
  -\sigma'_1dq^2\bigr)(\sigma_2cq^{2n}+d)\n
  &\phantom{\quad\times\Bigl(}
  +(q+q^{-1})d\bigl(\sigma_1\sigma_2c+\sigma_2(1-c)\sigma'_1q
  -\sigma_1\sigma'_2q^2\bigr)q^n\Bigr),\\
  r_{n,-1}^{X,\mathcal{D}}&=
  \frac{(1+q)(1-q^n)(1-a^{-1}\tilde{d}q^n)(1-b^{-1}\tilde{d}q^n)
  (1-cq^n)(1-c^{-1}\tilde{d}q^{n-2})(1-c^{-1}\tilde{d}q^n)}
  {\sigma_2
  (1-\tilde{d}q^{2n-3})(\tilde{d}q^{2n-1};q)_3}\n
  &\quad\times
  \Bigl(-\bigl(\sigma_2\sigma'_1
  +\sigma_1(1-c)dq-\sigma'_1dq^2\bigr)(\sigma_2cq^{2n-1}+dq)\n
  &\phantom{\quad\times\Bigl(}
  +(q+q^{-1})d\bigl(\sigma_1\sigma_2c+\sigma_2(1-c)\sigma'_1q
  -\sigma_1\sigma'_2q^2\bigr)q^n
  \Bigr),\n
  r_{n,0}^{X,\mathcal{D}}&=
  -\sum_{k=1}^2\bigl(r_{n,-k}^{X,\mathcal{D}}+r_{n,k}^{X,\mathcal{D}}\bigr).
  \nonumber
\end{align}
Direct calculation shows that $I(z)$ \eqref{Iz} is a polynomial of degree
4 in $z$. 
Its explicit form is somewhat lengthy and we omit it.

We have also obtained 7-term recurrence relations for $\mathcal{D}=\{2\}$
with $X(\eta)=X_{\text{min}}(\eta)$.
Since the explicit forms of $r_{n,k}^{X,\mathcal{D}}$ are somewhat lengthy,
we do not write down them here.

\section{Generalized Closure Relations and Creation and Annihilation Operators}
\label{sec:gcr}

In this section we discuss the generalized closure relations and the creation
and annihilation operators of the multi-indexed ($q$-)Racah rdQM systems
described by $\mathcal{H}_{\mathcal{D}}$ \eqref{HD}.

First let us recapitulate the essence of the (generalized) closure relation
\cite{rrmiop4}.
The closure relation of order $K$ is an algebraic relation between a
Hamiltonian $\mathcal{H}$ and some operator
$X$ ($=X(\eta(x))=\check{X}(x)$) \cite{rrmiop4}:
\begin{equation}
  (\text{ad}\,\mathcal{H})^KX
  =\sum_{i=0}^{K-1}(\text{ad}\,\mathcal{H})^iX\cdot R_i(\mathcal{H})
  +R_{-1}(\mathcal{H}),
  \label{crK}
\end{equation}
where $(\text{ad}\,\mathcal{H})X=[\mathcal{H},X]$,
$(\text{ad}\,\mathcal{H})^0X=X$ and $R_i(z)=R^X_i(z)$ is
a polynomial in $z$.
The original closure relation \cite{os7,os12} corresponds to $K=2$.
Since the closure relation of order $K$ implies that of order $K'>K$, 
we are interested in the smallest integer $K$ satisfying \eqref{crK}.
We assume that the matrix $A=(a_{ij})_{1\leq i,j\leq K}$
($a_{i+1,i}=1$ ($1\leq i\leq K-1$), $a_{i+1,K}=R_i(z)$ ($0\leq i\leq K-1$),
$a_{ij}=0$ (others)) has $K$ distinct real non-vanishing eigenvalues
$\alpha_i=\alpha_i(z)$ for $z\geq 0$, which are indexed in decreasing order
$\alpha_1(z)>\alpha_2(z)>\cdots>\alpha_K(z)$.
Then we obtain the exact Heisenberg solution of $X$,
\begin{equation}
  X_{\text{H}}(t)\eqdef e^{i\mathcal{H}t}Xe^{-i\mathcal{H}t}
  =\sum_{n=0}^{\infty}\frac{(it)^n}{n!}(\text{ad}\,\mathcal{H})^nX
  =\sum_{j=1}^Ka^{(j)}e^{i\alpha_j(\mathcal{H})t}
  -R_{-1}(\mathcal{H})R_0(\mathcal{H})^{-1}.
  \label{X(t)}
\end{equation}
Here $a^{(j)}=a^{(j)}(\mathcal{H},X)$ ($1\leq j\leq K$) are creation or
annihilation operators,
\begin{equation}
  a^{(j)}=\Bigl(\sum_{i=1}^K(\text{ad}\,\mathcal{H})^{i-1}X\cdot
  p_{ij}(\mathcal{H})
  +R_{-1}(\mathcal{H})\alpha_j(\mathcal{H})^{-1}\Bigr)
  \prod_{\genfrac{}{}{0pt}{1}{k=1}{k\neq j}}^K
  (\alpha_j(\mathcal{H})-\alpha_k(\mathcal{H}))^{-1},
  \label{a(j)}
\end{equation}
where $p_{ij}(z)$ ($1\leq i,j\leq K$) are
\begin{equation}
  p_{ij}(z)=\alpha_j(z)^{K-i}-\sum_{k=1}^{K-i}R_{K-k}(z)\,\alpha_j(z)^{K-i-k}.
\end{equation}

Let us consider the rdQM systems described by the multi-indexed ($q$-)Racah
polynomials.
The Hamiltonian is $\mathcal{H}_{\mathcal{D}}$ $\eqref{HD}$ and a candidate
of the operator $X$ is a polynomial $X(\eta(x))=\check{X}(x)$ discussed in
\S\,\ref{sec:rr_const}.
The closure relation \eqref{crK} is now
\begin{equation}
  (\text{ad}\,\mathcal{H}_{\mathcal{D}})^K\underline{X}
  =\sum_{i=0}^{K-1}(\text{ad}\,\mathcal{H}_{\mathcal{D}})^i
  \underline{X}\cdot R_i(\mathcal{H}_{\mathcal{D}})
  +R_{-1}(\mathcal{H}_{\mathcal{D}}),
  \label{crHD}
\end{equation}
where $\underline{X}$ is a diagonal matrix
$\underline{X}=(\check{X}(x)\delta_{x,y})_{0\leq x,y\leq x_{\text{max}}}$.
(In the notation used in \eqref{H}, this matrix $\underline{X}$ is expressed
as $\check{X}(x){\bf 1}$ or simply $\check{X}(x)$.)
{}From the tridiagonal form of $\mathcal{H}_{\mathcal{D}}$
and by assuming that $N$ is sufficiently large, 
polynomials $R_i(z)=R^X_i(z)$ have the following degrees,
\begin{equation}
  R_i(z)=\sum_{j=0}^{K-i}r_i^{(j)}z^j
  \ \ (0\leq i\leq K-1),\quad
  R_{-1}(z)=\sum_{j=0}^Kr_{-1}^{(j)}z^j,
\end{equation}
where $r_i^{(j)}=r_i^{X(j)}$ are coefficients.

The method of \cite{rrmiop4} is
(\romannumeral1) Find $X$ and $R_i(z)$ satisfying \eqref{crK},
(\romannumeral2) Calculate the eigenvalues $\alpha_j(z)$,
(\romannumeral3) Heisenberg solution $X_{\text{H}}(t)$ is obtained,
(\romannumeral4) Creation/annihilation operators $a^{(j)}$ are obtained.
Here we reverse a part of the logic, namely exchange (\romannumeral1)
and (\romannumeral2).
First we define functions $\alpha_j(z)$ by guess work.
Next, polynomials $R_i(z)$ are defined by using $\alpha_j(z)$ and
Conjecture \ref{conj_rnk2}.
Then we check the closure relation \eqref{crHD} for these $R_i(z)$
and $\underline{X}$ with $\check{X}(x)$ given in \S\,\ref{sec:rr_const}.

Let us define $\alpha_j(z)$ ($1\leq j\leq 2L$) as follows:
\begin{align}
  \text{R}:\ \ &\alpha_j(z)=\left\{
  \begin{array}{ll}
  (L+1-j)^2+(L+1-j)\sqrt{4z+\tilde{d}^2}&(1\leq j\leq L)\\[4pt]
  (j-L)^2-(j-L)\sqrt{4z+\tilde{d}^2}&(L+1\leq j\leq 2L)
  \end{array}\right.,
  \label{alphaj_R}\\
  \text{$q$R}:\ \ &\alpha_j(z)=\left\{
  \begin{array}{ll}
  \tfrac12\bigl((q^{-\frac12(L+1-j)}-q^{\frac12(L+1-j)})^2
  (z+1+\tilde{d})\\[4pt]
  \quad
  +(q^{-(L+1-j)}-q^{L+1-j})\sqrt{(z+1+\tilde{d})^2-4\tilde{d}}\,\bigr)
  &(1\leq j\leq L)\\[4pt]
  \tfrac12\bigl((q^{-\frac12(j-L)}-q^{\frac12(j-L)})^2
  (z+1+\tilde{d})\\[4pt]
  \quad
  -(q^{-(j-L)}-q^{j-L})\sqrt{(z+1+\tilde{d})^2-4\tilde{d}}\,\bigr)
  &\!\!\!\!\!\!\!(L+1\leq j\leq 2L)
  \end{array}\right..
  \label{alphaj_qR}
\end{align}
The pair of $\alpha_j(z)$ and $\alpha_{2L+1-j}(z)$ ($1\leq j\leq L$) satisfies
\begin{align}
  \alpha_j(z)+\alpha_{2L+1-j}(z)&=\left\{
  \begin{array}{ll}
  2(L+1-j)^2&:\text{R}\\[2pt]
  (q^{-\frac12(L+1-j)}-q^{\frac12(L+1-j)})^2(z+1+\tilde{d})&:\text{$q$R}\\
  \end{array}\right.,
  \label{al+al}\\[2pt]
  \alpha_j(z)\alpha_{2L+1-j}(z)&=\left\{
  \begin{array}{ll}
  (L+1-j)^2\bigl((L+1-j)^2-4z-\tilde{d}^2\bigr)&:\text{R}\\[2pt]
  (q^{-\frac12(L+1-j)}-q^{\frac12(L+1-j)})^2\\[2pt]
  \quad\times
  \bigl((q^{-\frac12(L+1-j)}+q^{\frac12(L+1-j)})^2\tilde{d}
  -(z+1+\tilde{d})^2\bigr)&:\text{$q$R}
  \end{array}\right..
  \label{al*al}
\end{align}
These $\alpha_j(z)$ satisfy
\begin{equation}
  \alpha_1(z)>\alpha_2(z)>\cdots>\alpha_L(z)>0
  >\alpha_{L+1}(z)>\alpha_{L+2}(z)>\cdots>\alpha_{2L}(z)\ \ (z\geq 0),
  \label{alphajorder}
\end{equation}
for $\tilde{d}>2L-1$ (R) and $\tilde{d}<q^{2L-1}$ ($q$R).
We remark that $\alpha_j(\mathcal{E}_n)$ is square root free,
$\sqrt{4\mathcal{E}_n+\tilde{d}^2}=2n+\tilde{d}$ for R and
$\sqrt{(\mathcal{E}_n+1+\tilde{d})^2-4\tilde{d}^2}=q^{-n}-\tilde{d}q^n$
for $q$R.
It is easy to show the following:
\begin{equation}
  \alpha_{j}(\mathcal{E}_n)=\left\{
  \begin{array}{ll}
  \mathcal{E}_{n+L+1-j}-\mathcal{E}_n>0&(1\leq j\leq L)\\[2pt]
  \mathcal{E}_{n-(j-L)}-\mathcal{E}_n<0&(L+1\leq j\leq 2L)
  \end{array}\right..
  \label{alphajEn}
\end{equation}

Like the Wilson and Askey-Wilson cases \cite{rrmiop2}, we conjecture the
following.
\begin{conj}
\label{conj_cr}
Take $X(\eta)$ as Theorem \ref{thm:rr_a=-N} and take
$R_i(z)$ $(-1\leq i\leq 2L-1)$ as follows:
\begin{align}
  R_i(z)&=(-1)^{i+1}\!\!\!\!\!\!\!\!\!\!\!
  \sum_{1\leq j_1<j_2<\cdots<j_{2L-i}\leq 2L}\!\!\!\!\!\!\!\!\!\!\!\!\!\!
  \alpha_{j_1}(z)\alpha_{j_2}(z)\cdots\alpha_{j_{2L-i}}(z)
  \ \ (0\leq i\leq 2L-1),
  \label{Riz}\\
  R_{-1}(z)&=-I(z),
\end{align}
where $I(z)$ is given by \eqref{Iz}.
Then the closure relation of order $K=2L$ \eqref{crHD} holds.
\end{conj}
We remark that $R_i(z)$ in \eqref{Riz} are indeed polynomials in $z$,
because RHS of \eqref{Riz} are symmetric under the exchange of $\alpha_j$
and $\alpha_{2L+1-j}$ and their sum and product are polynomials in $z$,
\eqref{al+al}--\eqref{al*al}.
Since $R_i(z)$ ($0\leq i\leq 2L-1$) are expressed in terms of $\alpha_j(z)$,
they do not depend on $\mathcal{D}$ and $X$ (except for $\deg X=L$).
Only $R_{-1}(z)$ depends on $\mathcal{D}$ and $X$.
For `$L=1$ case', namely the original system ($\mathcal{D}=\emptyset$,
$\ell_{\mathcal{D}}=0$, $\Xi_{\mathcal{D}}(\eta)=1$,
$X(\eta)=X_{\text{min}}(\eta)=\eta$), this generalized closure relation
reduces to the original closure relation \cite{os12}.
Direct verification of this conjecture is straightforward for lower
$M$ and smaller $d_j$, $\deg Y$ and $N$, by a computer algebra system.

Let us assume $\tilde{d}>2L-1$ for R and $\tilde{d}<q^{2L-1}$ for $q$R.
If Conjecture \ref{conj_cr} is true,
we have the exact Heisenberg operator solution $X_{\text{H}}(t)$ \eqref{X(t)}
and the creation/annihilation operators $a^{(j)}=a^{\mathcal{D},X(j)}$
\eqref{a(j)}.
Action of \eqref{X(t)} on $\phi_{\mathcal{D}\,n}(x)$ \eqref{phiDn} is
\begin{equation*}
  e^{i\mathcal{H}_\mathcal{D}t}\underline{X}
  e^{-i\mathcal{H}_\mathcal{D}t}\phi_{\mathcal{D}\,n}(x)
  =\sum_{j=1}^{2L}e^{i\alpha_j(\mathcal{E}_n)t}a^{(j)}\phi_{\mathcal{D}\,n}(x)
  -R_{-1}(\mathcal{E}_n)R_0(\mathcal{E}_n)^{-1}\phi_{\mathcal{D}\,n}(x).
\end{equation*}
On the other hand the LHS turns out to be
\begin{align*}
  e^{i\mathcal{H}_\mathcal{D}t}\underline{X}
  e^{-i\mathcal{H}_\mathcal{D}t}\phi_{\mathcal{D}\,n}(x)
  &=e^{i\mathcal{H}_\mathcal{D}t}\underline{X}
  e^{-i\mathcal{E}_nt}\phi_{\mathcal{D}\,n}(x)
  =e^{-i\mathcal{E}_nt}e^{i\mathcal{H}_\mathcal{D}t}
  \sum_{k=-L}^Lr_{n,k}^{X,\mathcal{D}}\phi_{\mathcal{D}\,n+k}(x)\n
  &=\sum_{k=-L}^Le^{i(\mathcal{E}_{n+k}-\mathcal{E}_n)t}\,
  r_{n,k}^{X,\mathcal{D}}\phi_{\mathcal{D}\,n+k}(x),
\end{align*}
where we have used \eqref{cXcP2}.
Comparing these $t$-dependence, we obtain \eqref{alphajEn} and
\begin{align}
  &a^{(j)}\phi_{\mathcal{D}\,n}(x)=\left\{
  \begin{array}{ll}
  r_{n,L+1-j}^{X,\mathcal{D}}\phi_{\mathcal{D}\,n+L+1-j}(x)
  &(1\leq j\leq L)\\[6pt]
  r_{n,-(j-L)}^{X,\mathcal{D}}\phi_{\mathcal{D}\,n-(j-L)}(x)
  &(L+1\leq j\leq 2L)
  \end{array}\right.,
  \label{ajphiDn}\\[2pt]
  &-R_{-1}(\mathcal{E}_n)R_0(\mathcal{E}_n)^{-1}
  =r_{n,0}^{X,\mathcal{D}},
  \label{Rm1En}
\end{align}
where $r_{n,k}^{X,\mathcal{D}}=0$ unless $0\leq n+k\leq n_{\text{max}}$.
Note that \eqref{Rm1En} is consistent with Conjecture \ref{conj_rnk2} and
\eqref{rn0}.
Therefore $a^{(j)}$ ($1\leq j\leq L$) and $a^{(j)}$ ($L+1\leq j\leq 2L$)
are creation and annihilation operators, respectively.
Among them, $a^{(L)}$ and $a^{(L+1)}$ are fundamental,
$a^{(L)}\phi_{\mathcal{D},n}(x)\propto\phi_{\mathcal{D}\,n+1}(x)$ and
$a^{(L+1)}\phi_{\mathcal{D},n}(x)\propto\phi_{\mathcal{D}\,n-1}(x)$.
Furthermore, $X=X_{\text{min}}$ case is the most basic.

By the similarity transformation (see \eqref{tHD}), 
the closure relation \eqref{crHD} becomes
\begin{equation}
  (\text{ad}\,\widetilde{\mathcal{H}}_{\mathcal{D}})^K\underline{X}
  =\sum_{i=0}^{K-1}(\text{ad}\,\widetilde{\mathcal{H}}_{\mathcal{D}})^i
  \underline{X}\cdot R_i(\widetilde{\mathcal{H}}_{\mathcal{D}})
  +R_{-1}(\widetilde{\mathcal{H}}_{\mathcal{D}}),
  \label{crtHD}
\end{equation}
and the creation/annihilation operators for eigenpolynomials can be obtained,
\begin{align}
  &\tilde{a}^{(j)}\eqdef\psi_{\mathcal{D}}(x)^{-1}\circ
  a^{(j)}(\mathcal{H}_{\mathcal{D}},\underline{X})\circ\psi_{\mathcal{D}}(x)
  =a^{(j)}(\widetilde{\mathcal{H}}_{\mathcal{D}},\underline{X}),\\
  &\tilde{a}^{(j)}\check{P}_{\mathcal{D},n}(x)=\left\{
  \begin{array}{ll}
  r_{n,L+1-j}^{X,\mathcal{D}}\check{P}_{\mathcal{D},n+L+1-j}(x)
  &(1\leq j\leq L)\\[6pt]
  r_{n,-(j-L)}^{X,\mathcal{D}}\check{P}_{\mathcal{D},n-(j-L)}(x)
  &(L+1\leq j\leq 2L)
  \end{array}\right..
  \label{ajtPDn}
\end{align}

\section{Summary and Comments}
\label{sec:summary}

Following the preceding papers on the multi-indexed Laguerre and Jacobi
polynomials in oQM
\cite{rrmiop,rrmiop2,rrmiop3} and the multi-indexed Wilson and Askey-Wilson
polynomials in idQM \cite{rrmiop,rrmiop2}, we have discussed the recurrence
relations for the multi-indexed Racah and $q$-Racah polynomials in rdQM.
The $3+2M$ term recurrence relations with variable dependent coefficients
\eqref{RRPDn1} are derived (Theorem\,\ref{thm:rr_var}).
They provide an efficient method to calculate
the multi-indexed ($q$-)Racah polynomials.
Two different kinds of
the $1+2L$ term ($L\geq M+1$) recurrence relations with constant coefficients
\eqref{XPthm} and \eqref{cXcP} are derived (Theorem\,\ref{thm:rr_indet_a},
\ref{thm:rr_a=-N}), and their examples are presented.
Since $Y(\eta)$ is arbitrary, we obtain infinitely many recurrence relations.
Not all of them are independent, but the relations among them are unclear.
To clarify their relations is an important problem.
The most basic one is the minimal degree one $X_{\text{min}}(\eta)$
\eqref{Xmin}, which gives $3+2\ell_{\mathcal{D}}$ term recurrence relations.
Corresponding to the recurrence relations with constant coefficients,
the rdQM systems described by the multi-indexed ($q$-)Racah polynomials
satisfy the generalized closure relations, from which the creation and
annihilation operators are obtained.
There are many creation and annihilation operators and it is an interesting
problem to study their relations.
A proof and some data of the recurrence relations with constant coefficients
for the multi-indexed Wilson and Askey-Wilson polynomials are presented in
Appendix \ref{app:proofAW} and \ref{app:exAW}, respectively.

In rdQM, dual polynomials are introduced naturally \cite{szego,os12,os22}.
The polynomial $\mathcal{P}_n(\eta(x))$ and its dual polynomial
$\mathcal{Q}_x(\mathcal{E}_n)$ are related as
$\mathcal{P}_n(\eta(x))=\mathcal{Q}_x(\mathcal{E}_n)$,
where the roles of the variable and the `degree' (the number of zeros) are
interchanged.
The multi-indexed ($q$-)Racah polynomials $P_{\mathcal{D},n}(\eta(x))$
satisfy the second order difference equations \cite{os26} and the $1+2L$ term
recurrence relations with constant coefficients derived in this paper.
Let us introduce dual polynomial $Q_{\mathcal{D},x}(\mathcal{E}_n)$ as
$P_{\mathcal{D},n}(\eta(x))\propto Q_{\mathcal{D},x}(\mathcal{E}_n)$.
Then dual polynomials $Q_{\mathcal{D},x}(\mathcal{E}_n)$ satisfy 
the three term recurrence relations and various $2L$-th order difference
equations which depend on the choice of $Y(\eta)$.
Therefore dual polynomials $Q_{\mathcal{D},x}(\mathcal{E}_n)$ are ordinary
orthogonal polynomials and they are the Krall-type.
It is an interesting problem to study these dual polynomials in detail.
We will report on this topic elsewhere \cite{dualmiopqR}.

The ($q$-)Racah polynomial $P^{\text{($q$)R}}_n(\eta)$ and the
(Askey-)Wilson polynomial $P^{\text{(A)W}}_n(\eta)$ are the `same'
polynomials \cite{kls}.
The replacement rule of this correspondence is
\begin{equation}
  ix^{\text{(A)W}}=\gamma\bigl(x^{\text{($q$)R}}
  +\tfrac12\lambda^{\text{($q$)R}}_4\bigr),\quad
  \bm{\lambda}^{\text{(A)W}}=\bm{\lambda}^{\text{($q$)R}}
  -\tfrac12\lambda^{\text{($q$)R}}_4\bm{\delta}^{\text{($q$)R}},
  \label{rule}
\end{equation}
which gives
\begin{align}
  P^{\text{R}}_n(\eta;\bm{\lambda}^{\text{R}})
  &=(a,b,c)_n^{-1}
  P^{\text{W}}_n\bigl(-\eta-\tfrac14d^2;\bm{\lambda}^{\text{W}}\bigr),\\
  P^{\text{$q$R}}_n(\eta;\bm{\lambda}^{\text{$q$R}})
  &=d^{\frac{n}{2}}(a,b,c;q)_n^{-1}
  P^{\text{AW}}_n\bigl(\tfrac12d^{-\frac12}(\eta+1+d);
  \bm{\lambda}^{\text{AW}}\bigr).
\end{align}
(The relation between $q$R and AW is given in \cite{casoidrdqm}.
See \cite{casoidrdqm} for notation.)
This property is inherited to the multi-indexed polynomials.
The multi-indexed ($q$-)Racah polynomial
$P^{\text{($q$)R}}_{\mathcal{D},n}(\eta)$ and the multi-indexed (Askey-)Wilson
polynomial $P^{\text{(A)W}}_{\mathcal{D},n}(\eta)$ with all type $\I$ indices
are the `same' polynomials.
The replacement rule of this correspondence is
\begin{equation}
  ix^{\text{(A)W}}=\gamma\bigl(x^{\text{($q$)R}}+\tfrac12M
  +\tfrac12\lambda^{\text{($q$)R}}_4\bigr),\quad
  \bm{\lambda}^{\text{(A)W}}=\bm{\lambda}^{\text{($q$)R}}
  -\tfrac12\lambda^{\text{($q$)R}}_4\bm{\delta}^{\text{($q$)R}},
  \label{ruleM}
\end{equation}
which gives
\begin{align}
  P^{\text{R}}_{\mathcal{D},n}(\eta;\bm{\lambda}^{\text{R}})
  &=(-1)^{\ell_{\mathcal{D}}+n}
  \frac{c^{P\,\text{R}}_{\mathcal{D},n}(\bm{\lambda}^{\text{R}})}
  {c^{P\,\text{W}}_{\mathcal{D},n}(\bm{\lambda}^{\text{W}})}
  P^{\text{W}}_{\mathcal{D},n}
  \bigl(-\eta-\tfrac14(d+M)^2;\bm{\lambda}^{\text{W}}\bigr),
  \label{miop:R=W}\\
  P^{\text{$q$R}}_{\mathcal{D},n}(\eta;\bm{\lambda}^{\text{$q$R}})
  &=\bigl(2d^{\frac12}q^{\frac12M}\bigr)^{\ell_{\mathcal{D}}+n}
  \frac{c^{P\,\text{$q$R}}_{\mathcal{D},n}(\bm{\lambda}^{\text{$q$R}})}
  {c^{P\,\text{AW}}_{\mathcal{D},n}(\bm{\lambda}^{\text{AW}})}
  P^{\text{AW}}_{\mathcal{D},n}
  \bigl(\tfrac12d^{-\frac12}q^{-\frac12M}(\eta+1+dq^M);
  \bm{\lambda}^{\text{AW}}\bigr).
  \label{miop:qR=AW}
\end{align}
Here $c^P_{\mathcal{D},n}(\bm{\lambda})$ is the coefficient of the highest
degree term of $P_{\mathcal{D},n}(\eta;\bm{\lambda})$ and they are given
by \eqref{cPDn} (eq.(3.59) in \cite{os26}) and eq.(A.7) in \cite{os27}.
Therefore the recurrence relations of the multi-indexed (Askey-)Wilson
polynomials give those of the multi-indexed ($q$-)Racah polynomials.
Conversely, the recurrence relations of the multi-indexed ($q$-)Racah
polynomials give those of the multi-indexed (Askey-)Wilson polynomials
with all type $\I$ indices.

\section*{Acknowledgments}

I thank R.\,Sasaki for discussion and useful comments on the manuscript.

\bigskip
\appendix
\section{Data for Multi-indexed ($q$-)Racah Polynomials}
\label{app:data}

In this appendix we present some data for the multi-indexed ($q$-)Racah
polynomials \cite{os12,os26}, which are not presented in the main text.

\noindent
$\bullet$ potential functions:
\begin{align}
  &B(x;\bm{\lambda})=
  \left\{
  \begin{array}{ll}
  {\displaystyle
  -\frac{(x+a)(x+b)(x+c)(x+d)}{(2x+d)(2x+1+d)}}&:\text{R}\\[8pt]
  {\displaystyle-\frac{(1-aq^x)(1-bq^x)(1-cq^x)(1-dq^x)}
  {(1-dq^{2x})(1-dq^{2x+1})}}&:\text{$q$R}
  \end{array}\right.,\n
  &D(x;\bm{\lambda})=
  \left\{
  \begin{array}{ll}
  {\displaystyle
  -\frac{(x+d-a)(x+d-b)(x+d-c)x}{(2x-1+d)(2x+d)}}&:\text{R}\\[8pt]
  {\displaystyle-\tilde{d}\,
  \frac{(1-a^{-1}dq^x)(1-b^{-1}dq^x)(1-c^{-1}dq^x)(1-q^x)}
  {(1-dq^{2x-1})(1-dq^{2x})}}&:\text{$q$R}
  \end{array}\right..
  \label{B,D}
\end{align}
$\bullet$ coefficients of the three term recurrence relations:
($A_{-1}(\bm{\lambda})\eqdef 0$)
\begin{align}
  B_n(\bm{\lambda})&=-A_n(\bm{\lambda})-C_n(\bm{\lambda}),\n
  A_n(\bm{\lambda})&=\left\{
  \begin{array}{ll}
  {\displaystyle
  \frac{(n+a)(n+b)(n+c)(n+\tilde{d})}{(2n+\tilde{d})(2n+1+\tilde{d})}}
  &:\text{R}\\[10pt]
  {\displaystyle
  \frac{(1-aq^n)(1-bq^n)(1-cq^n)(1-\tilde{d}q^n)}
  {(1-\tilde{d}q^{2n})(1-\tilde{d}q^{2n+1})}}&:\text{$q$R}
  \end{array}\right.,
  \label{AnBnCn}\\
  C_n(\bm{\lambda})&=\left\{
  \begin{array}{ll}
  {\displaystyle
  \frac{(n+\tilde{d}-a)(n+\tilde{d}-b)(n+\tilde{d}-c)n}
  {(2n-1+\tilde{d})(2n+\tilde{d})}}&:\text{R}\\[12pt]
  {\displaystyle
  d\,\frac{(1-a^{-1}\tilde{d}q^n)(1-b^{-1}\tilde{d}q^n)(1-c^{-1}\tilde{d}q^n)
  (1-q^n)}{(1-\tilde{d}q^{2n-1})(1-\tilde{d}q^{2n})}}&:\text{$q$R}
  \end{array}\right..\nonumber
\end{align}
$\bullet$ ground state eigenvector: $\phi_0(x;\bm{\lambda})>0$
\begin{equation}
  \phi_0(x;\bm{\lambda})^2=\left\{
  \begin{array}{ll}
  {\displaystyle
  \frac{(a,b,c,d)_x}{(d-a+1,d-b+1,d-c+1,1)_x}\,\frac{2x+d}{d}}
  &:\text{R}\\[12pt]
  {\displaystyle
  \frac{(a,b,c,d\,;q)_x}
  {(a^{-1}dq,b^{-1}dq,c^{-1}dq,q\,;q)_x\,\tilde{d}^x}\,\frac{1-dq^{2x}}{1-d}}
  &:\text{$q$R}
  \end{array}\right..
  \label{phi0}
\end{equation}
$\bullet$ normalization constant: $d_n(\bm{\lambda})>0$
\begin{align}
  &d_n(\bm{\lambda})^2
  =\left\{
  \begin{array}{ll}
  {\displaystyle
  \frac{(a,b,c,\tilde{d})_n}
  {(\tilde{d}-a+1,\tilde{d}-b+1,\tilde{d}-c+1,1)_n}\,
  \frac{2n+\tilde{d}}{\tilde{d}}
  }&\\[10pt]
  {\displaystyle
  \quad\times
  \frac{(-1)^N(d-a+1,d-b+1,d-c+1)_N}{(\tilde{d}+1)_N(d+1)_{2N}}
  }&:\text{R}\\[10pt]
  {\displaystyle
  \frac{(a,b,c,\tilde{d}\,;q)_n}
  {(a^{-1}\tilde{d}q,b^{-1}\tilde{d}q,c^{-1}\tilde{d}q,q\,;q)_n\,d^n}\,
  \frac{1-\tilde{d}q^{2n}}{1-\tilde{d}}
  }&\\[10pt]
  {\displaystyle
  \quad\times
  \frac{(-1)^N(a^{-1}dq,b^{-1}dq,c^{-1}dq\,;q)_N\,\tilde{d}^Nq^{\frac12N(N+1)}}
  {(\tilde{d}q\,;q)_N(dq\,;q)_{2N}}
  }&:\text{$q$R}
  \end{array}\right.\!.
  \label{dn}
\end{align}
$\bullet$ energy eigenvalue:
\begin{equation}
  \mathcal{E}_n(\bm{\lambda})=
  \left\{
  \begin{array}{ll}
  n(n+\tilde{d})&:\text{R}\\
  (q^{-n}-1)(1-\tilde{d}q^n)&:\text{$q$R}
  \end{array}\right..
  \label{En}
\end{equation}
$\bullet$ auxiliary functions:
(convention: $\prod\limits_{1\leq j<k\leq M}\!\!\!\!\!*=1$ for $M=0,1$)
\begin{align}
  \varphi(x;\bm{\lambda})&\eqdef
  \frac{\eta(x+1;\bm{\lambda})-\eta(x;\bm{\lambda})}{\eta(1;\bm{\lambda})}
  =\left\{
  \begin{array}{ll}
  {\displaystyle\frac{2x+d+1}{d+1}}&:\text{R}\\[6pt]
  {\displaystyle\frac{q^{-x}-dq^{x+1}}{1-dq}}&:\text{$q$R}
  \end{array}\right.,
  \label{varphi}\\
  \varphi_M(x;\bm{\lambda})&\eqdef\prod_{1\leq j<k\leq M}
  \frac{\eta(x+k-1;\bm{\lambda})-\eta(x+j-1;\bm{\lambda})}
  {\eta(k-j;\bm{\lambda})}\qquad
  (\varphi_0(x)=\varphi_1(x)=1)\n
  &=\prod_{1\leq j<k\leq M}
  \varphi\bigl(x+j-1;\bm{\lambda}+(k-j-1)\bm{\delta}\bigr).
  \label{varphiM}
\end{align}
$\bullet$ potential functions
$B'(x;\bm{\lambda})\eqdef B\bigl(x;\mathfrak{t}(\bm{\lambda})\bigr)$,
$D'(x;\bm{\lambda})\eqdef D\bigl(x;\mathfrak{t}(\bm{\lambda})\bigr)$ :
\begin{align}
  &B'(x;\bm{\lambda})=\left\{
  \begin{array}{ll}
  {\displaystyle
  -\frac{(x+d-a+1)(x+d-b+1)(x+c)(x+d)}{(2x+d)(2x+1+d)}}&:\text{R}\\[8pt]
  {\displaystyle
  -\frac{(1-a^{-1}dq^{x+1})(1-b^{-1}dq^{x+1})(1-cq^x)(1-dq^x)}
  {(1-dq^{2x})(1-dq^{2x+1})}}&:\text{$q$R}
  \end{array}\right.,\n
  &D'(x;\bm{\lambda})=\left\{
  \begin{array}{ll}
  {\displaystyle
  -\frac{(x+a-1)(x+b-1)(x+d-c)x}{(2x-1+d)(2x+d)}}&:\text{R}\\[8pt]
  {\displaystyle-\frac{cdq}{ab}\,
  \frac{(1-aq^{x-1})(1-bq^{x-1})(1-c^{-1}dq^x)(1-q^x)}
  {(1-dq^{2x-1})(1-dq^{2x})}}&:\text{$q$R}
  \end{array}\right..
  \label{B'D'}
\end{align}
$\bullet$ $\alpha(\bm{\lambda})$ and
virtual state energy $\tilde{\mathcal{E}}_{\text{v}}$:
\begin{equation}
  \alpha(\bm{\lambda})=\left\{
  \begin{array}{ll}
  1&:\text{R}\\
  abd^{-1}q^{-1}&:\text{$q$R}
  \end{array}\right.,\quad
  \tilde{\mathcal{E}}_{\text{v}}(\bm{\lambda})=\left\{
  \begin{array}{ll}
  -(c+\text{v})(\tilde{d}-c-\text{v})&:\text{R}\\[2pt]
  -(1-cq^{\text{v}})(1-c^{-1}\tilde{d}q^{-\text{v}})&:\text{$q$R}
  \end{array}\right..
  \label{Etv}
\end{equation}
$\bullet$ $r_j(x_j;\bm{\lambda},M)$ ($1\leq j\leq M+1$):
($x_j\eqdef x+j-1$)
\begin{equation}
  r_j(x_j;\bm{\lambda},M)=\left\{
  \begin{array}{ll}
  {\displaystyle
  \frac{(x+a,x+b)_{j-1}(x+d-a+j,x+d-b+j)_{M+1-j}}
  {(d-a+1,d-b+1)_M}}&:\text{R}\\[10pt]
  {\displaystyle
  \frac{(aq^x,bq^x;q)_{j-1}(a^{-1}dq^{x+j},b^{-1}dq^{x+j};q)_{M+1-j}}
  {(abd^{-1}q^{-1})^{j-1}q^{Mx}(a^{-1}dq,b^{-1}dq;q)_M}}&:\text{$q$R}
  \end{array}\right..
  \label{rj}
\end{equation}
$\bullet$ normalization constants $\mathcal{C}_{\mathcal{D}}(\bm{\lambda})$,
$\mathcal{C}_{\mathcal{D},n}(\bm{\lambda})$,
$\tilde{d}_{\mathcal{D},n}(\bm{\lambda})>0$ and
$d_{\mathcal{D},n}(\bm{\lambda})>0$ :
\begin{align}
  \mathcal{C}_{\mathcal{D}}(\bm{\lambda})&=
  \frac{1}{\varphi_M(0;\bm{\lambda})}
  \prod_{1\leq j<k\leq M}
  \frac{\tilde{\mathcal{E}}_{d_j}(\bm{\lambda})
  -\tilde{\mathcal{E}}_{d_k}(\bm{\lambda})}
  {\alpha(\bm{\lambda})B'(j-1;\bm{\lambda})},
  \label{CD}\\
  \mathcal{C}_{\mathcal{D},n}(\bm{\lambda})&=
  (-1)^M\mathcal{C}_{\mathcal{D}}(\bm{\lambda})
  \tilde{d}_{\mathcal{D},n}(\bm{\lambda})^2,
  \label{CDn}\\
  \tilde{d}_{\mathcal{D},n}(\bm{\lambda})^2&=
  \frac{\varphi_M(0;\bm{\lambda})}{\varphi_{M+1}(0;\bm{\lambda})}
  \prod_{j=1}^M\frac{\mathcal{E}_n(\bm{\lambda})
  -\tilde{\mathcal{E}}_{d_j}(\bm{\lambda})}
  {\alpha(\bm{\lambda})B'(j-1;\bm{\lambda})},
  \label{dtDn}\\
  d_{\mathcal{D},n}(\bm{\lambda})&=
  d_n(\bm{\lambda})\tilde{d}_{\mathcal{D},n}(\bm{\lambda}).
  \label{dDn}
\end{align}
$\bullet$ coefficients of the highest degree term:
\begin{alignat*}{2}
  P_n(\eta;\bm{\lambda})
  &=c_n(\bm{\lambda})\eta^n+(\text{lower order terms}),
  &\ \ P_{\mathcal{D}}(\eta;\bm{\lambda})
  &=c_{\mathcal{D},n}^{P}(\bm{\lambda})\eta^{\ell_{\mathcal{D}}+n}
  +(\text{lower order terms}),\\
  \xi_{\text{v}}(\eta;\bm{\lambda})
  &=\tilde{c}_{\text{v}}(\bm{\lambda})\eta^{\text{v}}
  +(\text{lower order terms}),
  &\ \ \Xi_{\mathcal{D}}(\eta;\bm{\lambda})
  &=c_{\mathcal{D}}^{\Xi}(\bm{\lambda})\eta^{\ell_{\mathcal{D}}}
  +(\text{lower order terms}),
\end{alignat*}
\begin{align}
  c_n(\bm{\lambda})&=\left\{
  \begin{array}{ll}
  {\displaystyle\frac{(\tilde{d}+n)_n}{(a,b,c)_n}}&:\text{R}\\[8pt]
  {\displaystyle\frac{(\tilde{d}q^n;q)_n}{(a,b,c;q)_n}}&:\text{$q$R}
  \end{array}\right.,\quad
  \tilde{c}_{\text{v}}(\bm{\lambda})=\left\{
  \begin{array}{ll}
  {\displaystyle\frac{(c+d-a-b+\text{v}+1)_{\text{v}}}
  {(d-a+1,d-b+1,c)_{\text{v}}}}&:\text{R}\\[8pt]
  {\displaystyle\frac{(a^{-1}b^{-1}cdq^{\text{v}+1};q)_{\text{v}}}
  {(a^{-1}dq,b^{-1}dq,c;q)_{\text{v}}}}&:\text{$q$R}
  \end{array}\right.,\\
  c_{\mathcal{D}}^{\Xi}(\bm{\lambda})&=
  \prod_{j=1}^M\tilde{c}_{d_j}(\bm{\lambda})\times\left\{
  \begin{array}{ll}
  {\displaystyle
  \frac{\prod_{j=1}^M(d-a+1,d-b+1,c)_{j-1}}
  {\prod\limits_{1\leq j<k\leq M}(c+d-a-b+d_j+d_k+1)}}
  &:\text{R}\\[22pt]
  {\displaystyle
  \frac{\prod_{j=1}^M(a^{-1}dq,b^{-1}dq,c;q)_{j-1}}
  {\prod\limits_{1\leq j<k\leq M}(1-a^{-1}b^{-1}cdq^{d_j+d_k+1})}}
  &:\text{$q$R}
  \end{array}\right.,
  \label{cXiD}\\
  c_{\mathcal{D},n}^{P}(\bm{\lambda})&=
  c_{\mathcal{D}}^{\Xi}(\bm{\lambda})c_n(\bm{\lambda})
  \times\left\{
  \begin{array}{ll}
  {\displaystyle\prod_{j=1}^M\frac{c+j-1}{c+d_j+n}}&:\text{R}\\[6pt]
  {\displaystyle\prod_{j=1}^M\frac{1-cq^{j-1}}{1-cq^{d_j+n}}}&:\text{$q$R}
  \end{array}\right..
  \label{cPDn}
\end{align}
$\bullet$ potential functions $B_{\mathcal{D}}(x;\bm{\lambda})$ and
$D_{\mathcal{D}}(x;\bm{\lambda})$ :
\begin{align}
  B_{\mathcal{D}}(x;\bm{\lambda})&=B(x;\bm{\lambda}+M\tilde{\bm{\delta}})\,
  \frac{\check{\Xi}_{\mathcal{D}}(x;\bm{\lambda})}
  {\check{\Xi}_{\mathcal{D}}(x+1;\bm{\lambda})}
  \frac{\check{\Xi}_{\mathcal{D}}(x+1;\bm{\lambda}+\bm{\delta})}
  {\check{\Xi}_{\mathcal{D}}(x;\bm{\lambda}+\bm{\delta})},\n
  D_{\mathcal{D}}(x;\bm{\lambda})&=D(x;\bm{\lambda}+M\tilde{\bm{\delta}})\,
  \frac{\check{\Xi}_{\mathcal{D}}(x+1;\bm{\lambda})}
  {\check{\Xi}_{\mathcal{D}}(x;\bm{\lambda})}
  \frac{\check{\Xi}_{\mathcal{D}}(x-1;\bm{\lambda}+\bm{\delta})}
  {\check{\Xi}_{\mathcal{D}}(x;\bm{\lambda}+\bm{\delta})}.
  \label{BD,DD}
\end{align}
$\bullet$ Casorati determinant (Casoratian) of a set of $n$ functions
$\{f_j(x)\}$ :
\begin{equation}
  \text{W}_{\text{C}}[f_1,f_2,\ldots,f_n](x)
  \eqdef\det\Bigl(f_k(x+j-1)\Bigr)_{1\leq j,k\leq n},
  \label{rdQM:Wdef}
\end{equation}
(for $n=0$, we set $\text{W}_{\text{C}}[\cdot](x)=1$).\\
$\bullet$ potential functions
$\hat{B}_{d_1\ldots d_s}(x;\bm{\lambda})$ and
$\hat{D}_{d_1\ldots d_s}(x;\bm{\lambda})$ :
\begin{align}
 \hat{B}_{d_1\ldots d_s}(x;\bm{\lambda})&\eqdef\alpha B'(x+s-1;\bm{\lambda})
  \frac{\text{W}_{\text{C}}[\check{\xi}_{d_1},\ldots,\check{\xi}_{d_{s-1}}]
  (x;\bm{\lambda})}
  {\text{W}_{\text{C}}[\check{\xi}_{d_1},\ldots,\check{\xi}_{d_{s-1}}]
  (x+1;\bm{\lambda})}\,
  \frac{\text{W}_{\text{C}}[\check{\xi}_{d_1},\ldots,\check{\xi}_{d_s}]
  (x+1;\bm{\lambda})}
  {\text{W}_{\text{C}}[\check{\xi}_{d_1},\ldots,\check{\xi}_{d_s}]
  (x;\bm{\lambda})},\n
  \hat{D}_{d_1\ldots d_s}(x;\bm{\lambda})&\eqdef\alpha D'(x;\bm{\lambda})
  \frac{\text{W}_{\text{C}}[\check{\xi}_{d_1},\ldots,\check{\xi}_{d_{s-1}}]
  (x+1;\bm{\lambda})}
  {\text{W}_{\text{C}}[\check{\xi}_{d_1},\ldots,\check{\xi}_{d_{s-1}}]
  (x;\bm{\lambda})}\,
  \frac{\text{W}_{\text{C}}[\check{\xi}_{d_1},\ldots,\check{\xi}_{d_s}]
  (x-1;\bm{\lambda})}
  {\text{W}_{\text{C}}[\check{\xi}_{d_1},\ldots,\check{\xi}_{d_s}]
  (x;\bm{\lambda})}.
  \label{BdsDdsform}
\end{align}

\section{Proof of Conjecture 2 in Ref.\,\cite{rrmiop2}}
\label{app:proofAW}

In \cite{rrmiop2} we discussed the recurrence relations with constant
coefficients for the multi-indexed Wilson (W) and Askey-Wilson (AW)
polynomials and presented Conjecture 2,
\begin{quote}
{\bf Conjecture 2 in \cite{rrmiop2}}\,
{\it For any polynomial $Y(\eta)$, we take $X(\eta)$ as}\\
\hspace*{15mm}
$X(\eta)=I[\Xi_{\mathcal{D}}Y](\eta),\quad
\text{deg}\,X(\eta)=L=\ell_{\mathcal{D}}+\text{deg}\,Y(\eta)+1.$\\
{\it Then the multi-indexed Wilson and Askey-Wilson polynomials
$P_{\mathcal{D},n}(\eta)$ satisfy $1+2L$ term recurrence relations with
constant coefficients:}\\[2pt]
\hspace*{15mm}
$X(\eta)P_{\mathcal{D},n}(\eta)
=\sum\limits_{k=-L}^Lr_{n,k}^{X,\mathcal{D}}P_{\mathcal{D},n+k}(\eta)
\ \ (\forall n\in\mathbb{Z}_{\geq 0})$.
\end{quote}
Here we prove this conjecture by the same method used in \S\,\ref{sec:a:indet}.
The `step 0' was given in \cite{rrmiop2}.
We follow the notation of \cite{os27,rrmiop2}.
(Many same symbols are used for ($q$-)R and (A)W cases, but all the
quantities used in this appendix correspond to (A)W cases.)

\subsection{Step 1}
\label{app:step_1}

The sinusoidal coordinates are $\eta(x)=x^2$ (W) and $\eta(x)=\cos x$ (AW),
and the parameters $a_i$ ($i=1,2,3,4$) satisfy $\{a_1^*,a_2^*\}=\{a_1,a_2\}$
(as a set) and $\{a_3^*,a_4^*\}=\{a_3,a_4\}$ (as a set).
The denominator polynomials $\Xi_{\mathcal{D}}(\eta)$ and the multi-indexed
(Askey-)Wilson polynomials $P_{\mathcal{D},n}(\eta)$
($n\in\mathbb{Z}_{\geq 0}$) are
\begin{alignat}{3}
  \check{\Xi}_{\mathcal{D}}(x;\bm{\lambda})
  &\eqdef \Xi_{\mathcal{D}}\bigl(\eta(x);\bm{\lambda}\bigr),
  &\deg\Xi_{\mathcal{D}}(\eta)&=\ell_{\mathcal{D}},
  &\Xi_{\mathcal{D}}^*(\eta)&=\Xi_{\mathcal{D}}(\eta),
  \label{AW:XiD}\\
  \check{P}_{\mathcal{D},n}(x;\bm{\lambda})
  &\eqdef P_{\mathcal{D},n}\bigl(\eta(x);\bm{\lambda}\bigr),
  &\ \ \deg P_{\mathcal{D},n}(\eta)&=\ell_{\mathcal{D}}+n,
  &\ \ P_{\mathcal{D},n}^*(\eta;\bm{\lambda})
  &=P_{\mathcal{D},n}(\eta;\bm{\lambda}),
  \label{AW:PDn}
\end{alignat}
and
$P_{\mathcal{D},0}(\eta;\bm{\lambda})\propto
\Xi_{\mathcal{D}}(\eta;\bm{\lambda}+\bm{\delta})$.
Let us define the set of finite linear combinations of
$P_{\mathcal{D},n}(\eta)$, $\mathcal{U}_{\mathcal{D}}\subset\mathbb{C}[\eta]$,
by
\begin{equation}
  \mathcal{U}_{\mathcal{D}}
  \eqdef\text{Span}\{P_{\mathcal{D},n}(\eta)\bigm|n\in\mathbb{Z}_{\geq 0}\}.
  \label{AW:UD}
\end{equation}
Since the degree of $P_{\mathcal{D},n}(\eta)$ is $\ell_{\mathcal{D}}+n$,
it is trivial that $p(\eta)\in\mathcal{U}_{\mathcal{D}}\Rightarrow
\deg\,p\geq\ell_{\mathcal{D}}$, except for $p(\eta)=0$.
The multi-indexed (Askey-)Wilson polynomials
$\check{P}_{\mathcal{D},n}(x)$ with $x\in\mathbb{C}$ satisfy second order
difference equations,
\begin{equation}
  \widetilde{\mathcal{H}}_{\mathcal{D}}(\bm{\lambda})
  \check{P}_{\mathcal{D},n}(x;\bm{\lambda})
  =\mathcal{E}_n(\bm{\lambda})\check{P}_{\mathcal{D},n}(x;\bm{\lambda})
  \ \ (n\in\mathbb{Z}_{\geq 0}),
  \label{AW:cont:tHDcPDn=}
\end{equation}
where $\widetilde{\mathcal{H}}_{\mathcal{D}}(\bm{\lambda})$ is
\begin{align}
  \widetilde{\mathcal{H}}_{\mathcal{D}}(\bm{\lambda})
  &=V(x;\bm{\lambda}^{[M_{\I},M_{\II}]})\,
  \frac{\check{\Xi}_{\mathcal{D}}(x+i\frac{\gamma}{2};\bm{\lambda})}
  {\check{\Xi}_{\mathcal{D}}(x-i\frac{\gamma}{2};\bm{\lambda})}
  \biggl(e^{\gamma p}
  -\frac{\check{\Xi}_{\mathcal{D}}(x-i\gamma;\bm{\lambda}+\bm{\delta})}
  {\check{\Xi}_{\mathcal{D}}(x;\bm{\lambda}+\bm{\delta})}\biggr)\n
  &\quad+V^*(x;\bm{\lambda}^{[M_{\I},M_{\II}]})\,
  \frac{\check{\Xi}_{\mathcal{D}}(x-i\frac{\gamma}{2};\bm{\lambda})}
  {\check{\Xi}_{\mathcal{D}}(x+i\frac{\gamma}{2};\bm{\lambda})}
  \biggl(e^{-\gamma p}
  -\frac{\check{\Xi}_{\mathcal{D}}(x+i\gamma;\bm{\lambda}+\bm{\delta})}
  {\check{\Xi}_{\mathcal{D}}(x;\bm{\lambda}+\bm{\delta})}\biggr).
  \label{AW:conttHD}
\end{align}
For $p(\eta)\in\mathbb{C}[\eta]$,
$\widetilde{\mathcal{H}}_{\mathcal{D}}(\bm{\lambda})$ acts on
$\check{p}(x)\eqdef p\bigl(\eta(x)\bigr)$ as
\begin{align}
  \widetilde{\mathcal{H}}_{\mathcal{D}}(\bm{\lambda})
  \check{p}(x)
  &=V(x;\bm{\lambda}^{[M_{\I},M_{\II}]})\,
  \frac{\check{\Xi}_{\mathcal{D}}(x+i\frac{\gamma}{2};\bm{\lambda})}
  {\check{\Xi}_{\mathcal{D}}(x-i\frac{\gamma}{2};\bm{\lambda})}
  \biggl(\check{p}(x-i\gamma)
  -\frac{\check{\Xi}_{\mathcal{D}}(x-i\gamma;\bm{\lambda}+\bm{\delta})}
  {\check{\Xi}_{\mathcal{D}}(x;\bm{\lambda}+\bm{\delta})}\,
  \check{p}(x)\biggr)\n
  &\quad+V^*(x;\bm{\lambda}^{[M_{\I},M_{\II}]})\,
  \frac{\check{\Xi}_{\mathcal{D}}(x-i\frac{\gamma}{2};\bm{\lambda})}
  {\check{\Xi}_{\mathcal{D}}(x+i\frac{\gamma}{2};\bm{\lambda})}
  \biggl(\check{p}(x+i\gamma)
  -\frac{\check{\Xi}_{\mathcal{D}}(x+i\gamma;\bm{\lambda}+\bm{\delta})}
  {\check{\Xi}_{\mathcal{D}}(x;\bm{\lambda}+\bm{\delta})}\,
  \check{p}(x)\biggr).\!
  \label{AW:conttHDp}
\end{align}
Let zeros of $\Xi_{\mathcal{D}}(\eta;\bm{\lambda})$ and
$\Xi_{\mathcal{D}}(\eta;\bm{\lambda}+\bm{\delta})$ be $\beta^{(\eta)}_j$ and
$\beta^{\prime\,(\eta)}_j$ ($j=1,2,\ldots,\ell_{\mathcal{D}}$), respectively,
which are simple for generic parameters.
We define $\beta_j$ and $\beta'_j$ as
$\beta^{(\eta)}_j=\eta(\beta_j)$ and
$\beta^{\prime\,(\eta)}_j=\eta(\beta'_j)$.
(For $x\in\mathbb{C}$, $\eta=\eta(x)$ are not one-to-one functions, but it
does not cause any problems in the following argument.)

Let us consider the condition such that
$\widetilde{\mathcal{H}}_{\mathcal{D}}(\bm{\lambda})
\check{p}(x)$ \eqref{AW:conttHDp} is a polynomial in $\eta(x)$.
The poles at $x=\beta'_j,\beta_j\pm i\frac{\gamma}{2}$ in \eqref{AW:conttHDp}
should be canceled.
First we consider $x=\beta'_j$.
Since $\check{\Xi}_{\mathcal{D}}(x;\bm{\lambda}+\bm{\delta})
\propto\check{P}_{\mathcal{D},0}(x;\bm{\lambda})$ and
$\check{P}_{\mathcal{D},n}(x;\bm{\lambda})$ ($n>0$) do not have common roots
for generic parameters, \eqref{AW:conttHDp} with
$\check{p}(x)=\check{P}_{\mathcal{D},n}(x)$ implies that the poles at
$x=\beta'_j$ are canceled, namely,
\begin{align}
  &\quad V(\beta'_j;\bm{\lambda}^{[M_{\I},M_{\II}]})\,
  \frac{\check{\Xi}_{\mathcal{D}}(\beta'_j+i\frac{\gamma}{2};\bm{\lambda})}
  {\check{\Xi}_{\mathcal{D}}(\beta'_j-i\frac{\gamma}{2};\bm{\lambda})}\,
  \check{\Xi}_{\mathcal{D}}(\beta'_j-i\gamma;\bm{\lambda}+\bm{\delta})\n
  &+V^*(\beta'_j;\bm{\lambda}^{[M_{\I},M_{\II}]})\,
  \frac{\check{\Xi}_{\mathcal{D}}(\beta'_j-i\frac{\gamma}{2};\bm{\lambda})}
  {\check{\Xi}_{\mathcal{D}}(\beta'_j+i\frac{\gamma}{2};\bm{\lambda})}\,
  \check{\Xi}_{\mathcal{D}}(\beta'_j+i\gamma;\bm{\lambda}+\bm{\delta})
  =0\ \ (j=1,2,\ldots,\ell_{\mathcal{D}}).
  \label{AW:beta'j=0}
\end{align}
This relation implies that we do not need bother the poles at $x=\beta'_j$
in \eqref{AW:conttHDp} for general $p(\eta)$.
Next we consider $x=\beta_j\pm i\frac{\gamma}{2}$. For generic parameters,
$\check{\Xi}_{\mathcal{D}}(x-i\frac{\gamma}{2};\bm{\lambda})$ and
$\check{\Xi}_{\mathcal{D}}(x+i\frac{\gamma}{2};\bm{\lambda})$ do not common
roots,
and the numerators of $V(x;\bm{\lambda}^{[M_{\I},M_{\II}]})$ and
$V^*(x;\bm{\lambda}^{[M_{\I},M_{\II}]})$ do not cancel the poles coming
from $\check{\Xi}_{\mathcal{D}}(x\pm i\frac{\gamma}{2};\bm{\lambda})$,
and zeros of the denominators of $V(x;\bm{\lambda}^{[M_{\I},M_{\II}]})$ and
$V^*(x;\bm{\lambda}^{[M_{\I},M_{\II}]})$ do not coincide with
$\beta_j\pm i\frac{\gamma}{2}$.
The residue of the first term of \eqref{AW:conttHDp} at
$x=\beta_j+i\frac{\gamma}{2}$ is
\begin{equation*}
  V(\beta_j+i\tfrac{\gamma}{2};\bm{\lambda}^{[M_{\I},M_{\II}]})\,
  \frac{\check{\Xi}_{\mathcal{D}}(\beta_j+i\gamma;\bm{\lambda})}
  {\frac{d}{dx}\check{\Xi}_{\mathcal{D}}(x-i\frac{\gamma}{2};\bm{\lambda})
  |_{x=\beta_j+i\frac{\gamma}{2}}}
  \biggl(\check{p}(\beta_j-i\tfrac{\gamma}{2})
  -\frac{\check{\Xi}_{\mathcal{D}}(\beta_j-i\frac{\gamma}{2};
  \bm{\lambda}+\bm{\delta})}
  {\check{\Xi}_{\mathcal{D}}(\beta_j+i\frac{\gamma}{2};
  \bm{\lambda}+\bm{\delta})}\,
  \check{p}(\beta_j+i\tfrac{\gamma}{2})\biggr),
\end{equation*}
and that of the second term of \eqref{AW:conttHDp} at
$x=\beta_j-i\frac{\gamma}{2}$ is
\begin{equation*}
  V^*(\beta_j-i\tfrac{\gamma}{2};\bm{\lambda}^{[M_{\I},M_{\II}]})\,
  \frac{\check{\Xi}_{\mathcal{D}}(\beta_j-i\gamma;\bm{\lambda})}
  {\frac{d}{dx}\check{\Xi}_{\mathcal{D}}(x+i\frac{\gamma}{2};\bm{\lambda})
  |_{x=\beta_j-i\frac{\gamma}{2}}}
  \biggl(\check{p}(\beta_j+i\tfrac{\gamma}{2})
  -\frac{\check{\Xi}_{\mathcal{D}}(\beta_j+i\frac{\gamma}{2};
  \bm{\lambda}+\bm{\delta})}
  {\check{\Xi}_{\mathcal{D}}(\beta_j-i\frac{\gamma}{2};
  \bm{\lambda}+\bm{\delta})}\,
  \check{p}(\beta_j-i\tfrac{\gamma}{2})\biggr).
\end{equation*}
These residues should be vanished. So we obtain the conditions:
\begin{equation}
  \frac{\check{\Xi}_{\mathcal{D}}(\beta_j-i\frac{\gamma}{2};
  \bm{\lambda}+\bm{\delta})}
  {\check{\Xi}_{\mathcal{D}}(\beta_j+i\frac{\gamma}{2};
  \bm{\lambda}+\bm{\delta})}\,
  \check{p}(\beta_j+i\tfrac{\gamma}{2})
  =\check{p}(\beta_j-i\tfrac{\gamma}{2})
  \ \ (j=1,2,\ldots,\ell_{\mathcal{D}}).
  \label{AW:condcp}
\end{equation}

Let us assume $\deg p(\eta)<\ell_{\mathcal{D}}$.
Without loss of generality, we take $p(\eta)$ is a monic polynomial.
Then the number of adjustable coefficients of $p(\eta)$ is $\deg p(\eta)$.
On the other hand, the number of conditions \eqref{AW:condcp} is
$\ell_{\mathcal{D}}$. Therefore the conditions \eqref{AW:condcp} can not be
satisfied for generic parameters, except for $p(\eta)=0$.

Since any polynomial $p(\eta)$ is expanded as
\begin{equation*}
  p(\eta)=\!\!\!\sum_{n=0}^{\deg p-\ell_{\mathcal{D}}}\!\!\!
  a_nP_{\mathcal{D},n}(\eta)+r(\eta),
  \ \ \deg r(\eta)<\ell_{\mathcal{D}}
  \ \ \bigl(\text{$p(\eta)=r(\eta)$ for $\deg\,p<\ell_{\mathcal{D}}$}\bigr),
\end{equation*}
we have
\begin{align*}
  &\phantom{\Leftrightarrow\ \,}
  \text{$\widetilde{\mathcal{H}}_{\mathcal{D}}(\bm{\lambda})
  \check{p}(x)$ : a polynomial in $\eta(x)$}\\
  &\Leftrightarrow
  \text{$\widetilde{\mathcal{H}}_{\mathcal{D}}(\bm{\lambda})
  \check{r}(x)$ : a polynomial in $\eta(x)$}
  \Leftrightarrow
  r(\eta)=0.
\end{align*}
Therefore we obtain the following proposition:

\begin{prop}
\label{AW:prop:pinUD}
For $p(\eta)\in\mathbb{C}[\eta]$, the following holds:
\begin{equation}
  p(\eta)\in\mathcal{U}_{\mathcal{D}}\Leftrightarrow
  \widetilde{\mathcal{H}}_{\mathcal{D}}(\bm{\lambda})
  \check{p}(x):\text{a polynomial in $\eta(x)$}.
  \label{AW:pinUD}
\end{equation}
\end{prop}

\subsection{Step 2}
\label{app:step_2}

Let us consider a polynomial $X(\eta)$ giving the following
recurrence relations with constant coefficients,
\begin{equation}
  X(\eta)P_{\mathcal{D},n}(\eta)
  =\sum\limits_{k=-L}^Lr_{n,k}^{X,\mathcal{D}}P_{\mathcal{D},n+k}(\eta)
  \ \ (\forall n\in\mathbb{Z}_{\geq 0}),
  \label{AW:XP}
\end{equation}
where $P_{\mathcal{D},n}(\eta)=0$ ($n<0$).
For $X(\eta)$, $\check{X}(x)$ is defined by
\begin{equation}
  \check{X}(x)\eqdef X\bigl(\eta(x)\bigr).
  \label{AW:cX}
\end{equation}
{}From Proposition\,\ref{AW:prop:pinUD}, $X(\eta)$ in \eqref{AW:XP} should
satisfy
\begin{equation*}
  \text{$\widetilde{\mathcal{H}}_{\mathcal{D}}(\bm{\lambda})
  \bigl(\check{X}(x)\check{P}_{\mathcal{D},n}(x;\bm{\lambda})\bigr)$
  : a polynomial in $\eta(x)$}.
\end{equation*}
Action of $\widetilde{\mathcal{H}}_{\mathcal{D}}(\bm{\lambda})$
on $\check{X}(x)\check{P}_{\mathcal{D},n}(x;\bm{\lambda})$ is
\begin{align*}
  &\quad\widetilde{\mathcal{H}}_{\mathcal{D}}(\bm{\lambda})
  \bigl(\check{X}(x)\check{P}_{\mathcal{D},n}(x;\bm{\lambda})\bigr)\\
  &=\check{X}(x)
  \widetilde{\mathcal{H}}_{\mathcal{D}}(\bm{\lambda})
  \check{P}_{\mathcal{D},n}(x;\bm{\lambda})\n
  &\quad+V(x;\bm{\lambda}^{[M_{\I},M_{\II}]})\,
  \frac{\check{\Xi}_{\mathcal{D}}(x+i\frac{\gamma}{2};\bm{\lambda})}
  {\check{\Xi}_{\mathcal{D}}(x-i\frac{\gamma}{2};\bm{\lambda})}
  \bigl(\check{X}(x-i\gamma)-\check{X}(x)\bigr)
  \check{P}_{\mathcal{D},n}(x-i\gamma;\bm{\lambda})\\
  &\quad+V^*(x;\bm{\lambda}^{[M_{\I},M_{\II}]})\,
  \frac{\check{\Xi}_{\mathcal{D}}(x-i\frac{\gamma}{2};\bm{\lambda})}
  {\check{\Xi}_{\mathcal{D}}(x+i\frac{\gamma}{2};\bm{\lambda})}
  \bigl(\check{X}(x+i\gamma)-\check{X}(x)\bigr)
  \check{P}_{\mathcal{D},n}(x+i\gamma;\bm{\lambda}),
\end{align*}
namely,
\begin{equation}
  \widetilde{\mathcal{H}}_{\mathcal{D}}(\bm{\lambda})
  \bigl(\check{X}(x)\check{P}_{\mathcal{D},n}(x;\bm{\lambda})\bigr)
  =\mathcal{E}_n(\bm{\lambda})\check{X}(x)
  \check{P}_{\mathcal{D},n}(x;\bm{\lambda})+F(x).
  \label{AW:tHDcXcPDn}
\end{equation}
Here $F(x)$ is
\begin{align}
  F(x)&=V(x;\bm{\lambda}^{[M_{\I},M_{\II}]})\,
  \frac{\check{\Xi}_{\mathcal{D}}(x+i\frac{\gamma}{2};\bm{\lambda})}
  {\check{\Xi}_{\mathcal{D}}(x-i\frac{\gamma}{2};\bm{\lambda})}
  \bigl(\check{X}(x-i\gamma)-\check{X}(x)\bigr)
  \check{P}_{\mathcal{D},n}(x-i\gamma;\bm{\lambda})\n
  &\quad+V^*(x;\bm{\lambda}^{[M_{\I},M_{\II}]})\,
  \frac{\check{\Xi}_{\mathcal{D}}(x-i\frac{\gamma}{2};\bm{\lambda})}
  {\check{\Xi}_{\mathcal{D}}(x+i\frac{\gamma}{2};\bm{\lambda})}
  \bigl(\check{X}(x+i\gamma)-\check{X}(x)\bigr)
  \check{P}_{\mathcal{D},n}(x+i\gamma;\bm{\lambda}).
  \label{AW:F}
\end{align}
Equations (3.6) in \cite{rrmiop2} and \eqref{AW:cX} imply
\begin{equation*}
  \check{X}(x-i\gamma)-\check{X}(x)
  =\bigl(\eta(x-i\gamma)-\eta(x)\bigr)
  \times\bigl(\text{a polynomial in $\eta(x-i\tfrac{\gamma}{2})$}\bigr).
\end{equation*}
In order to cancel the zeros of
$\check{\Xi}_{\mathcal{D}}(x-i\frac{\gamma}{2};\bm{\lambda})
=\Xi_{\mathcal{D}}\bigl(\eta(x-i\frac{\gamma}{2});\bm{\lambda}\bigr)$
in \eqref{AW:F}, the polynomial appeared in the above
expression should have the following form,
\begin{equation}
  \check{X}(x-i\gamma)-\check{X}(x)
  =\bigl(\eta(x-i\gamma)-\eta(x)\bigr)
  \check{\Xi}_{\mathcal{D}}(x-i\tfrac{\gamma}{2};\bm{\lambda})
  Y\bigl(\eta(x-i\tfrac{\gamma}{2})\bigl),
  \label{AW:X-X}
\end{equation}
where $Y(\eta)$ is an arbitrary polynomial in $\eta$.
Note that this $X(\eta)$ can be expressed in terms of the map
$I$ eq.(3.10) in \cite{rrmiop2} by eq.(3.12) in \cite{rrmiop2},
\begin{equation}
  X(\eta)=I\bigl[\Xi_{\mathcal{D}}Y\bigr](\eta).
\end{equation}
Then $F(x)$ \eqref{AW:F} becomes
\begin{align}
  F(x)&=V(x;\bm{\lambda}^{[M_{\I},M_{\II}]})
  \bigl(\eta(x-i\gamma)-\eta(x)\bigr)
  \check{\Xi}_{\mathcal{D}}(x+i\tfrac{\gamma}{2};\bm{\lambda})
  Y\bigl(\eta(x-i\tfrac{\gamma}{2})\bigl)
  \check{P}_{\mathcal{D},n}(x-i\gamma;\bm{\lambda})
  \label{AW:F2}\\
  &\quad+V^*(x;\bm{\lambda}^{[M_{\I},M_{\II}]})
  \bigl(\eta(x+i\gamma)-\eta(x)\bigr)
  \check{\Xi}_{\mathcal{D}}(x-i\tfrac{\gamma}{2};\bm{\lambda})
  Y\bigl(\eta(x+i\tfrac{\gamma}{2})\bigl)
  \check{P}_{\mathcal{D},n}(x+i\gamma;\bm{\lambda}).
  \nonumber
\end{align}
{}From the explicit forms of $V(x;\bm{\lambda})$, we have
\begin{align}
  &\quad V(x;\bm{\lambda}^{[M_{\I},M_{\II}]})
  \bigl(\eta(x-i\gamma)-\eta(x)\bigr)\n
  &=\left\{
  \begin{array}{ll}
  {\displaystyle
  -\frac{\prod_{j=1}^2(a_j-\frac12M'+ix)\cdot
  \prod_{j=3}^4(a_j+\frac12M'+ix)}{2ix}}&:\text{W}\\[10pt]
  {\displaystyle
  -\frac{(q^{-1}-1)\prod_{j=1}^2(1-a_jq^{-\frac12M'}e^{ix})
  \cdot\prod_{j=3}^4(1-a_jq^{\frac12M'}e^{ix})}
  {2e^{2ix}(e^{ix}-e^{-ix})}}&:\text{AW}
  \end{array}\right.,
  \label{AW:B(eta-eta)}\\[4pt]
  &\quad V^*(x;\bm{\lambda}^{[M_{\I},M_{\II}]})
  \bigl(\eta(x+i\gamma)-\eta(x)\bigr)\n
  &=\left\{
  \begin{array}{ll}
  {\displaystyle
  \frac{\prod_{j=1}^2(a_j-\frac12M'-ix)\cdot
  \prod_{j=3}^4(a_j+\frac12M'-ix)}{2ix}}&:\text{W}\\[10pt]
  {\displaystyle
  \frac{(q^{-1}-1)\prod_{j=1}^2(1-a_jq^{-\frac12M'}e^{-ix})
  \cdot\prod_{j=3}^4(1-a_jq^{\frac12M'}e^{-ix})}
  {2e^{-2ix}(e^{ix}-e^{-ix})}}&:\text{AW}
  \end{array}\right.,
  \label{AW:D(eta-eta)}
\end{align}
where $M'=M_{\I}-M_{\II}$.
For AW case, they are rational functions of $z=e^{ix}$.
Residues of \eqref{AW:B(eta-eta)}--\eqref{AW:D(eta-eta)} at $x=0$ (W) or
$z=\pm1$ (AW) are related as
\begin{align*}
  \text{W}:\ \ &
  \text{Res}_{x=0}\Bigl(V(x;\bm{\lambda}^{[M_{\I},M_{\II}]})
  \bigl(\eta(x-i\gamma)-\eta(x)\bigr)\Bigr)\\
  &=-\text{Res}_{x=0}\Bigl(V^*(x;\bm{\lambda}^{[M_{\I},M_{\II}]})
  \bigl(\eta(x+i\gamma)-\eta(x)\bigr)\Bigr),\\
  \text{AW}:\ \ &
  \text{Res}_{z=\pm1}\Bigl(V(x;\bm{\lambda}^{[M_{\I},M_{\II}]})
  \bigl(\eta(x-i\gamma)-\eta(x)\bigr)\Bigr)\\
  &=-\text{Res}_{z=\pm1}\Bigl(V^*(x;\bm{\lambda}^{[M_{\I},M_{\II}]})
  \bigl(\eta(x+i\gamma)-\eta(x)\bigr)\Bigr).
\end{align*}
At $x=0$ or $z=\pm1$, we have
\begin{alignat*}{3}
  \text{W}:\ \ &&\eta(x-i\tfrac{\gamma}{2})\bigl|_{x=0}
  &=\eta(x+i\tfrac{\gamma}{2})\bigl|_{x=0}\,,\quad
  &\eta(x-i\gamma)\bigl|_{x=0}&=\eta(x+i\gamma)\bigl|_{x=0}\,,\\
  \text{AW}:\ \ &&\eta(x-i\tfrac{\gamma}{2})\bigl|_{z=\pm1}
  &=\eta(x+i\tfrac{\gamma}{2})\bigl|_{z=\pm1},\quad
  &\eta(x-i\gamma)\bigl|_{z=\pm1}&=\eta(x+i\gamma)\bigl|_{z=\pm1}.
\end{alignat*}
Combining these and \eqref{AW:F2}, we obtain
\begin{equation*}
  \text{W}:\ \text{Res}_{x=0}F(x)=0,\quad
  \text{AW}:\ \text{Res}_{z=\pm1}F(x)=0.
\end{equation*}
Therefore $F(x)$ \eqref{AW:F2} is a polynomial in $x$ for W, a Laurent
polynomial in $z$ for AW.
By introducing an involution $\mathcal{I}:x\to-x$ ($\Rightarrow z\to z^{-1}$),
we have
\begin{align*}
  &\mathcal{I}\Bigl(V(x;\bm{\lambda}^{[M_{\I},M_{\II}]})
  \bigl(\eta(x-i\gamma)-\eta(x)\bigr)\Bigr)
  =V^*(x;\bm{\lambda}^{[M_{\I},M_{\II}]})
  \bigl(\eta(x+i\gamma)-\eta(x)\bigr),\\
  &\mathcal{I}\bigl(\eta(x-i\tfrac{\gamma}{2})\bigr)
  =\eta(x+i\tfrac{\gamma}{2}),\quad
  \mathcal{I}\bigl(\eta(x-i\gamma)\bigr)=\eta(x+i\gamma).
\end{align*}
Hence $F(x)$ \eqref{AW:F2} satisfies $\mathcal{I}\bigl(F(x)\bigr)=F(x)$,
which implies that $F(x)$ is a polynomial in $\eta(x)$.
Therefore, from \eqref{AW:tHDcXcPDn}, we have shown that
$\widetilde{\mathcal{H}}_{\mathcal{D}}(\bm{\lambda})
\bigl(\check{X}(x)\check{P}_{\mathcal{D},n}(x;\bm{\lambda})\bigr)$
is a polynomial in $\eta(x)$.

\subsection{Step 3}
\label{app:step_3}

Let us summarize the result.
For the denominator polynomial
$\Xi_{\mathcal{D}}(\eta)=\Xi_{\mathcal{D}}(\eta;\bm{\lambda})$ and
a polynomial in $\eta$, $Y(\eta)(\neq 0)$, we set $X(\eta)=X(\eta;\bm{\lambda})
=X^{\mathcal{D},Y}(\eta;\bm{\lambda})$ as
\begin{equation}
  X(\eta)=I\bigl[\Xi_{\mathcal{D}}Y\bigr](\eta),\quad
  \deg X(\eta)=L=\ell_{\mathcal{D}}+\deg Y(\eta)+1,
  \label{AW:X=I[XiY]}
\end{equation}
where $\Xi_{\mathcal{D}}Y$ means a polynomial
$(\Xi_{\mathcal{D}}Y)(\eta)=\Xi_{\mathcal{D}}(\eta)Y(\eta)$.
Note that $L\geq M+1$ because of $\ell_{\mathcal{D}}\geq M$.
The minimal degree one, which corresponds to $Y(\eta)=1$, is
\begin{equation}
  X_{\text{min}}(\eta)=I\bigl[\Xi_{\mathcal{D}}\bigr](\eta),\quad
  \deg X_{\text{min}}(\eta)=\ell_{\mathcal{D}}+1.
  \label{AW:Xmin}
\end{equation}
Then we have the following theorem.
\begin{thm}
\label{AW:thm:rr_indet_a}
For any polynomial $Y(\eta)(\neq 0)$, we take
$X(\eta)=X^{\mathcal{D},Y}(\eta)$ as \eqref{AW:X=I[XiY]}.
Then the multi-indexed (Askey-)Wilson polynomials $P_{\mathcal{D},n}(\eta)$
satisfy $1+2L$ term recurrence relations with constant coefficients:
\begin{equation}
  X(\eta)P_{\mathcal{D},n}(\eta)
  =\sum_{k=-L}^Lr_{n,k}^{X,\mathcal{D}}P_{\mathcal{D},n+k}(\eta)
  \ \ (n\in\mathbb{Z}_{\geq 0}).
  \label{AW:XPthm}
\end{equation}
\end{thm}
{\bf Remark 1}\,
We have assumed the convention $P_{\mathcal{D},n}(\eta)=0$ ($n<0$).
If we replace $\sum\limits_{k=-L}^L$ with $\sum\limits_{k=-\min(L,n)}^L$,
it is unnecessary.

\vspace*{1mm}
\noindent
{\bf Remark 2}\,
As shown near \eqref{AW:X-X}, any polynomial $X(\eta)$ giving the recurrence
relations with constant coefficients must have the form \eqref{AW:X=I[XiY]}.

\noindent
{\bf Remark 3}\,
If $Y(\eta)$ satisfies $Y^*(\eta)=Y(\eta)$, we have $X^*(\eta)=X(\eta)$ and
$r_{n,k}^{X,\mathcal{D}\,*}=r_{n,k}^{X,\mathcal{D}}$.

\section{$r_{n,0}^{X,\mathcal{D}}$ in (B.11) and (B.12) of
Ref.\,\cite{rrmiop2}}
\label{app:exAW}

In \cite{rrmiop2} we discussed the recurrence relations with constant
coefficients for the multi-indexed Wilson (W) and Askey-Wilson (AW)
polynomials.
As examples, the explicit forms of $r_{n,k}^{X,\mathcal{D}}$ for
$\mathcal{D}=\{1^{\I}\}$ (type $\I$) and $X(\eta)=X_{\text{min}}(\eta)$
are presented in Appendix B.3 and B.4 of \cite{rrmiop2}.
However, we did not write down $r_{n,0}^{X,\mathcal{D}}$ explicitly due to
their lengthy expressions.
Here we present concise expressions of $r_{n,0}^{X,\mathcal{D}}$.
We follow the notation of \cite{rrmiop2}.

For W, $r_{n,0}^{X,\mathcal{D}}$ in eq.(B.11) of \cite{rrmiop2} is
expressed as
\begin{align}
  r_{n,0}^{X,\mathcal{D}}&=X_0
  -\frac{\sigma_1+n}{\sigma_1+n-2}
  (\sigma'_1+n+3)(\sigma'_1+n)
  \prod_{j=1}^2(a_j+a_4+n)_2\cdot
  r_{n,2}^{X,\mathcal{D}}\n
  &\quad-\frac{\sigma_1+n-1}{\sigma_1+n-2}
  \frac{\sigma'_1+n+2}{\sigma'_1+n+1}(\sigma'_1+n)
  \prod_{j=1}^2(a_j+a_4+n)\cdot
  r_{n,1}^{X,\mathcal{D}}\n
  &\quad-\frac{\sigma_1+n-3}{\sigma_1+n-2}
  \frac{\sigma'_1+n}{\sigma'_1+n+1}
  \frac{r_{n,-1}^{X,\mathcal{D}}}
  {(\sigma'_1+n-1)\prod_{j=1}^2(a_j+a_4+n-1)}\n
  &\quad-\frac{\sigma_1+n-4}{\sigma_1+n-2}
  \frac{r_{n,-2}^{X,\mathcal{D}}}
  {(\sigma'_1+n+1)(\sigma'_1+n-2)
  \prod_{j=1}^2(a_j+a_4+n-2)_2},
\end{align}
where $X_0$ is given by
\begin{align}
  X_0&\eqdef X\bigl(-(a_4+\tfrac12)^2\bigr)\n
  &=\frac{1}{32}(1+2a_4)^2
  \Bigl(2-7\sigma'_1-20\sigma'_2-8a_4-4(3+\sigma'_1)a_4^2\n
  &\phantom{=\frac{1}{32}(1+2a_4)^2\Bigl(}
  +\bigl(8(\sigma'_1+\sigma'_2)-1+4a_4(1+a_4)\bigr)\sigma_1
  -8\sigma'_1\sigma_2\Bigr).
\end{align}
For AW, $r_{n,0}^{X,\mathcal{D}}$ in eq.(B.12) of \cite{rrmiop2} is
expressed as
\begin{align}
  r_{n,0}^{X,\mathcal{D}}&=X_0
  -a_4^{-2}q^{-2}\frac{1-\sigma_2q^n}{1-\sigma_2q^{n-2}}
  (1-\sigma'_2q^{n+3})(1-\sigma'_2q^n)
  \prod_{j=1}^2(a_ja_4q^n;q)_2\cdot
  r_{n,2}^{X,\mathcal{D}}\n
  &\quad-a_4^{-1}q^{-1}\frac{1-\sigma_2q^{n-1}}{1-\sigma_2q^{n-2}}
  \frac{1-\sigma'_2q^{n+2}}{1-\sigma'_2q^{n+1}}(1-\sigma'_2q^n)
  \prod_{j=1}^2(1-a_ja_4q^n)\cdot
  r_{n,1}^{X,\mathcal{D}}\n
  &\quad-a_4q\frac{1-\sigma_2q^{n-3}}{1-\sigma_2q^{n-2}}
  \frac{1-\sigma'_2q^n}{1-\sigma'_2q^{n+1}}
  \frac{r_{n,-1}^{X,\mathcal{D}}}
  {(1-\sigma'_2q^{n-1})\prod_{j=1}^2(1-a_ja_4q^{n-1})}\n
  &\quad-a_4^2q^2\frac{1-\sigma_2q^{n-4}}{1-\sigma_2q^{n-2}}
  \frac{r_{n,-2}^{X,\mathcal{D}}}
  {(1-\sigma'_2q^{n+1})(1-\sigma'_2q^{n-2})
  \prod_{j=1}^2(a_ja_4q^{n-2};q)_2},
\end{align}
where $X_0$ is given by
\begin{align}
  X_0&\eqdef X\bigl(\tfrac12(a_4q^{\frac12}+a_4^{-1}q^{-\frac12})\bigr)\n
  &=\frac{1+a_4^2q}{2q^{\frac12}(1+q)\sigma_2a_4^2}
  \Bigl(\sigma'_1q^3a_4+(1-\sigma'_2q)a_4^2q^2
  -q(1+q)(1-\sigma'_2)\sigma_1a_4\n
  &\phantom{=\frac{1+a_4^2q}{2q^{\frac12}(1+q)\sigma_2a_4^2}\Bigl(}
  +\bigl(1-a_4^2-(1+q)\sigma'_2\bigr)\sigma_2\Bigr).
\end{align}
These expressions are obtained by the following simple observation.
Substituting some specific value $\eta_0$ for $\eta$ in the recurrence
relations, we have
$X(\eta_0)P_{\mathcal{D},n}(\eta_0)
=\sum\limits_{k=-L}^Lr_{n,k}^{X,\mathcal{D}}P_{\mathcal{D},n+k}(\eta_0)$,
which gives
\begin{equation}
  r_{n,0}^{X,\mathcal{D}}=X(\eta_0)
  -\sum_{\genfrac{}{}{0pt}{1}{k=-L}{k\neq 0}}^L
  \frac{P_{\mathcal{D},n+k}(\eta_0)}{P_{\mathcal{D},n}(\eta_0)}
  r_{n,k}^{X,\mathcal{D}}.
\end{equation}

Let the index set $\mathcal{D}$ be
$\mathcal{D}=\{d_1,d_2,\ldots,d_M\}=\{d^{\I}_1,\ldots,d^{\I}_{M_{\I}},
d^{\II}_1,\ldots,d^{\II}_{M_{\II}}\}$
($0\leq d^{\I}_1<\cdots<d^{\I}_{M_{\I}}$,
$0\leq d^{\II}_1<\cdots<d^{\II}_{M_{\II}}$,
$M=M_{\I}+M_{\II}$).
Let $x_0$ and $\eta_0$ be
\begin{align}
  x_0&\eqdef -i\gamma\bigl(\lambda_4+\tfrac12(M_{\I}-M_{\II})\bigr),\n
  \eta_0&\eqdef\eta(x_0)=\left\{
  \begin{array}{ll}
  -\bigl(a_4+\frac12(M_{\I}-M_{\II})\bigr)^2&:\text{W}\\[2pt]
  \frac12(a_4q^{\frac12(M_{\I}-M_{\II})}+a_4^{-1}q^{-\frac12(M_{\I}-M_{\II})})
  &:\text{AW}
  \end{array}\right..
\end{align}
Note that, as coordinates $x$ and $\eta$, these values $x_0$ and $\eta_0$ are
unphysical ($x_0$ is imaginary, $\eta_0$ is out of the range of $\eta$).
The multi-indexed (Askey-)Wilson polynomials
$P_{\mathcal{D},n}(\eta)$ take `simple' values at these `unphysical' values
$\eta_0$:
\begin{align}
  &\quad P_{\mathcal{D},n}(\eta_0)\ \ :\text{W}\n
  &=(-1)^{\ell_{\mathcal{D}}+n}c^P_{\mathcal{D},n}(\bm{\lambda})\n
  &\quad\times
  \prod_{j=1}^{M_{\I}}\frac{(a_4-a_1+1,a_4-a_2+1,a_3+a_4)_{d^{\I}_j}}
  {(a_3+a_4-a_1-a_2+d^{\I}_j+1)_{d^{\I}_j}}\cdot
  \frac{\prod\limits_{1\leq j<k\leq M_{\I}}
  (a_3+a_4-a_1-a_2+d^{\I}_j+d^{\I}_k+1)}
  {\prod\limits_{j=1}^{M_{\I}}(a_4-a_1+1,a_4-a_2+1,a_3+a_4)_{j-1}}\n
  &\quad\times
  \prod_{j=1}^{M_{\II}}\frac{(a_1-a_4+1,a_2-a_4+1,1-a_3-a_4)_{d^{\II}_j}}
  {(a_1+a_2-a_3-a_4+d^{\II}_j+1)_{d^{\II}_j}}\cdot
  \frac{\prod\limits_{1\leq j<k\leq M_{\II}}
  (a_1+a_2-a_3-a_4+d^{\II}_j+d^{\II}_k+1)}
  {\prod\limits_{j=1}^{M_{\II}}(a_1-a_4+1,a_2-a_4+1,1-a_3-a_4)_{j-1}}\n
  &\quad\times
  \prod_{\genfrac{}{}{0pt}{1}{1\leq j\leq M_{\I}}{1\leq k\leq M_{\II}}}
  \frac{(a_4-a_1+j-k)(a_4-a_2+j-k)(a_3+a_4+j-k)}
  {a_3+a_4-a_1-a_2+d^{\I}_j-d^{\II}_k}\n
  &\quad\times
  \frac{(a_1+a_4,a_2+a_4,a_3+a_4)_n}{(a_1+a_2+a_3+a_4+n-1)_n}
  \prod_{j=1}^{M_{\I}}\frac{a_3+a_4+d^{\I}_j+n}{a_3+a_4+j-1}\cdot
  \prod_{j=1}^{M_{\II}}\frac{d^{\II}_j+1-a_3-a_4}{d^{\II}_j+1-n-a_3-a_4},\\
  &\quad P_{\mathcal{D},n}(\eta_0)\ \ :\text{AW}\n
  &=\bigl(2a_4q^{\frac12(M_{\I}-M_{\II})}\bigr)^{-\ell_{\mathcal{D}}-n}
  c^P_{\mathcal{D},n}(\bm{\lambda})\n
  &\quad\times
  \prod_{j=1}^{M_{\I}}\frac{(a_1^{-1}a_4q,a_2^{-1}a_4q,a_3a_4;q)_{d^{\I}_j}}
  {(a_1^{-1}a_2^{-1}a_3a_4q^{d^{\I}_j+1};q)_{d^{\I}_j}}\cdot
  \frac{\prod\limits_{1\leq j<k\leq M_{\I}}
  (1-a_1^{-1}a_2^{-1}a_3a_4q^{d^{\I}_j+d^{\I}_k+1})}
  {\prod\limits_{j=1}^{M_{\I}}(a_1^{-1}a_4q,a_2^{-1}a_4q,a_3a_4;q)_{j-1}}\n
  &\quad\times
  \prod_{j=1}^{M_{\II}}\frac{(a_1a_4^{-1}q,a_2a_4^{-1}q,
  a_3^{-1}a_4^{-1}q;q)_{d^{\II}_j}}
  {(a_1a_2a_3^{-1}a_4^{-1}q^{d^{\II}_j+1};q)_{d^{\II}_j}}\cdot
  \frac{\prod\limits_{1\leq j<k\leq M_{\II}}
  (1-a_1a_2a_3^{-1}a_4^{-1}q^{d^{\II}_j+d^{\II}_k+1})} 
  {\prod\limits_{j=1}^{M_{\II}}(a_1a_4^{-1}q,a_2a_4^{-1}q,
  a_3^{-1}a_4^{-1}q;q)_{j-1}}\n
  &\quad\times
  \bigl(a_4q^{\frac12(M_{\I}-M_{\II})}
  \bigr)^{2\sum_{j=1}^{M_{\II}}d^{\II}_j-M_{\II}(M_{\II}-1)}
  q^{-M_{\I}M_{\II}(M_{\II}-1)}\n
  &\quad\times
  \prod_{\genfrac{}{}{0pt}{1}{1\leq j\leq M_{\I}}{1\leq k\leq M_{\II}}}
  \frac{(a_4q^{j-1}-a_1q^{k-1})(a_4q^{j-1}-a_2q^{k-1})
  (a_3a_4q^{j-1}-q^{k-1})}
  {a_3a_4q^{d^{\I}_j}-a_1a_2q^{d^{\II}_k}}\n
  &\quad\times
  q^{-M_{\II}n}
  \frac{(a_1a_4,a_2a_4,a_3a_4;q)_n}{(a_1a_2a_3a_4q^{n-1};q)_n}
  \prod_{j=1}^{M_{\I}}\frac{1-a_3a_4q^{d^{\I}_j+n}}{1-a_3a_4q^{j-1}}\cdot
  \prod_{j=1}^{M_{\II}}\frac{1-a_3^{-1}a_4^{-1}q^{d^{\II}_j+1}}
  {1-a_3^{-1}a_4^{-1}q^{d^{\II}_j+1-n}},
\end{align}
where $c^P_{\mathcal{D},n}(\bm{\lambda})$ is given by eq.(A.7) in \cite{os27}
and $\ell_{\mathcal{D}}=\sum_{j=1}^Md_j-\frac12M(M-1)+2M_{\I}M_{\II}$.
For $M_{\II}=0$ (type $\I$ only), this is a consequence of
\eqref{miop:R=W}--\eqref{miop:qR=AW}.


\end{document}